\documentclass[twocolumn]{aastex7}
\providecommand{\sorthelp}[1]{}
\usepackage{amsmath}
\usepackage[T1]{fontenc}

\begin{document}

\title{Full-sky Models of Galactic Microwave Emission and Polarization at Sub-arcminute Scales for the Python Sky Model}
\shorttitle{Full-Sky Models of Galactic Microwave Emission and Polarization}
\shortauthors{The PanEx GS Group}

\correspondingauthor{Giuseppe Puglisi}
\email{Giuseppe.puglisi@dfa.unict.it, \\ susanclark@stanford.edu, bhensley@jpl.nasa.gov}

\collaboration{100}{The Pan-Experiment Galactic Science Group}

\author[0000-0001-5104-7122]{Julian Borrill}
\affiliation{Computational Cosmology Center \& Physics Division, Lawrence Berkeley National Laboratory, 1 Cyclotron Road, Berkeley, CA 94720, USA}
\affiliation{Space Sciences Laboratory, University of California at Berkeley, 7 Gauss Way, Berkeley, CA 94720, USA}
\affiliation{Department of Physics and Astronomy, University of Pennsylvania, 209 South 33rd Street, Philadelphia, PA 19104, USA}
\email{jdborrill@lbl.gov}

\author[0000-0002-7633-3376]{Susan E. Clark}
\affiliation{Department of Physics, Stanford University, Stanford, CA 94305, USA}
\affiliation{Kavli Institute for Particle Astrophysics \& Cosmology, P.O. Box 2450, Stanford University, Stanford, CA 94305, USA}
\email{susanclark@stanford.edu}

\author[0000-0002-7217-4689]{Jacques Delabrouille}
\affiliation{CNRS-UCB International Research Laboratory, Centre Pierre Bin\'etruy, IRL 2007, CPB-IN2P3, Berkeley, CA 94720, USA}
\affiliation{Lawrence Berkeley National Laboratory, 1 Cyclotron Road, Berkeley, CA 94720, USA}
\email{jdelabrouille@lbl.gov}

\author[0000-0002-1984-8234]{Andrei V. Frolov}
\affiliation{Department of Physics, Simon Fraser University, 8888 University Drive, Burnaby, V5A 1S6, BC, Canada}
\email{frolov@sfu.ca}

\author[0000-0002-7546-0509]{Shamik Ghosh}
\affiliation{Computational Cosmology Center \& Physics Division, Lawrence Berkeley National Laboratory, 1 Cyclotron Road, Berkeley, CA 94720, USA}
\affiliation{CNRS-UCB International Research Laboratory, Centre Pierre Bin\'etruy, IRL 2007, CPB-IN2P3, Berkeley, CA 94720, USA}
\email{shamik@lbl.gov}

\author[0000-0001-7449-4638]{Brandon S. Hensley}
\affiliation{Jet Propulsion Laboratory, California Institute of Technology, 4800 Oak Grove Drive, Pasadena, CA 91109, USA}
\email{bhensley@jpl.nasa.gov}

\author{Monica D. Hicks}
\affiliation{Department of Physics, Stanford University, Stanford, CA 94305, USA}
\affiliation{Kavli Institute for Particle Astrophysics \& Cosmology, P.O. Box 2450, Stanford University, Stanford, CA 94305, USA}
\affiliation{Department of Computer Science, Stanford University, Stanford, CA 94305, USA}
\email{mdhicks@stanford.edu}

\author[0000-0002-5501-8449]{Nicoletta Krachmalnicoff}
\affiliation{International School for Advanced Studies (SISSA), Via Bonomea 265, 34136 Trieste, Italy}
\affiliation{Institute for Fundamental Physics of the Universe (IFPU), Via Beirut, 2, 34151 Grignano, Trieste, Italy} 
\affiliation{Istituto Nazionale di Fisica Nucleare (INFN), Sezione di Trieste, Via Valerio 2, 34127 Trieste, Italy}
\email{nkrach@sissa.it}

\author[0000-0002-6445-2407]{King Lau}
\affiliation{Department of Physics, California Institute of Technology, 1200 E. California Boulevard, Pasadena, CA 91125, USA}
\email{kennylau@caltech.edu}

\author[0009-0003-8536-5323]{Myra M. Norton}
\affiliation{Department of Astrophysical Sciences, Princeton University, Princeton, NJ 08544, USA}
\email{mmnorton@princeton.edu}

\author[0000-0003-3983-6668]{Clement Pryke}
\affiliation{Minnesota Institute for Astrophysics, University of Minnesota, Minneapolis, Minnesota 55455, USA}
\affiliation{School of Physics and Astronomy, University of Minnesota, Minneapolis, Minnesota 55455, USA}
\email{pryke@symmetryone.net}

\author[0000-0002-0689-4290]{Giuseppe Puglisi}
\affiliation{Dipartimento di Fisica e Astronomia, Universit\`a degli Studi di Catania, via S. Sofia, 64, 95123, Catania, Italy}
\affiliation{INFN - Sezione di Catania, Via S. Sofia 64, 95123 Catania, Italy}
\affiliation{INAF - Osservatorio Astrofisico di Catania, via S. Sofia 78, 95123 Catania, Italy}
\email{Giuseppe.puglisi@dfa.unict.it}

\author[0000-0001-9126-6266]{Mathieu Remazeilles}
\affiliation{Instituto de Física de Cantabria (CSIC-UC), Avda. de los Castros s/n, 39005 Santander, Spain}
\email{remazeilles@ifca.unican.es}

\author[0009-0005-3268-1044]{Elisa Russier}
\affiliation{Lawrence Berkeley National Laboratory, 1 Cyclotron Road, Berkeley, CA 94720, USA}
\affiliation{CNRS-UCB International Research Laboratory, Centre Pierre Bin\'etruy, IRL 2007, CPB-IN2P3, Berkeley, CA 94720, USA}
\email{erussier@lbl.gov}

\author[0000-0002-0457-0153]{Benjamin Thorne}
\affiliation{Lawrence Berkeley National Laboratory, 1 Cyclotron Road, Berkeley, CA 94720, USA}
\affiliation{Department of Physics and Astronomy, University of California, Davis, CA 95616, USA}
\email{bn.thorne@gmail.com}

\author[0000-0003-0813-9480]{Jian Yao}
\affiliation{International School for Advanced Studies (SISSA), Via Bonomea 265, 34136 Trieste, Italy}
\email{jyao@sissa.it}

\author[0000-0001-6841-1058]{Andrea Zonca}
\affiliation{San Diego Supercomputer Center, University of California San Diego, La Jolla, CA 92093, USA}
\email{zonca@ucsd.edu}

\date{\today}

\begin{abstract}
Polarized foreground emission from the Galaxy is one of the biggest challenges facing current and upcoming cosmic microwave background (CMB) polarization experiments. We develop new models of polarized Galactic dust and synchrotron emission at CMB frequencies that draw on the latest observational constraints, that employ the ``polarization fraction tensor'' framework to couple intensity and polarization in a physically motivated way, and that allow for stochastic realizations of small-scale structure at sub-arcminute angular scales currently unconstrained by full-sky data. We implement these models into the publicly available Python Sky Model (\texttt{PySM}) software and additionally provide \texttt{PySM} interfaces to select models of dust and CO emission from the literature. We characterize the behavior of each model by quantitatively comparing it to observational constraints in both maps and power spectra, demonstrating an overall improvement over previous \texttt{PySM} models. Finally, we synthesize models of the various Galactic foreground components into a coherent suite of three plausible microwave skies that span a range of astrophysical complexity allowed by current data.\footnote{A supplement describing author contributions to this paper can\\ be found at \url{https://pysm3.readthedocs.io/en/latest/pysm_methods_author_contributions.html}.}
\end{abstract}

\section{Introduction}
One of the principal challenges for current and future cosmic microwave background (CMB) polarization experiments is mitigating contaminating emission from the Galaxy. Polarized Galactic emission is brighter than current upper limits on a primordial $B$-mode signal at all frequencies, even in particularly clean patches of the high Galactic latitude sky \citep{planck2016-l11A, Ade:2021}. Many ongoing and upcoming surveys such as the Simons Observatory \citep{Ade:2019}, CMB-S4 \citep{Abazajian:2022}, and LiteBIRD \citep{LiteBIRDCollaboration:2023} observe large sky fractions and will need to mitigate foreground emission potentially many times brighter than the targeted cosmological signals. Understanding the potential complexities of Galactic emission and designing analyses that are robust to these complexities are of paramount importance for constraining the physics of the early Universe.

Galactic emission is constrained by current data across the full sky over a range of angular scales and frequencies. However, it is not known perfectly, and a primary role of modeling is to provide a suite of possible extrapolations to unmeasured scales and frequencies that reflect the current level of uncertainties in the spatial distribution and frequency dependence of each contributing emission mechanism. At the same time, models should accord with observational constraints as closely as possible.

To this end, tools have been developed to simulate full-sky, multi-frequency realizations of Galactic emission drawing both on data-driven constraints and theoretical models. The Planck Sky Model \citep[PSM;][]{delabrouille2012} was originally built to develop data analysis tools for the Planck mission, using pre-existing data sets. The PSM evolved throughout the Planck mission timeline to include Planck observations and adjust to Planck data analysis requirements, and it has been widely used in various data challenges and for planning future CMB and 21\,cm line mapping experiments \citep[e.g.,][]{Remazeilles:2018, Fornazier:2022, Ghosh:2022}. Building on the PSM, the Python Sky Model \citep[\texttt{PySM};][]{Thorne:2017} provides a Python interface to expanded classes of foreground models. \texttt{PySM} has been employed for both forecasting \citep[e.g.,][]{Abazajian:2022, Hensley:2022, CCAT-PrimeCollaboration:2023, Wolz:2024} and data analysis \citep[e.g.,][]{BICEP2Collaboration:2018, Vacher:2023, Ade:2025b}.

In this work, we develop techniques to overcome a number of limitations of previous generations of Galactic emission models. We first introduce a ``polarization fraction tensor'' formalism as a physically motivated way to model Galactic polarization at small angular scales. We demonstrate the use of this framework in generating ensembles of sky realizations in which each realization has a different spatial morphology but the same statistical properties at small angular scales while retaining the well-measured large-scale features of the Galaxy. In addition to algorithmic development, we employ data products from more recent analyses of microwave total intensity and polarization data than used in previous models. Finally, we implement a number of additional models of dust and CO emission from the literature to provide a more complete range of models, consistent with current data, that reflect the present uncertainties on the behavior of microwave foregrounds. We validate and characterize all of the models, and assess their relative abilities to capture various physical effects and to accord with available data-driven constraints.

Synthesizing the suite of models implemented in this work and elsewhere, we define three full-sky models of Galactic foreground emission in total intensity and polarization. These models are driven by the existing data where observational constraints are unambiguous and span a range of complexities (labeled low, medium, and high) in their approach to modeling emission in the parameter space that is not well-constrained by data. This suite enables individual experiments to optimize their designs over a range of uncertainties in our current understanding of the Galactic foreground sky, and supports self-consistent comparative and---especially---joint analyses of multiple experiments.

We organize the paper as follows: Section~\ref{sec:methods} is an overview of how Galactic emission models are implemented in \texttt{PySM}; Section~\ref{sec:small_scales} presents our methodology for generating stochastic simulations of small-scale emission using the polarization fraction tensor formalism; Section~\ref{sec:other_models} describes our implementation of an alternative dust model and of a suite of CO emission models; Section~\ref{sec:validation} details a collection of validation metrics for the new models; Section~\ref{sec:discussion} discusses limitations of the models presented here and future directions for development; Section \ref{sec:modelsuite} presents our proposed suite of three sky models that span a range of complexity; and Section~\ref{sec:summary} concludes with a summary.

\section{Modeling Framework} \label{sec:methods}

\subsection{The PySM Software}
The \texttt{PySM} software\footnote{\url{https://pysm3.readthedocs.io/en/latest/index.html}} provides a convenient interface for generating full-sky maps of total and polarized emission at far-infrared through radio frequencies. Users can select one or more emission mechanisms to simulate, including the CMB, dust, synchrotron radiation, free-free emission, and CO line emission. Most emission mechanisms have more than one model to select from, where here a model refers both to the spatial morphology and frequency dependence of the emission. Stokes $I$, $Q$, and $U$ maps are provided in HEALPix\footnote{\url{https://healpix.sourceforge.io}}  \citep{Gorski:2005} format at the requested $N_{\rm side}$ at a user-specified frequency or integrated over a user-specified bandpass.

One of the aims of this work is to extend the highest angular resolution of the maps that \texttt{PySM} can generate. The first and most stringent challenge of such high resolutions is the sheer size of the templates: a single $I$, $Q$, or $U$ map at a pixel size of $0.4\arcmin$ ($N_{\rm side}=8192$) occupies ~3\,GB in single precision and is 256 times larger than an original \texttt{PySM}~2 map. The groundwork that allowed the implementation of this new generation of Galactic models started in 2019 with the rewrite of \texttt{PySM} and its release as \texttt{PySM}~3 \citep[see][for details]{Zonca:2021}. The problem of distribution has been solved by hosting all the input templates at NERSC\footnote{\url{https://portal.nersc.gov/project/cmb/pysm-data}}, with templates downloaded and cached by \texttt{PySM} as needed using the facilities included in \texttt{astropy.data} \citep{AstropyCollaboration:2013, AstropyCollaboration:2018}. Moreover, all the members of the CMB community using NERSC for computing are able to directly access the same folders locally.

The second issue is memory usage. \texttt{PySM}~3 leverages \texttt{numba} \citep{Lam:2015} to compile Python on-the-fly to machine code so that the evaluation and bandpass integration of each model avoids the temporary arrays allocated by \texttt{numpy} and reduces memory consumption at least by a factor of two. Moreover, \texttt{numba} supports multi-threading so it can make use of all the cores available in the system. Thanks to these improvements, foreground models with a resolution up to $N_{\rm side}=8192$ can be generated while keeping the disk, memory, and CPU requirements manageable. We defer to Section~\ref{sec:small_scales} a discussion of the algorithmic improvements that enable models to be constructed at these small angular scales.

\subsection{Overview of Galactic Emission Mechanisms} \label{sec:emission_mechanisms}
The Galactic interstellar medium (ISM) consists of matter and radiation in various forms. At the microwave wavelengths relevant for CMB observations there are several relevant emission mechanisms, each with their own frequency dependence, and different mechanisms dominate the emission from different regions of the ISM. The spatial distribution of microwave emission thus varies with frequency \citep[e.g.,][]{planck2014-a12}.

Cold ($\sim$10--30~K) grains of interstellar dust emit thermal vibrational radiation with a spectrum that peaks at $\sim$2~THz. The dust emission spectrum shows an excess over expectations from thermal vibrational emission below $\nu=100$~GHz: this excess is dubbed anomalous microwave emission (AME). AME is thought to be electric dipole radiation from rapidly spinning, sub-nanometer dust grains \citep{Draine:1998b}. Relativistic cosmic ray electrons spiraling in the Galactic magnetic field emit synchrotron radiation, while warm ionized gas emits free-free radiation (or \emph{bremsstrahlung}) through the interaction of free electrons with ions. Synchrotron and free-free emission typically dominate the Galactic ISM emission at frequencies $\sim$10--100~GHz. Finally, atoms and molecules in the Galactic ISM, through vibrational and rotational shifts in energy levels, emit radiation in the form of a rich spectrum of discrete lines. Most notable among these is a bright comb of carbon monoxide (CO) emission lines at integer multiples of the $J = 1\rightarrow 0$ ($\nu = 115$~GHz) rotational transition.

All of these emission mechanisms are capable of producing polarized radiation. Dust and synchrotron polarization have both been robustly detected and are the principal polarized Galactic foregrounds at microwave frequencies. Dust grains rotate about their short axis---the axis of greatest moment of inertia---and the rotation axis in turn aligns with the local magnetic field due to the magnetic moment of the grain. The aggregate emission from a population of such grains is polarized perpendicular to the projected field orientation at levels up to $\sim$20\% \citep{planck2014-XIX}. Synchrotron emission is inherently polarized at the $\sim$75\% level and, like dust emission, is polarized perpendicular to the projection of the local magnetic field onto the plane of the sky \citep{Rybicki:1986}. %Moreover, \citet{planck2014-XIX} and \citet{Miville-Deschenes:2016} showed that the angular power spectra of both synchrotron and thermal dust emissions are characterized by a scale invariant spectrum which could be approximated as a power-law.

The relevance of the other mechanisms is less clear. CO line emission can be linearly polarized via the Goldreich-Kylafis effect \citep{Goldreich:1981}, though there have been few detections to date \citep[e.g.,][]{Greaves:1999, Greaves:2002, Cortes:2008, Houde:2013}. If the sub-nanometer dust grains believed to be responsible for the AME are able to align with the local magnetic field, then AME will be polarized \citep{Draine:1998a}. Searches for AME polarization in specific objects and over large sky areas have so far yielded only upper limits \citep[e.g.,][]{Genova-Santos:2017, Herman:2023}. Finally, free-free emission is inherently unpolarized, but a small level of polarization can be produced near the edges of \ion{H}{2} regions from Thomson scattering \citep{Rybicki:1986}. However, this effect is relevant only at much higher angular resolution than typically employed in CMB observations.

Here we describe our approach to modeling each of these components. Note that we express all Stokes parameters in specific intensity units (e.g., MJy\,sr$^{-1}$).

\subsubsection{Dust Emission} \label{subsubsec:dust_model}
In all dust models developed in this work, the frequency dependence of the dust emission is described by a modified blackbody emission law:

\begin{equation} \label{eq:dust-emission-law}
    S_\nu = A_d^S \left(\frac{\nu}{\nu_0}\right)^{\beta_d} \, B_\nu(T_d)
    ~~~,
\end{equation}
where $S$ is any of $I$, $Q$, or $U$ and $B_\nu\left(T\right)$ is the Planck function. The parameter $A_d^S$ is the dust intensity $S_\nu$ at the reference frequency $\nu_0$, here taken to be 353\,GHz. We refer to the $A_d^S$ as ``amplitude'' parameters, with each pixel in a map having its own $A_d$ value. 

The parameters $\beta_d$ and dust temperature $T_d$ describe the frequency dependence of the emission. We refer to them as ``spectral'' parameters. Typical values of the spectral parameters for dust emission in both total intensity and polarization are $\beta_d = 1.5$ and $T_d = 20$\,K \citep{planck2016-l11A}, with variations of order 10\% observed in total intensity in the high Galactic latitude sky \citep[e.g.,][]{planck2014-a12, planck2016-XLVIII}.

In this work, we present the new dust models \texttt{d9}, \texttt{d10}, \texttt{d11} (Section~\ref{sec:small_scales}) and an implementation of the \citet{Martinez-Solaeche:2018} (``MKD'') model as \texttt{d12} (Section~\ref{sec:layers}). We employ previous \texttt{PySM} dust models only for purposes of comparison.

\subsubsection{Synchrotron Emission} \label{subsubsec:synch_model}
In all synchrotron models developed in this work, the frequency dependence of the synchrotron emission is described by a power law with possible curvature:

\begin{equation} \label{eq:synch-emission-law}
    S_\nu = A_s^S \left(\frac{\nu}{\nu_0}\right)^{\beta_s + c_s \ln\left(\nu/\nu_c\right)}
    ~~~,
\end{equation}
where $S$ is any of $I$, $Q$, or $U$. The parameter $A_s^S$ is the synchrotron intensity $S_\nu$ at the reference frequency $\nu_0$, here taken to be 23\,GHz. As with dust, we refer to the $A_s$ as ``amplitude'' parameters that vary from pixel to pixel in a map. 

The parameters $\beta_s$, $c_s$, and $\nu_c$ describe the frequency dependence of the emission. Nonzero $c_s$ has been found in both simulations of Galactic synchrotron emission \citep{Jaffe:2011} and in observational data \citep{Kogut:2012}. It arises from a non-power law energy distribution of the cosmic ray electrons responsible for the emission, due for instance to radiative energy losses. In all models, we take $\nu_c = 23$\,GHz. When expressing $S_\nu$ in specific intensity units, $\beta_s \simeq -1$.

In this work, we present the new synchrotron models \texttt{s5}, \texttt{s6}, and \texttt{s7} (Section~\ref{sec:small_scales}). We employ previous \texttt{PySM} synchrotron models only for purposes of comparison.

\subsubsection{Free-free Emission}
We do not develop new models of free-free emission in this work, but rather rely on the existing \texttt{f1} model \citep{Thorne:2017}. The frequency dependence of free-free emission is known from theory \citep[][and references therein]{Draine:2011}, but over the range of frequencies relevant for microwave observations, it can be approximated as a simple power law. The \texttt{f1} model assumes a sky-constant power-law behavior $S_\nu \propto \nu^{-0.14}$ in specific intensity units. The amplitude is given by the free-free amplitude map from the \texttt{Commander} component separation analysis \citep{planck2014-a12}, which has an angular resolution of $1^\circ$. At angular scales $<1^\circ$, the emission is based on Gaussian random fluctuations modulated by the intensity map \citep[see][for details]{Thorne:2017}. Free-free emission is assumed to be unpolarized.

\subsubsection{Anomalous Microwave Emission (AME)}
We do not develop new models of AME in this work, but rather rely on the existing \texttt{a1} and \texttt{a2} models \citep{Thorne:2017}. These models are based on a \texttt{Commander} component separation analysis that produced full-sky maps of AME amplitude and spectral parameters at $1^\circ$ resolution \citep{planck2014-a12}. The AME frequency dependence is modeled as a sum of two components each having spectra based on theoretical models computed by the \texttt{SpDust} software \citep{Ali-Haimoud:2009, Silsbee:2011}. While each component is described by an amplitude and a peak frequency, one component was required to have a sky-constant peak frequency, fit to a value of 33.35\,GHz \citep{planck2014-a12}. Thus, the AME model is based upon three maps: the amplitude of each of the two components and the peak frequency of one component. Emission at scales $<1^\circ$ is generated using the higher-resolution observations of thermal dust emission, where the spatially varying scaling factor is determined by convolving the thermal dust emission map to $1^\circ$ resolution and taking the ratio with the AME map \citep{Thorne:2017}.

The \texttt{a1} and \texttt{a2} models differ only in their treatment of polarization. AME polarization has not been detected, and recent analysis places an upper limit on the intrinsic polarization fraction of $\lesssim3$\% \citep{Herman:2023}. The \texttt{a1} model assumes that AME is unpolarized, whereas the \texttt{a2} model assumes a constant polarization fraction of 2\%. The polarization angle of the AME in the \texttt{a2} model is based on the polarization angle of the 353\,GHz polarized dust emission determined by component separation with \texttt{Commander} \citep{planck2014-a12}. As with total intensity, the \texttt{a2} AME polarization map inherits the small-scale fluctuations introduced to the thermal dust polarization map \citep[see][for details]{Thorne:2017}.

\subsubsection{CO Emission}
In all CO models presented in this work, emission from the $J = 1\rightarrow0, 2\rightarrow1$, and $3\rightarrow2$ transitions at 115.3, 230.5, and 345.8\,GHz, respectively, is modeled as a delta function at the rest frequency. Therefore, a model is fully defined by maps of $I$, $Q$, and $U$ at the reference frequency. We present the new \texttt{PySM} CO models \texttt{co1}, \texttt{co2}, and \texttt{co3} in Section~\ref{subsec:co_models}.

\subsubsection{Other Emission Mechanisms}
The current suite of \texttt{PySM} models encompasses most of the microwave emission and polarization mechanisms observed from the diffuse ISM of the Milky Way. However, it is not exhaustive, and we identify here a few potential targets for future work.

Other isotopologues of CO emit at frequencies near the $^{12}$C$^{18}$O lines. However, these species produce much weaker emission and reside in even denser gas than does $^{12}$C$^{18}$O, and so have a more minor role as a CMB foreground. Likewise, HCN has been identified as a potential contributor to the \texttt{Commander} sky model, but mostly toward the Galactic Center \citep{planck2014-a12}. Line emission from \ion{C}{2} at 158\,$\mu$m and \ion{N}{2} at 122 and 205\,$\mu$m from the diffuse ISM was mapped by COBE/FIRAS \citep{Bennett:1994}, but there are currently no CMB experiments operating at such high frequencies. COBE/FIRAS also detected weaker line emission from \ion{C}{1} (370 and 609\,$\mu$m) and CO transitions beyond those modeled in this work, though these lines were mostly seen toward the Galactic Center \citep{Bennett:1994}. 

Our focus in this work is on emission from the ISM of the Milky Way, but \texttt{PySM} also includes models of extragalactic emission (e.g., the Cosmic Infrared Background, the CMB). Given the tight coupling between extragalactic signals through mechanisms like CMB lensing, the thermal Sunyaev-Zeldovich effect, and the kinetic Sunyaev-Zeldovich effect, extragalactic modeling requires a separate, dedicated effort beyond our scope.

Finally, emission from the Solar System is detected at microwave wavelengths in the form of Zodiacal light as well as thermal emission from planets and other rocky bodies. Solar System signals are necessarily time-dependent and thus are not suited for the map-based framework employed by \texttt{PySM}. We therefore do not consider them here.

\section{Dust and Synchrotron Models: Stochastic Emission at Small Angular Scales} \label{sec:small_scales}
The methods presented here aim to preserve the well-measured large-scale information in existing observations of dust and synchrotron emission (the ``template'' maps), filter out the noisy small-scale emission in the maps, and replace those small scales with a stochastic realization that has a reasonable correspondence with the large-scale emission. Specifically, the synthetic small-scale structure should have a power spectrum that connects smoothly to the power spectrum of the real data at large scales. We rely on the fact that the angular power spectra of both synchrotron and thermal dust emission are well-approximated by a power-law in $\ell$ \citep{planck2014-XIX, Miville-Deschenes:2016}. Our approach is to generate stochastic realizations of small-scale emission that is modulated by the large-scale signal.

We divide our discussion into amplitude parameters ($A_d^S$ and $A_s^S$ in Equations~\ref{eq:dust-emission-law} and \ref{eq:synch-emission-law}, respectively) in Section~\ref{sec:amp_params} and spectral parameters ($\beta_d$ and $T_d$ in Equation~\ref{eq:dust-emission-law} and $\beta_s$ and $c_s$ in Equation~\ref{eq:synch-emission-law}) in Section~\ref{sec:spec_params} as the methodology differs for each. With maps of both sets of parameters, foreground model maps at any desired frequency can be generated from the amplitude maps at the reference frequency using Equations~\ref{eq:dust-emission-law} and \ref{eq:synch-emission-law}.

\subsection{Small-Scale Fluctuations in Amplitude Parameters} \label{sec:amp_params}
To implement the parametric models of dust and synchrotron emission described in Section~\ref{sec:emission_mechanisms}, we require maps of $A^I$, $A^Q$, and $A^U$ for each mechanism, i.e., the total emission and polarization at a reference frequency. We start from template $I$, $Q$, and $U$ maps derived from data, to which we add fluctuations at scales that are noise-dominated.

\subsubsection{The Polarization Fraction Tensor Formalism} \label{sec:polfrac}
A principal challenge for generating realizations of Galactic emission at small angular scales over the full sky is that the amplitude of the fluctuations is strongly dependent on proximity to the Galactic plane. Further, Gaussian random fluctuations are a poor representation of the typically filamentary structure of the Galactic ISM \citep[e.g.,][]{Hacar:2023}. We address each of these challenges through use of a polarization fraction tensor framework.

While the mathematical model of this framework can be applied to any emission mechanism, it is motivated by a simple model of dust emission. In this picture, the morphology of the dust polarized intensity $P$ at large angular scales is set primarily by the density structure of the dust, probed in projection by the total intensity $I$, and then secondarily by the large-scale morphology of the Galactic magnetic field, which modulates the polarization fraction of the dust, $p \equiv P/I$. Random fluctuations of the turbulent component of the Galactic magnetic field lead to fluctuations on top of this smooth, large-scale polarized intensity distribution. The amplitude of these fluctuations is much more spatially isotropic than fluctuations in the $P$ map itself. Further, total and polarized dust emission are coupled through the angle $\gamma$ between the local magnetic field and the line of sight, such that the sum $I+P$ is independent of $\gamma$ and proportional to the dust column density \citep{Hensley:2019}. Finally, Galactic dust emission has a large dynamic range in $I$, whereas $\ln I$ not only varies much less but is better described by a Gaussian distribution over the sky.

A two-dimensional sky map of intensity and linear polarization is described by the symmetric rank-2 tensor \citep[e.g.,][]{Landau:1975}

\begin{equation}
    P_{ab} = \frac{1}{2} \left[ \begin{matrix} I+Q & U \\ U & I-Q \end{matrix} \right]
\end{equation}
whose components transform under local coordinate changes. This suggests that any transformation that attempts to normalize emission should be applied to the polarization tensor, not its components. The simplest such transformation is logarithmic, with $p_{ab} \equiv \ln 2 P_{ab}$ in the matrix sense:
\begin{equation}\label{eq:logP}
p_{ab} \equiv \ln \left[ \begin{matrix} I+Q & U \\ U & I-Q \end{matrix} \right] \ = \ 
    \left[ \begin{matrix} i+q & u \\ u & i-q \end{matrix} \right].
\end{equation}
The (arbitrary) factor of two multiplying $P_{ab}$ enables a more physical interpretation of the $i$, $q$, and $u$ parameters, as we shall see.

The transformation can be computed analytically by taking the logarithm of the eigenvalue decomposition of $P_{ab}$. While the polarization direction is preserved, the Stokes parameters $I$, $Q$, and $U$ are compressed into their polarization fraction tensor analogues $i$, $q$, and $u$:
\begin{align}\label{eq:real2pt}
    i &\equiv \frac{1}{2} \ln (I^2 - P^2)\nonumber  \\
    q &\equiv  \frac{1}{2}\frac{Q}{P} \ln \frac{I+P}{I-P} \\
    u &\equiv  \frac{1}{2}\frac{U}{P} \ln \frac{I+P}{I-P}\nonumber 
    ~~~,
\end{align}
where $P \equiv \sqrt{Q^2 + U^2}$ is the usual polarized intensity independent of coordinate system. For small polarization fractions, these reduce to familiar quantities $i\simeq\ln I$, $q\simeq Q/I$, and $u\simeq U/I$, motivating our designation of $p_{ab}$ as the ``polarization fraction tensor.''

The inverse transformations are
\begin{align}\label{eq:pt2real}
    I &= e^i \cosh \xi \nonumber \\
    Q &= \frac{q}{\xi}\,e^i\sinh \xi  \\
    U &= \frac{u}{\xi}\,e^i\sinh \xi \nonumber
    ~~~,
\end{align}
where $\xi \equiv \sqrt{q^2 + u^2}$. A number of useful features of this transformation are evident. First, $I$ is guaranteed to be positive for any values of $i$, $q$, and $u$. Further,

\begin{equation}
    p = \tanh\xi
    ~~~,
\end{equation}
and so $0 \leq p \leq 1$. Since the transformation is non-linear, Gaussian fluctuations introduced in $i$, $q$, and $u$ maps will necessarily become non-Gaussian when transformed back to $I$, $Q$, and $U$.

\subsubsection{General Methodology}\label{subsec:methodology}

\begin{figure*}
    \includegraphics[width=0.85\textwidth]{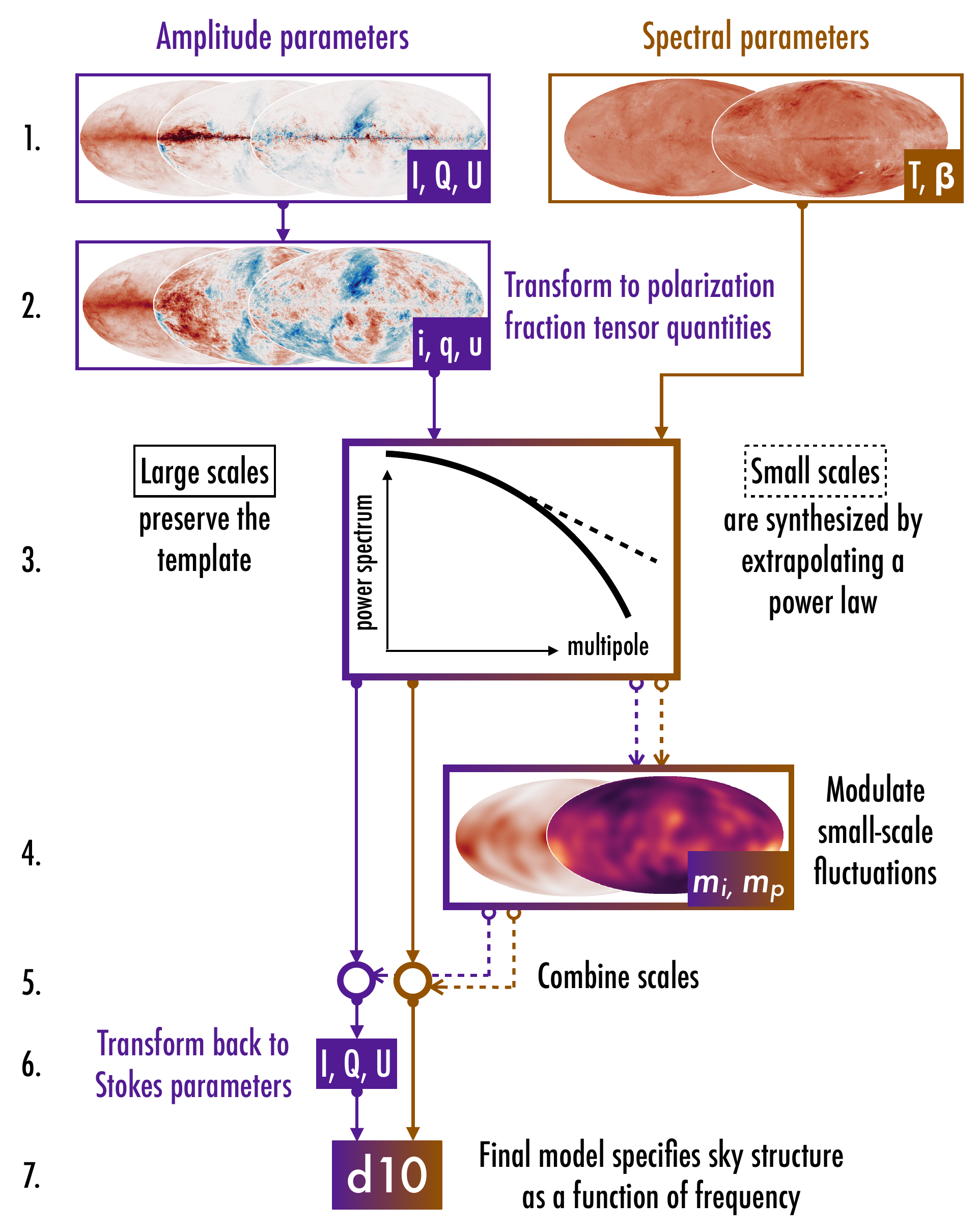}
    \caption{An overview of the steps for building model \texttt{d10}. 1. We begin with GNILC-derived $I$, $Q$, $U$ maps of the polarized dust emission at 353 GHz (``amplitude parameters'', purple) and maps of the GNILC-derived dust temperature and spectral index (``spectral parameters'', brown).  2. The amplitude parameters are transformed via the polarization tensor fraction formalism to $i$, $q$, $u$. 3. Both sets of maps are filtered by scale: large scales are untouched, and small-scale map structure is replaced by structure synthesized from a power spectrum extrapolated from a power-law fit to the large-scale data. 4. The small-scale structure maps are modulated by maps of the large-scale structure in either total intensity or polarization: the fluctuations in $i$ and the spectral parameters are modulated by $m_i$, and the fluctuations in $q$ and $u$ are modulated by $m_p$. 5. The large- and small-scale maps are combined. 6. The $i$, $q$, $u$ maps are transformed back to $I$, $Q$, and $U$. 7. Together, the final amplitude maps at 353 GHz, and the final spectral parameter maps, which define how the sky varies with frequency, fully specify the \texttt{d10} model.}
    \label{fig:flowchart}
\end{figure*}

Our method for generating small-scale fluctuations is summarized as follows (Figure~\ref{fig:flowchart}): 
\begin{enumerate}
    \item We first transform the $I$, $Q$, and $U$ templates into $i$, $q$, and $u$ templates via Equation~\eqref{eq:real2pt}.
    \item We low-pass filter these template $i$, $q$, and $u$ maps with cut-off multipole $\ell_1$ to yield the final large-scale maps.
    \item We then compute the $tt$, $te$, $ee$, and $bb$ full-sky power spectra from the $i$, $q$, and $u$ maps in analogy with how $TT$, $TE$, $EE$, and $BB$ spectra are computed from $I$, $Q$, and $U$ maps.
    \item We model the $\ell$-dependence of each spectrum as a broken power law in $\ell$. To estimate the power spectrum at scales smaller than some scale $\ell_1$, we first fit the spectrum over the range $\ell_0 < \ell < \ell_1$ with a free amplitude and a fixed power law index from the literature (Table~\ref{tab:smallscale_par}). We then extrapolate from $\ell_1$ to a second pivot scale $\ell_2$ using the fit power law. Finally, the $ee$ and $bb$ spectra are extended to $\ell > \ell_2$ using the steeper power law index of the $tt$ spectrum, while the $te$ spectrum retains its fit index. The $tb$ and $eb$ spectra are assumed to be zero for $\ell > \ell_1$.
    \item  We synthesize $i$, $q$, and $u$ maps using the constructed $tt$, $te$, $ee$, and $bb$ spectra. These maps are then high-pass filtered at cut-off multipole $\ell_1$.
    \item We construct modulation maps $m_i$ and $m_p$ for intensity and polarization, respectively (see Figure~\ref{fig:flowchart}). The synthesized maps are then multiplied by the modulation maps to yield the final small-scale maps.
    \item Finally, we combine the large-scale and small-scale maps (from Steps~2 and 6, respectively) and transform back to $I$, $Q$, and $U$ maps via Equation~\eqref{eq:pt2real}.
\end{enumerate}

This prescription has several free parameters that require tuning. The pivot scale $\ell_1$ governs up to which multipole information is taken from the input template maps versus what is generated randomly. The scale $\ell_0$ avoids using the largest scales in the power spectrum fit as they are subject to large sample variance. The pivot scale $\ell_2$ prevents the $EE$ and $BB$ spectra from exceeding the $TT$ spectrum at any scale. Finally, the modulation maps $m_i$ and $m_p$ ensure that the injected small-scale fluctuations are larger in bright regions (e.g., at low Galactic latitude) and smaller in faint regions (e.g., at high Galactic latitude).

To construct the modulation maps $m_i$ and $m_p$, we generally follow the method outlined in \citet{Thorne:2017}. For each pixel at the location $\hat{\theta}$ in an $N_{\rm side} = 8$ HEALPix map, we define a circular mask centered on $\hat{\theta}$ of radius $11.3^\circ$  and apodize it with a $2^\circ$ Gaussian profile. We compute the $tt$ and $ee$ power spectra within each circle using NaMaster \citep{Alonso:2019}, to which we fit simple power laws in $\ell$. We then evaluate these best-fit power-law spectra at the $\ell_*$ multipole (Table \ref{tab:smallscale_par}). Finally, the modulation maps are constructed from the ratios:

\begin{align}
\label{eq:mod_maps}
    m_i\left(\hat{\theta}\right) &= \left(\frac{C^{tt}_{\ell_*,circ}}{C^{tt}_{\ell_*,full}}\right)^{1/2} \\
    m_p\left(\hat{\theta}\right) &= \left(\frac{C^{ee}_{\ell_*,circ}}{C^{ee}_{\ell_*,full}}\right)^{1/2} \label{eq:mod_maps2}
\end{align}
where $C^{tt}_{\ell_*,full}$ and $C^{ee}_{\ell_*,full}$ are the full-sky $tt$ and $ee$ power spectra evaluated at $\ell = \ell_*$, respectively. We choose $\ell_* \lesssim \ell_1$ in order to have reliable estimates of the power spectrum at $\ell_*$. Finally, we smooth $m_i$ and $m_p$ with a kernel equivalent to the apodized mask described above. 

Whereas previous \texttt{PySM} models used modulation maps constructed in $N_{\rm side} = 2$ superpixels and then smoothed with a $10^\circ$ beam \citep{Thorne:2017}, we employ $N_{\rm side} = 8$ superpixels and overlapping circular regions. This is motivated by improvements to the Planck polarization maps since the earlier study. Thus the models developed in this work incorporate more small-scale information into the modulation maps than previous \texttt{PySM} models.
  
The adopted values of $\ell_1$ and $\ell_*$ are driven both by the angular resolution and the signal-to-noise ratio of the template maps. We employ $\ell_1=100$ and $\ell_*=80$ for dust and $\ell_1=38$ and $\ell_*=36$ for synchrotron. The $\ell_2$ parameter is chosen to be high enough that, for $\ell > \ell_2$, dust and synchrotron emission are relatively unconstrained by current full-sky data. We adopt  $\ell_2=2000$ for dust and $\ell_2 = 400$ for synchrotron. For dust we employ a power law index of $\alpha_{tt} = -0.80$ based on the range of fit $\alpha_{TT}$ in Planck, WISE, and MegaCam data by \citet{Miville-Deschenes:2016}. We adopt $\alpha_{ee} = -0.42$, $\alpha_{te} = -0.50$, and $\alpha_{bb} = -0.54$ from fits to Planck $TT$, $TE$, and $BB$ spectra, respectively, over 71\% of the sky \citep{planck2016-l11A}. For synchrotron we adopt $\alpha_{tt} = -1.00$ and $\alpha_{te} = -1.00$ based on the observed spectra of the synchrotron template map. We adopt $\alpha_{ee} = -0.84$ and $\alpha_{bb} = -0.76$ following fits to synchrotron maps over 78\% of the sky produced with the \texttt{Commander} pipeline \citep{planck2016-l04}. The $\ell_1$, $\ell_2$, $\ell_*$, and $\alpha$ values used to construct the small-scale maps are all summarized in Table~\ref{tab:smallscale_par}.

The final maps are obtained by combining the large and small scales: 
\begin{equation} \label{eq:filter}
    a_{\ell m }^{x, out}=  a_{\ell m }^{x, \mathcal{T}} \left(1-\sigma_\ell\right)^{\gamma} + a_{\ell m }^{x, ss} \sigma_\ell
    ~~~, 
\end{equation}
where $a_{\ell m }^{x, \mathcal{T}}$ are the spherical harmonic coefficients of the large-scale templates, with $x$ any of $t$, $e$, and $b$, and $a_{\ell m }^{x, ss}$ is derived from the synthesized small-scale fluctuation maps. The exponent of $\gamma = 0.2$ is chosen empirically to smooth the transition between our data-driven templates and the generated small scales. We found that this choice minimizes artifacts at $\ell \simeq \ell_1$ in power spectra computed over large sky areas while maintaining the correct asymptotic behavior. The filter $\sigma_\ell$ is given by

\begin{equation} \label{eq:filter2}
\sigma_\ell  = \left[1+  e^{ -c_1 (\ell/ \ell_1  -c_2 )}\right]^{-1}  
~~~,
\end{equation}
where the parameters $c_1$ and $c_2$ govern the width of the filter in multipoles. We adopt $c_1=40$ and $c_2=1.05$ throughout this work.

Since the generation of the small scales depends only upon a fixed input power spectrum, we can generate a different map realization on the fly each time a sky is simulated. Thus, the methodology presented here enables generation of an ensemble of sky realizations that have the same well-measured large scales but stochastic realizations of the poorly-measured small scales. Further, the small-scale fluctuations can be generated at arbitrarily small scales set only by the resolution of the map, since all power spectra can be extended indefinitely in $\ell$. In practice, small scales are generated up to $\ell_{\rm max} = 3N_{\rm side}-1$.

We do not intend for the above procedure to be useful for the inner Galactic plane, which has been imaged at high signal-to-noise ratio at relatively small scales. We thus do not apply the small-scale amplitude extrapolation to the $3\%$ of the sky in the Galactic plane as defined by the Planck \texttt{GAL097} mask\footnote{\texttt{HFI\_Mask\_GalPlane-apo2\_2048\_R2.00.fits}}, where we instead use the observed templates as-is at all scales. To ensure a smooth transition between the template used in this region and synthesized small scales employed over the rest of the map, we apodize the mask with a $5^\circ$ Gaussian taper. This reflects our expectation that our procedure is most useful for making physically motivated synthetic realizations of diffuse Galactic emission at high Galactic latitudes rather than in the Galactic plane.

Another challenge for our prescription is pixels that, whether from noise or systematic errors, have $I^2 < P^2$, rendering the transformation in Equation~\ref{eq:real2pt} invalid. We identify 527 such pixels in our dust template and 5 in our synchrotron template, all in the vicinity of the Crab Nebula. To address this, we set the $i$, $q$, and $u$ values at the location of these pixels to fiducial values of 4.51, 0.07, and 0.01 in the dust maps and 4.55, 0.017, $-0.001$ in the synchrotron maps. Given these complications, we do not advise using the simulations in the vicinity of the Crab Nebula.

Finally, we emphasize that our procedures are optimized for producing polarization rather than total intensity maps. For instance, we could in principle retain much smaller angular scales from the intensity templates, however we require that the $\ell_0$, $\ell_1$, $\ell_2$, and $\ell_*$ parameters are the same in total intensity and polarization in order to implement the polarization fraction tensor framework. Other choices, such as the value of $\gamma$, are based on polarization spectra. Some of the imperfections of the maps in total intensity produced as a consequence of these choices are documented in Section~\ref{sec:validation}.

\begin{deluxetable*}{lcccccccc} \label{tab:smallscale_par}
    \caption{Model parameters for synthesizing small-scale emission}
   \tablehead{& \colhead{$\ell_0$} &   \colhead{$\ell_1$} & \colhead{$\ell_2$} & \colhead{$\ell_*$} & \colhead{$\alpha_{tt}$}  & \colhead{$\alpha_{ee}$} & \colhead{$\alpha_{te}$} & \colhead{$\alpha_{bb}$}}
   \startdata
   Dust & 50 & 100 & 2000 & 80 & -0.80 & -0.42& -0.50 & -0.54 \\ 
   Synchrotron & 10 & 38 & 400 & 36 & -1.00& -0.84 & -1.00 & -0.76 \\
   \enddata
    \tablecomments{Spectra are parameterized assuming $D_{\ell}^{xy} \propto \ell^{\alpha_{xy}}$.}
\end{deluxetable*}

%\vspace{-1.1cm}

\subsubsection{Dust Amplitude}\label{sec:dustamplitude}
The Planck~2015 component separation results in total intensity remain state of the art despite updates in polarization in the 2018 data release \citep{planck2016-l04}. Previous \texttt{PySM} models, e.g., \texttt{d0} and \texttt{d1}, employed the dust templates from the \texttt{Commander} component separation analysis \citep{planck2014-a11}. However, the model fitting employed in \citet{planck2014-a11} did not differentiate between Galactic dust emission and the Cosmic Infrared Background (CIB), and so the component separated dust maps retain CIB signal that should not be included in simulations of Galactic emission (see Section~\ref{sec:CIBcontamination} for a detailed discussion). To address this issue, we instead use dust templates from analyses that separated Galactic dust emission from the CIB using the Generalized Needlet Internal Linear Combination (GNILC) algorithm \citep{Remazeilles:2011}. 

In total intensity we employ the Planck GNILC 2015 component separated map\footnote{\texttt{COM\_CompMap\_Dust-GNILC-F353\_2048\_R2.00.fits}} at $353$\,GHz \citep{planck2016-XLVIII}. This map has variable angular resolution, ranging from $21.8\arcmin$ to $5\arcmin$ depending on the sky region, which complicates its use in our analysis. Therefore, we reprocess the map to a spatially uniform resolution of $21.8\arcmin$. To do so, we first note that the map was synthesized from ten needlet maps of different, but spatially uniform, resolutions \citep[][Figure~A.2]{planck2016-XLVIII}. Each of the $j=1,\cdots,10$ needlet maps can be reconstructed from the Planck GNILC 2015 dust map as follows:

\begin{equation}
\gamma^{j}(\hat{n})
=\sum_{\substack{\ell,m}} \left(a_{\ell m}\, h^{j}_\ell\right)Y_{\ell m}(\hat{n})\,,
\label{eq:needlets_maps}
\end{equation}
where $a_{\ell m}$ are the spherical harmonic coefficients of the Planck GNILC 2015 dust map,  $h^{j}_\ell$ is the $j^\text{th}$ needlet window function \citep[see][Figure~A.2]{planck2016-XLVIII}, and $Y_{\ell m}(\hat{n})$ denotes spherical harmonics. We produce a map of uniform resolution by retaining only the first six needlet maps\footnote{Available at NERSC: \url{https://portal.nersc.gov/project/cmb/pysm-data/dust_gnilc/inputs/}}, which contain full-sky information and probe the dust intensity from the largest scales down to $21.8\arcmin$:

 \begin{equation}
 d^{\text{GNILC,}\, 21.8'}(\hat{n})
=\sum_{j=1}^6 h^{j}_\ell \gamma^{j}(\hat{n})\,.
\label{eq:trucated_synthesis}
\end{equation}
This yields an $N_{\rm side} = 2048$ map that reproduces the Planck GNILC 2015 dust intensity template at all scales $\geq 21.8\arcmin$. Finally, we subtract the CIB monopole of $0.13\, \text{MJy}\,\text{sr}^{-1}$ present in the map \citep[][Section~2.2]{planck2016-l11B}.

For the dust $Q$ and $U$ maps we employ the dust maps\footnote{\texttt{COM\_CompMap\_IQU-thermaldust-gnilc-varres\_2048\_R3.00.fits}} produced by the GNILC component separation from the Planck Public Release 3 \citep{planck2016-l04,planck2016-l11B}. While these maps have variable angular resolution, the resolution varies over the sky much more smoothly than in the $I$ map. Given this, and that we wish to retain as much polarization information in our analysis as possible, we employ these variable resolution maps as-is. Their resolution ranges from $5\arcmin$ in the Galactic plane to $80\arcmin$ at high Galactic latitudes. The maps are pixelated at $N_{\rm side} = 2048$. 

To produce dust emission templates at a monochromatic frequency of 353\,GHz, we divide each of the $I$, $Q$, and $U$ maps by a factor of 1.098 to correct for the Planck bandpass \citep[][Table~2]{planck2016-l11A}.

\subsubsection{Synchrotron Amplitude}
Deriving a full-sky template for synchrotron emission in total intensity is challenging, both due to the paucity of low-frequency surveys with full sky coverage and the difficulty in disentangling synchrotron emission from free-free emission and AME at frequencies above $\sim$10\,GHz. We follow \citet{Thorne:2017} in basing our template on the Haslam 408\,MHz survey \citep{Haslam:1982} and in particular the reprocessing of the original maps by \citet{Remazeilles:2015}. We rescale the 408\,MHz map to 23\,GHz assuming a sky-constant scaling of the brightness temperature with $\nu^{-3.1}$. Finally, we smooth the resulting map with a $2^\circ$ Gaussian beam.

A key limitation of the current synchrotron total intensity template is that, despite the point source masking, we still find excess power from Galactic point sources at low Galactic latitudes near the Galactic center. This excess power can then leak into our final polarization maps through the polarization fraction tensor transformations (Equations~\eqref{eq:real2pt} and \eqref{eq:pt2real}). Our models are unaffected by this issue at $\ell \lesssim 3000$ and outside the Galactic plane. Addressing small-scale structures in bright regions will be deferred to a future release.

Constructing a synchrotron template in polarization is more straightforward than total intensity because both free-free emission and AME have very low levels of polarization. Thus, we directly employ the WMAP K-band (23\,GHz) $Q$ and $U$ maps\footnote{\url{https://lambda.gsfc.nasa.gov/data/map/dr5/skymaps/9yr/raw/wmap_band_iqumap_r9_9yr_K_v5.fits}} from Data Release~5 as our templates. As with the $I$ map, we smooth the $Q$ and $U$ templates with a $2^\circ$ Gaussian beam.

\subsection{Small-Scale Fluctuations in Spectral Parameters} \label{sec:spec_params}

\subsubsection{Methods Overview} \label{subsec:spec_params_overview}

Just as a map of Galactic emission at a single frequency is expected to have fluctuations at smaller angular scales than have been measured, maps of parameters governing its frequency dependence should also have small-scale fluctuations. We therefore introduce small-scale fluctuations to our spectral parameter template maps in a manner analogous to the amplitude template maps. This more realistically captures the complexity of small-scale Galactic emission than spatially uniform spectral parameters would.

As there is no sense of polarization in the spectral parameter maps, we do not utilize the polarization fraction tensor framework, but rather work with standard power spectra. Given a template map $\mathcal{T}$ of some spectral parameter $X$, we first fit the power spectrum of $\mathcal{T}$ with a power law $\mathcal{D}_\ell^{XX} \propto \ell^\alpha$ over a range of scales $\ell_0 < \ell < \ell_1$ where it is well-measured. We then generate a map of small-scale fluctuations using the power law fit. We multiply the resulting map by a modulation map, in analogy with the amplitude modulations described in Section~\ref{subsec:methodology}. Where there is more dust there is likely to be both more discrete environments (either within a beam or along the line of sight) and more variation of dust properties within a region (e.g., the interior versus exterior of a molecular cloud, a photo-dissociation region, etc.).

Finally, we combine the template map with the synthesized small scales using the filter function in Equation~\ref{eq:filter}. As with the amplitude parameters, we use the input template maps for the spectral parameters as-is in the inner Galactic plane, defined by the Planck \texttt{GAL097} mask.

Model maps are not guaranteed to resemble data far from the reference frequency. This is especially true in cases where random fluctuations have been introduced to the spectral parameter maps. As a rule of thumb, we recommend that the models presented here be used in the frequency range $\sim$1--1000\,GHz.

The application of this procedure to the dust and synchrotron spectral parameters is described in the following sections. The adopted fit parameters for each spectral parameter are listed in Table~\ref{tab:smallscale_specpar}.

\begin{deluxetable}{lccc}
    \caption{Model parameters for synthesizing spectral parameter maps at small scales}
   \tablehead{& \colhead{$\ell_0$} & \colhead{$\ell_1$} & \colhead{$\alpha$}}
   \startdata
   $\beta_d$ & 200 & 400 & 0.04 \\ 
   $T_d$ & 100 & 400  & -0.47\\
    $\beta_s$ & 10 & 36 & -0.61\\
    $c_s$ & 10 & 36 & -0.61  \\ 
    \enddata
    \tablecomments{Spectra are parameterized assuming $\mathcal{D}_\ell^{XX} \propto \ell^{\alpha_{XX}}$.}
    \label{tab:smallscale_specpar}
\end{deluxetable}

%\vspace{-.5cm}

\subsubsection{Dust Spectral Parameters}\label{subsec:dust_spec_params}

As described in Section~\ref{subsubsec:dust_model}, the dust emission in all models developed here is governed by the spectral parameters $\beta_d$ and $T_d$. For the $\beta_d$ and $T_d$ template maps, we employ the $\beta_d$ and $T_d$ maps\footnote{\texttt{COM\_CompMap\_Dust-GNILC-Model-Spectral-Index\_2048\_R2.00.fits}, \texttt{COM\_CompMap\_Dust-GNILC-Model-Temperature\_2048\_R2.00.fits}} derived from 2015 Planck GNILC component separation analysis \citep{planck2016-XLVIII}. These maps have lower CIB contamination than the \texttt{Commander} $\beta_d$ and $T_d$ maps \citep{planck2014-a12} as the methodology employed both spatial and spectral information to disentangle the CIB contribution to the total emission (see discussion in Section~\ref{sec:CIBcontamination}). The GNILC spectral parameter fits show a large variability over the sky, when compared with the \texttt{Commander} spectral parameters, especially at high Galactic latitudes \citep{planck2016-XLVIII}. The informative priors on spectral parameters used in the \texttt{Commander} analysis restricts spatial variation of the best fit of spectral parameters in regions of low signal-to-noise, with a distribution that peaks close to the prior \citep{planck2014-a12}. The spatial variability seen in the GNILC spectral parameters correlates to structures seen in the dust intensity. 

We fit the power spectra of the $\beta_d$ and $T_d$ template maps over the multipole ranges $200 < \ell < 400$ and $100 < \ell < 400$, respectively. We find that $\alpha_{\beta_d}= 0.04$ and $\alpha_{T_d} = -0.47$ over this range, both flatter than the $\alpha_{tt} = -0.80$ found for the dust amplitude. The adopted pivot multipole $\ell_1$ is larger than that used for dust amplitudes (see Table~\ref{tab:smallscale_par}) since the template maps employed here are derived from intensity-only data rather than a combination of total and polarized intensity. Thus, they remain signal-dominated at $\sim$4 times smaller angular scales. We employ the same modulation map as was used for the dust $i$ map (see Section~\ref{subsec:methodology}). The small-scale structure added to the spectral parameter maps induces $\beta_d$ and $T_d$ fluctuations that are typically $<10\%$ relative to the input templates at $N_\text{side}=2048$ ($\sim$a few percent in high Galactic latitude regions).

\subsubsection{Synchrotron Spectral Parameters}\label{sec:beta_s}
As described in Section~\ref{subsubsec:synch_model}, the synchrotron emission in all models developed here is governed by the spectral parameters $\beta_s$ and $c_s$. To build the large-scale template for the spatial variation of the synchrotron spectral index $\beta_s$, we begin with the full-sky $\beta_s$ map of \citet{Miville-Deschenes:2008}, obtained by combining the Haslam map in total intensity at 408\,MHz \citep{Haslam:1982} and WMAP 3\,yr K-band data \citep{Hinshaw:2007}. The map has an angular resolution of about $7^{\circ}$ and was employed by previous \texttt{PySM} synchrotron models \citep{Thorne:2017}.

Incorporating newer constraints on synchrotron emission at 2.3\,GHz from S-PASS, \citet{Krachmalnicoff:2018} determined that this $\beta_s$ map underestimates the true level of spatial variations in $\beta_s$ across the Southern Galactic Hemisphere. Thus, we follow \citet{Krachmalnicoff:2018} and rescale the $\beta_s$ map by first subtracting its mean value ($\bar{\beta} = -3.00$), multiplying the resulting map by a factor of $1.572$, and then adding back the mean value. This rescaling is further motivated by recent results in the Northern Galactic Hemisphere from QUIJOTE, which find even higher $\beta_s$ variation \citep{delaHoz:2023, Rubino-Martin:2023}. 

We next fit a power law to the power spectrum of this new template map over the multipole range $10 < \ell < 36$, finding $\alpha_{\beta_s}=-0.61$, then construct a map from this power spectrum extrapolated to high multipoles. The $\beta_s$ small scales are modulated employing the same modulation map as the synchrotron intensity map (Section~\ref{subsec:methodology}). As with the dust spectral parameter maps (Section~\ref{subsec:dust_spec_params}), we combine this high-$\ell$ map with the low-$\ell$ template following the filter function of Equation~\ref{eq:filter}.
 
For the synchrotron curvature parameter $c_s$, there are no readily available template maps. The existing \texttt{s3} model implements curvature as a sky-constant $c_s = -0.052$, consistent with the measurements from ARCADE \citep[$c_s=-0.052 \pm 0.005$,][]{Kogut:2012}. To model a reasonable range of spatial variability, we assume that fluctuations in the value of $c_s$ follow the synchrotron intensity at large angular scales. Specifically, we start from the Haslam map 408\,MHz $I$ map of \citet{Remazeilles:2015}, smoothed to a resolution of $5^\circ$. We then rescale this map with a linear transformation such that the (dimensionless) minimum and maximum pixel values over the full sky are $-0.076$ and $-0.041$ respectively, corresponding to the approximate range of $c_s$ measurements from ARCADE \citep[][Figure~6]{Kogut:2012}. The resulting $c_s$ map has a mean and standard deviation of $-0.0517$ and $0.0054$, respectively, within the ARCADE footprint, in good agreement with the ARCADE measurements.

To extend our $c_s$ map to angular scales $\ell > \ell_1 = 36$, we first generate a map of Gaussian random fluctuations with the same power law index as the $\beta_s$ map, i.e., $\alpha _{c_s}=\alpha _{\beta_s} = -0.61$. We modulate the resulting map with the Haslam 408\,GHz $I$ map, smoothed to a $5^\circ$ FWHM resolution. We renormalize the map through a linear transformation such that the (dimensionless) pixel values range from 0.1--2. Finally, the modulated high-$\ell$ map is combined with the low-$\ell$ template following the filter function of Equation~\ref{eq:filter}.

\subsection{Summary of New Dust and Synchrotron Models}

{\footnotesize
\setlength{\tabcolsep}{3pt} % Reduces space between columns (default is 6pt)
\begin{deluxetable*}{c l l l l l c}
\tablecaption{Summary of the \texttt{PySM}~3.4 models --- Dust. \label{table:summarydust}}
\tablewidth{0pt}
\tablehead{
\colhead{Tag} & \colhead{Spectrum Model} & \colhead{Templates} & \colhead{Templates} & \colhead{Frequency scaling} & \colhead{Frequency scaling} & \colhead{Stochasticity} \\
& & \colhead{Large scale} & \colhead{Small scale} & \colhead{Large scale} & \colhead{Small scale} &
}
\startdata
\texttt{d9} & Modified blackbody & \begin{tabular}{@{}l@{}}GNILC PR2 $I$ + \\ GNILC PR3 $Q$/$U$ \\ 353 GHz\end{tabular} & \begin{tabular}{@{}l@{}}Modulated + \\ polarization \\ fraction tensor\end{tabular} & Uniform $\beta_d$, $T_d$ & Uniform $\beta_d$, $T_d$  & --- \\
\hline
\texttt{d10} & \multicolumn{1}{c}{\textquotedbl} & \multicolumn{1}{c}{\textquotedbl} & \multicolumn{1}{c}{\textquotedbl} & \begin{tabular}{@{}l@{}}$\beta_d, \, T_d$ from \\ GNILC PR2\end{tabular} & Modulated & --- \\
\hline
\texttt{d11} & \multicolumn{1}{c}{\textquotedbl} & \multicolumn{1}{c}{\textquotedbl} & \multicolumn{1}{c}{\textquotedbl} & \multicolumn{1}{c}{\textquotedbl} & \multicolumn{1}{c}{\textquotedbl} & \begin{tabular}{@{}l@{}}$I$, $Q$, $U$ \& \\ $\beta_d$, $T_d$\end{tabular} \\
\hline
\texttt{d12} & \begin{tabular}{@{}l@{}}Six layers, each with a \\ different modified \\ blackbody\end{tabular} & \begin{tabular}{@{}l@{}}GNILC PR2 $I$ + \\ GNILC $Q$/$U$ \\ 353 GHz\end{tabular} & \begin{tabular}{@{}l@{}}Modulated + \\ Gaussian\end{tabular} & \begin{tabular}{@{}l@{}}Random realization \\ for each layer\end{tabular} & \begin{tabular}{@{}l@{}}Random realization \\ for each layer\end{tabular} & ---
\enddata
\tablecomments{All models have a maximum $N_\text{side} = 8192$, except \texttt{d12} (2048). The symbol \textquotedbl\ represents ``ditto'', indicating that the value is the same as the one above. The symbol --- indicates that a specific feature is not available in a model.}
\end{deluxetable*}
} 

{\small 
\begin{deluxetable*}{c l l l l l c}
\tablecaption{Summary of the \texttt{PySM}~3.4 models --- Synchrotron. \label{table:summarysynch}}
\tablewidth{0pt}
\tablehead{
\colhead{Tag} & \colhead{Spectrum Model} & \colhead{Templates} & \colhead{Templates} & \colhead{Frequency scaling} & \colhead{Frequency scaling} & \colhead{Stochasticity} \\
& & \colhead{Large scale} & \colhead{Small scale} & \colhead{Large scale} & \colhead{Small scale} &
}
\startdata
\texttt{s4} & Power law & \begin{tabular}{@{}l@{}}Haslam $I$ 408 MHz + \\ WMAP $Q$/$U$ 23 GHz\end{tabular} & \begin{tabular}{@{}l@{}}Modulated + \\ polarization \\ fraction tensor\end{tabular} & Uniform $\beta_s$ & Uniform $\beta_s$ & --- \\
\hline
\texttt{s5} & \multicolumn{1}{c}{\textquotedbl} & \multicolumn{1}{c}{\textquotedbl} & \multicolumn{1}{c}{\textquotedbl} & \begin{tabular}{@{}l@{}}$\beta_s$ from Haslam, \\ S-PASS, WMAP\end{tabular} & Modulated & --- \\
\hline
\texttt{s6} & \multicolumn{1}{c}{\textquotedbl} & \multicolumn{1}{c}{\textquotedbl} & \multicolumn{1}{c}{\textquotedbl} & \multicolumn{1}{c}{\textquotedbl} & \multicolumn{1}{c}{\textquotedbl} & $I$, $Q$, $U$ \& $\beta_s$ \\
\hline
\texttt{s7} & \multicolumn{1}{c}{Curved power law} & \multicolumn{1}{c}{\textquotedbl} & \multicolumn{1}{c}{\textquotedbl} & \begin{tabular}{@{}l@{}}\textquotedbl + $c_s$ \\ from ARCADE\end{tabular} & \begin{tabular}{@{}l@{}}\textquotedbl + $c_s$ \\ fluctuations\end{tabular} & ---
\enddata
\tablecomments{All models have a maximum $N_\text{side} = 8192$.}
\end{deluxetable*}
}

The new dust and synchrotron models implemented here all improve on previous models through new data-driven templates, and use of the polarization fraction tensor framework to model small-scale fluctuations. Multiple models of dust and synchrotron are provided to explore a range of astrophysical complexity allowed by current constraints. In brief, the \texttt{d9} and \texttt{s4} models use the new data-driven templates and include $I$, $Q$, and $U$ fluctuations up to the largest supported $\ell$ values, but these models have sky-constant spectral parameters and thus no frequency decorrelation (see Section~\ref{subsec:decorrelation} for discussion). In contrast, \texttt{d10} and \texttt{s5} employ data-driven maps of spectral parameters to which small-scale fluctuations are added, inducing frequency decorrelation. The \texttt{s7} synchrotron model provides an extra spectral parameter---curvature of the frequency spectrum---and thus additional complexity. The \texttt{d11} and \texttt{s6} models allow many realizations of \texttt{d10} and \texttt{s5}, respectively, to be generated in which the large scales are fixed to the data-driven templates and the small scales have the same statistical properties but differing spatial morphologies. Tables~\ref{table:summarydust} and \ref{table:summarysynch} provide a high-level summary of these models. Extensive model comparisons are made in Section~\ref{sec:validation}.

\section{Other Models Implemented} \label{sec:other_models}

In this section, we discuss the construction of other new foreground models, including the dust \texttt{d12} and CO models (\texttt{co1}, \texttt{co2}, and \texttt{co3}), that employ different methods from those outlined in Section~\ref{sec:small_scales}.

\subsection{Dust Layer Model} \label{sec:layers}
Evidence for variation of Galactic foreground emission laws as a function of frequency across the sky \citep[e.g.,][]{Krachmalnicoff:2018, Ade:2025b} implies that emission laws must also vary along the line of sight. As a consequence, even if dust emission can be described locally by a modified blackbody (Equation~\ref{eq:dust-emission-law}), a superposition of emission regions (i.e., an integral along the line of sight) is not a modified blackbody. In addition, if along the line of sight different line elements emit polarized radiation with different polarization angles, the frequency scaling can vary between $I$, $Q$ and $U$ \citep{Tassis:2015}. Evidence for this ``line-of-sight frequency decorrelation'' has been found in Planck data \citep{Pelgrims:2021}.

To model the complexity of multiple layers of dust along the line of sight, we base our \texttt{PySM}~3 implementation \texttt{d12} on the approach of \cite{Martinez-Solaeche:2018} in the PSM software\footnote{See version 2.3.3 \url{https://apc.u-paris.fr/~delabrou/PSM/psm.html}.}. 

The PSM was used to produce six maps of dust emission $S_{\nu_0}^k(\theta)$ at $\nu_0 = 353$ GHz (intensity and polarization, for a total of 18 HEALPix maps at $N_{\rm side}$=2048). It was also used to generate six maps of dust spectral index $\beta_d^k(\theta)$ and dust temperature $T_d^k(\theta)$ at $N_{\rm side}$=2048, one per layer, following the approach described in \cite{Martinez-Solaeche:2018}. This set of maps is used to form six different layers of dust emission.

The \texttt{PySM} software uses these maps as inputs, and generates the dust Stokes parameters maps as:
\begin{equation}
    S_\nu(\theta) = \sum_{k=1}^6 S^{(k)}_{\nu_0}(\theta)
    \left( \frac{\nu}{\nu_0} \right)^{\beta^{(k)}_d(\theta)}
    \frac{B_\nu(T^{(k)}_d(\theta))}{B_{\nu_0}(T^{(k)}_d(\theta))},
\end{equation}
where $S_\nu(\theta)$ stands for any of the three Stokes parameters of interest, $I$, $Q$ and $U$, at frequency $\nu$ at sky location $\theta$, and superscripts ${(k)}$ indicate the layer, from one to six.

In the implementation used here, the 353~GHz templates for the six emission layers are slightly different from those of \cite{Martinez-Solaeche:2018}. They are generated at HEALPix $N_{\rm side} = 2048$ instead of 512, with an updated version of the \citet{Martinez-Solaeche:2018} pipeline that supports different $N_{\rm side}$ values. Intensity maps are based on the Planck PR2 GNILC maps \citep{planck2016-XLVIII}, while large scale polarized emission maps are the GNILC-derived maps from \cite{Martinez-Solaeche:2018}. These are complemented by small-scale realizations with scale dependence matching the $TT$, $EE$ and $BB$ dust spectra measured in \cite{planck2016-l11A}, modulated by the large-scale local intensity and polarized intensity level. Specifically, for each layer $k$, and for each of $T$, $E$ and $B$, we model the final emission in harmonic space as:
\begin{equation}
    S^{(k)}_{\nu_0} = X^{(k)}_{\nu_0} h_\ell^{1/2} + Y^{(k)}_{\nu_0} (1-h_\ell)^{1/2},
\end{equation}
where $X^{(k)}_{\nu_0}$ is the observed emission deconvolved from the instrumental beam, $Y^{(k)}_{\nu_0}$ is a randomly generated set of harmonic coefficients with harmonic spectrum following Planck constraints \citep{planck2016-l11A}, and $h_\ell$ is a window defining the transition between the two regimes. For intensity, $h_\ell$ corresponds to a $5\arcmin$ beam window function, while for polarization, we use a $150\arcmin$ beam for the three nearest layers (which dominate at high Galactic latitude) and a $120\arcmin$ beam for the three farthest layers (which dominate near and in the Galactic plane).

The construction of these maps does result in some filtering out of the real small scale power in intensity and polarization, which is replaced by random fluctuations. Thus we expect departures from the non-Gaussian and non-stationary properties of real dust emission at these scales.

In this model, maps of spectral parameters ($\beta_d$ and $T_d$) are randomly generated for each layer as described in \cite{Martinez-Solaeche:2018}. As a consequence, they are not constrained to match the observed temperature and spectral index maps. While their statistics (e.g., amplitude, correlation between $\beta_d$ and $T_d$) are compatible with those of real data, they provide a different realization, which results in increased differences between the real sky and the model at frequencies increasingly far from the reference frequency, $\nu_0 = 353$~GHz.

\subsection{CO Models}\label{subsec:co_models}

{\small 
\begin{deluxetable*}{c l l l c}
\tablecaption{Summary of the \texttt{PySM}~3.4 models --- CO. \label{table:summaryco}}
\tablewidth{0pt}
\tablehead{
\colhead{Tag} & \colhead{Spectrum Model} & \colhead{Templates} & \colhead{Templates} & \colhead{Stochasticity} \\ 
& & \colhead{Large scale} & \colhead{Small scale} &
}
\startdata
\texttt{co1} & \begin{tabular}{@{}l@{}}Single line emissions at \\ 115, 230, 346~GHz\end{tabular} & \begin{tabular}{@{}l@{}}Planck PR2 \texttt{Type-1} maps \\ smoothed to $1^{\circ}$\end{tabular} & \multicolumn{1}{c}{---} & \multicolumn{1}{c}{---} \\
\hline
\texttt{co2} & \begin{tabular}{@{}l@{}}\textquotedbl + $0.1\%$ polarized\end{tabular} & \multicolumn{1}{c}{\textquotedbl} & \multicolumn{1}{c}{---} & \multicolumn{1}{c}{---} \\
\hline
\texttt{co3} & \multicolumn{1}{c}{\textquotedbl} & \multicolumn{1}{c}{\textquotedbl} & \begin{tabular}{@{}l@{}}Simulated high \\ Galactic clouds\end{tabular} & \multicolumn{1}{c}{---}
\enddata
\tablecomments{All models have a maximum $N_\text{side} = 2048$.}
\end{deluxetable*}
}

Galactic CO line emission is strong enough that the $J = 1\rightarrow0$, $J = 2\rightarrow1$, $J = 3\rightarrow2$ transitions have been detected even in the broad photometric Planck bands \citep{planck2013-p03a, planck2014-a12}. CO emission can be polarized \citep{Goldreich:1981} and so may be a relevant foreground for CMB polarization analyses \citep{Puglisi:2017}. In this section, we describe the implementation of three CO models into the \texttt{PySM}~3 framework. These models are summarized in Table \ref{table:summaryco}. 

Because of the intrinsically low degree of CO polarization and the long integration time required to achieve a significant detection, it is difficult to carry out CO polarization surveys, and wide area measurements have not yet been made. In the absence of data-based templates, the approach to modeling CO has been to assume a small degree of polarization applied to total intensity maps. For instance, \citet{Puglisi:2017} presented a model to simulate the polarized emission of CO lines in molecular clouds at high Galactic latitudes by considering the 3D spatial distribution of CO in the Galaxy.

All CO models implemented here are based on the CO $J = 1\rightarrow0$, $J = 2\rightarrow1$, $J = 3\rightarrow2$ \texttt{Type-1} intensity maps\footnote{\url{HFI_CompMap_CO-Type1_2048_R2.00.fits}} derived by \citet{planck2013-p03a} from Planck data. The CO maps are obtained exploiting the mismatches in the detector bandpasses to recover the CO from the CMB and the other Galactic foregrounds using the Modified Internal Linear Combination Algorithm (\texttt{MILCA}). The templates produced from this analysis are preferred to direct measurements from existing spectroscopic surveys \citep[e.g.,][]{Dame:2001} because they are not limited to low Galactic latitudes and have a uniform angular resolution over the sky.

\citet{planck2013-p03a} also produced \texttt{Type-2} maps by using single-channel maps at multiple frequencies to separate the CO emission from the astrophysical and CMB signals. However, this method is prone to foreground mis-modeling errors and to instrumental systematics, and the resulting maps contain known discrepancies with other CO observations \citep{planck2014-a12}. We thus prefer the \texttt{Type-1} templates for our purposes.

The \texttt{Type-1} CO templates have a native resolution of $10\arcmin$. We convolve each of the $J = 1\rightarrow0$, $J = 2\rightarrow1$, and $J = 3\rightarrow2$ maps with a 1$^\circ$ Gaussian beam to reduce noise contamination, especially at intermediate and high Galactic latitudes. Finally, the templates are downgraded to a coarser pixelization of $N_{\rm side} = 512$. These templates are the foundation on which all three of the CO models are built.

The first CO model \texttt{co1} adopts the $J = 1\rightarrow0,$  $J = 2\rightarrow1$ and $J = 3\rightarrow2$ templates as-is and assumes no polarization. 

The second CO model \texttt{co2} introduces a simple implementation of polarization. The $I$ maps of the three transitions used in \texttt{co1} are converted to $P$ maps using a sky-constant, user-defined polarization fraction $p$, where the default is $p = 0.1$\%. The $Q$ and $U$ maps are made from the $P$ maps using the polarization angle of the thermal dust emission as determined by a component separation analysis with \texttt{Commander} \citep{planck2014-a12}.

Finally, the third CO model \texttt{co3} not only accounts for the polarized emission as in \texttt{co2} but also includes small-scale emission from high-Galactic latitude clouds. We perform a dedicated simulation with the {\bf\texttt{LogSpiral}} model from \citet{Puglisi:2017}, which provided the best fit to Planck data, to produce maps of the contribution to the total and polarized CO intensity at sub-degree scales from molecular clouds at high Galactic latitudes. The $I$, $Q$, and $U$ maps generated by the simulation are then added to the $I$, $Q$, and $U$ maps of \texttt{co2} to produce the final \texttt{co3} model. This injection introduces CO emission at scales $\lesssim1^\circ$ in regions where the CO templates have very low signal-to-noise ratio. 
 
\section{Validation and Characterization of Models} \label{sec:validation}

In this section, we compare the new foreground models presented in this work both to data and to previous models. In Section~\ref{subsec:maps}, map-based comparisons are made to visually highlight differences between the models and observational data. In Section~\ref{sec:PS-validation}, we compare power spectra of the dust and synchrotron models with observations. After developing the methodology in Section~\ref{subsubsec:methods}, we demonstrate that the two-point statistics of the stochastic small scales are properly modulated for different regions of sky defined by Galactic masks of varying size. Section~\ref{sec:dust_validation} compares the dust models to Planck NPIPE maps at 353\,GHz. Section~\ref{sec:sync_validation} compares the \texttt{PySM} synchrotron models with the synchrotron map from the BeyondPlanck analysis for intensity, and with the Planck Revisited reprocessed maps for polarization. Additional validation is performed in the sky patch observed by the BICEP/Keck telescopes in Section~\ref{sec:BK_validation}. Section~\ref{subsec:decorrelation} compares the level of frequency decorrelation in dust $BB$ emission across the models, demonstrating that all models respect current constraints from Planck observations. Section~\ref{sec:CIBcontamination} quantifies the level of extragalactic contamination in the maps. In Section~\ref{sec:nongaussianity}, we assess the level of non-Gaussianity introduced by the polarization fraction tensor formalism.

Since the \texttt{d10} model is one particular realization of the stochastic model \texttt{d11}, and likewise \texttt{s5} is one realization of \texttt{s6}, we use \texttt{d10} and \texttt{s5} as representative of \texttt{d11} and \texttt{s6} wherever we make comparisons to those models. We find that other realizations of \texttt{d11} and \texttt{s6} produce qualitatively consistent results.

\subsection{Maps}\label{subsec:maps}

We provide several map-level comparisons between \texttt{PySM} dust models and observations. Specifically, we compare data from the Planck third data release PR3 \cite{planck2016-l03} and \texttt{PySM} dust models \texttt{d1}, which was widely used as a reference in previous versions of the \texttt{PySM}, and is described in \cite{Thorne:2017}, {\tt d9}, which is similar to \texttt{d1} but uses different input templates and includes small-scale fluctuations of template maps and spectral parameters, as described in Sec.~\ref{sec:small_scales}, and {\tt d12}, which is conceptually quite different from the other two models. The goal is to provide a qualitative comparison of the spatial characteristics of the models at small scales. We select $16.7^\circ \times 16.7^\circ$ patches of the sky, and compare the data and models at 353~GHz in both total and polarized intensity. We focus on two patches, one close to the Galactic plane with $(l,b) =(180^\circ,-10^\circ)$, and the second centered on the BICEP/Keck field at $(l,b) =(318^\circ,-61^\circ)$. 

\begin{figure*}[t!]
    \centering
    \includegraphics[height=0.393\textwidth]{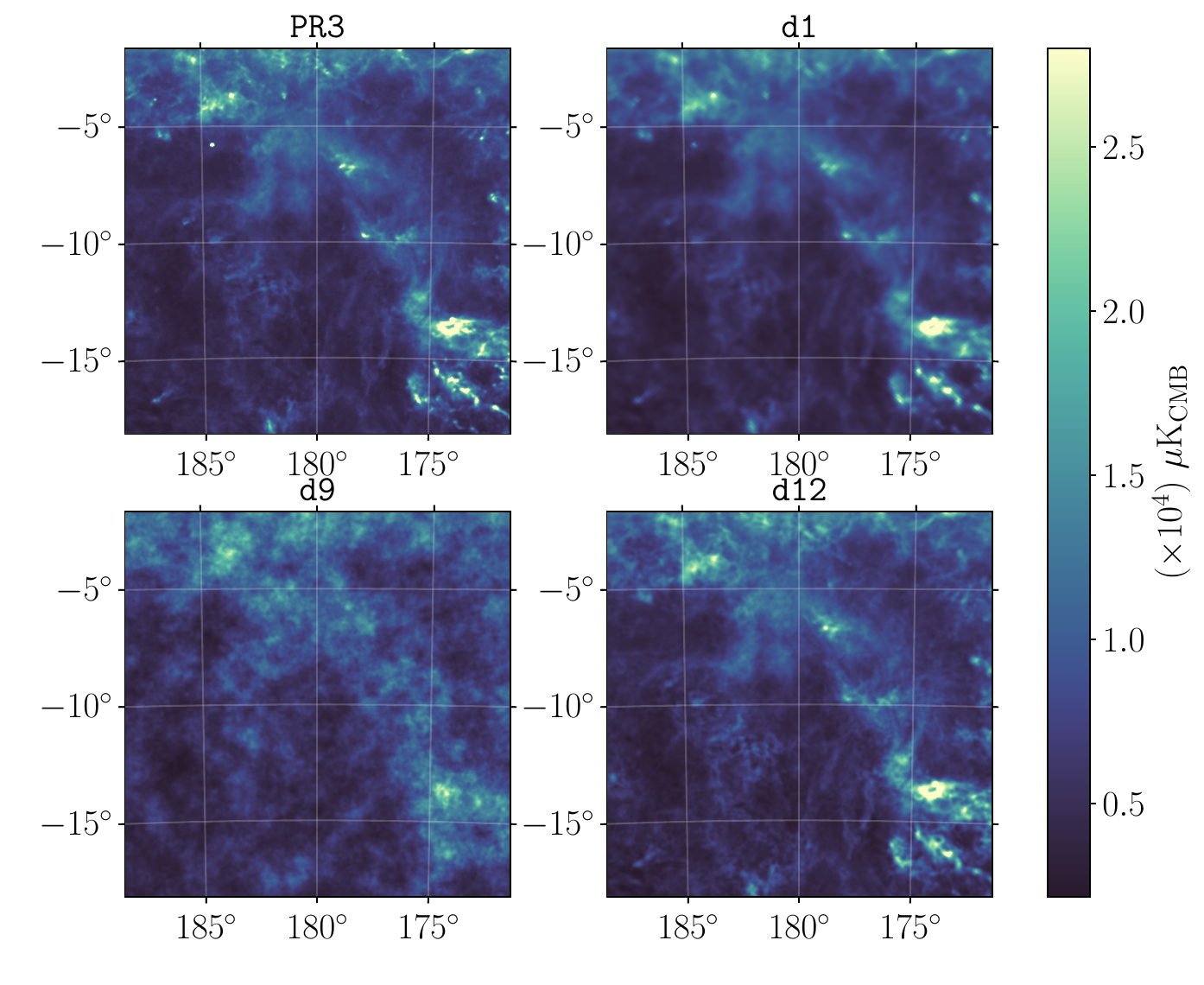}
    \includegraphics[height=0.393\textwidth]{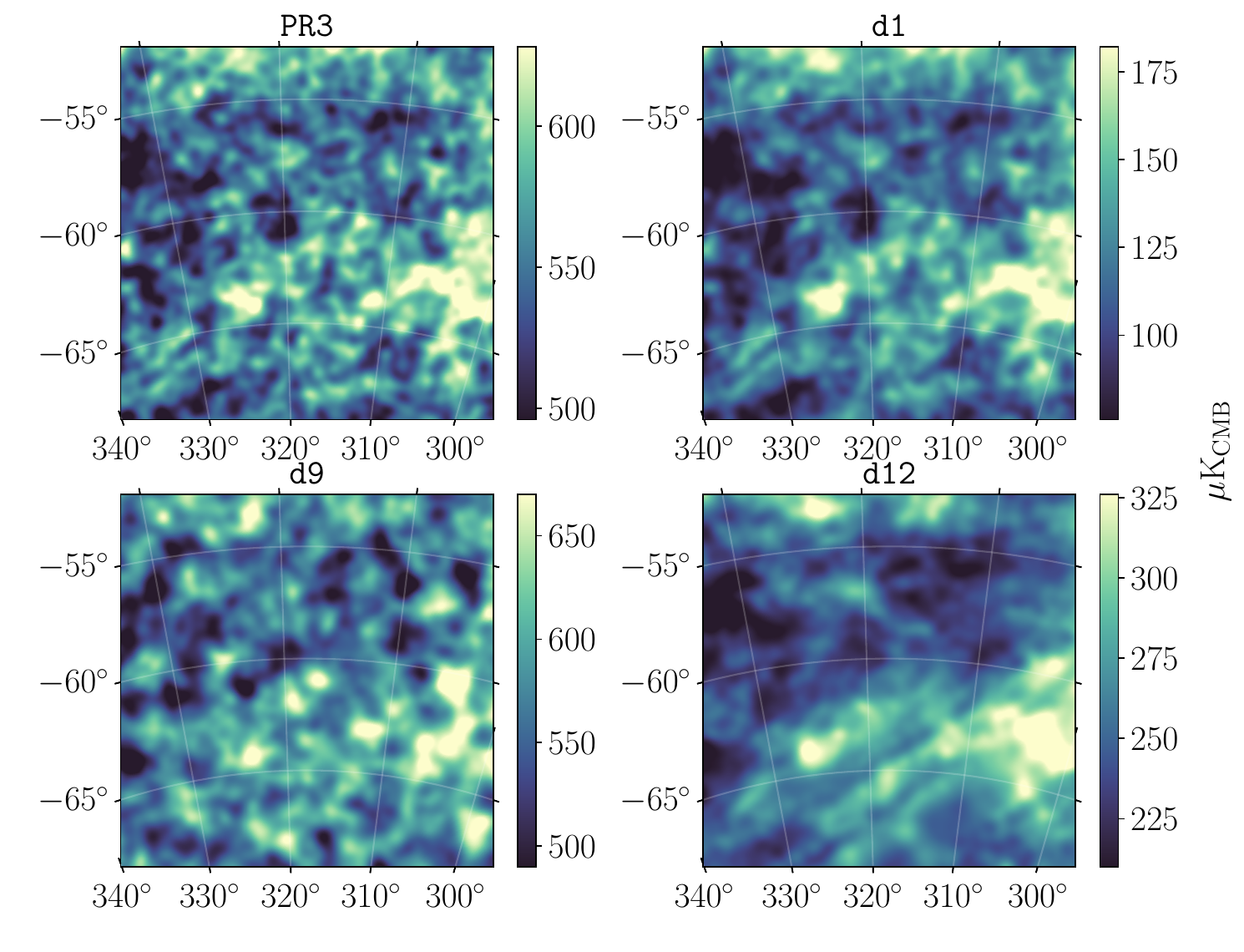}
\caption{Patches $(16.7^\circ\times 16.7^\circ)$ of dust intensity at 353\,GHz centered at $(l,b) =(180^\circ,-10^\circ)$ (left, ``low latitude'') and $(l,b) =(318^\circ,-61^\circ)$ (right, ``high latitude'' that is centered on the BICEP/Keck field) with an angular resolution of $4.94\arcmin$ for the {\tt d1}, {\tt d9}, and {\tt d12} dust models and Planck PR3 data.}    
\label{fig:353_int}
\end{figure*}

\begin{figure*}[t!]
    \centering
    \includegraphics[height=0.40\textwidth]{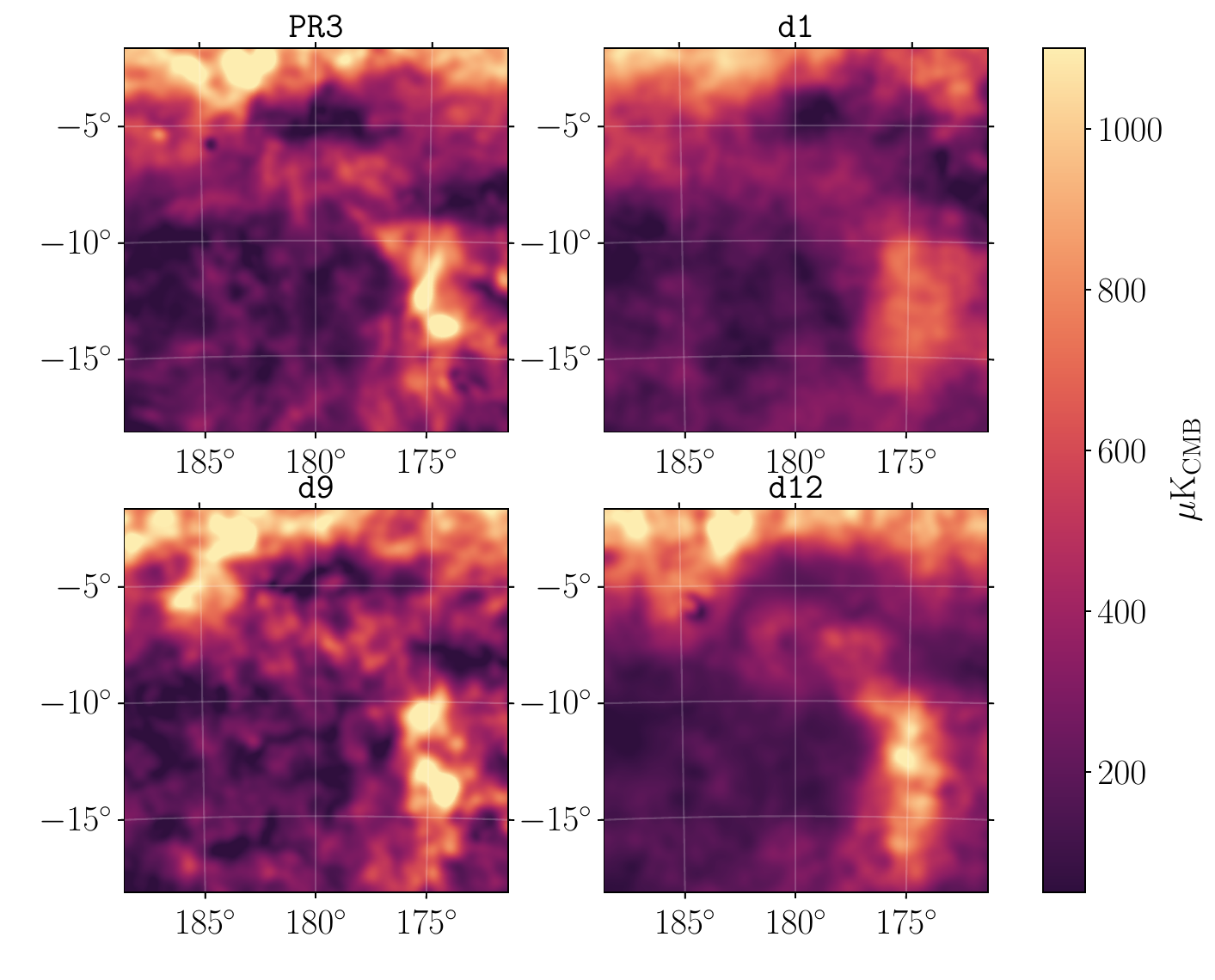}
    \includegraphics[height=0.40\textwidth]{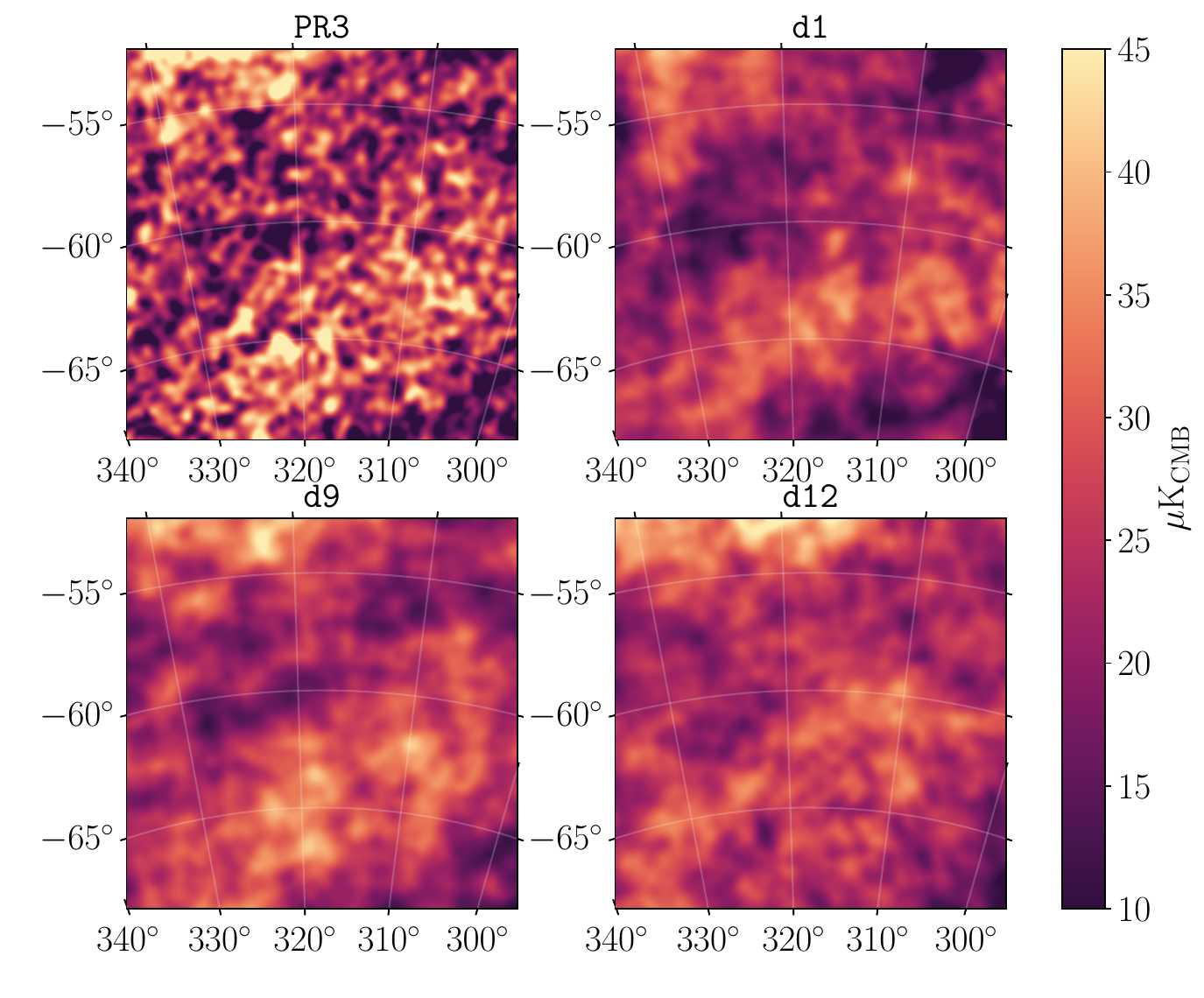}
    \caption{The same sky regions as Figure~\ref{fig:353_int} shown in 353\,GHz polarized dust intensity for the {\tt d1}, {\tt d9}, and {\tt d12} dust models and Planck PR3 data.}
    \label{fig:353_pol_int}
\end{figure*}

We integrate the dust models in the Planck passband \citep{planck2013-p03d}. For the comparison between total intensity maps, we subtract a Wiener-filtered CMB temperature anisotropy map from SMICA from the Planck map.

Intensity and polarization at 353~GHz are displayed in Figures~\ref{fig:353_int} and \ref{fig:353_pol_int}. In total intensity, model {\tt d9} filters more of the real data and generates more random small-scale structures than models {\tt d1} and {\tt d12}. The {\tt d12} model preserves the observed emission deconvolved from the instrumental beam up to 5\arcmin\,  in intensity (see Section~\ref{sec:layers}) whereas {\tt d9} preserves the dust template only at scales $\ell < \ell_1 = 100$, corresponding approximately to $1.8^\circ$. This is especially evident in high signal-to-noise regions near the Galactic plane, as shown in the left panel of Figure~\ref{fig:353_int}.

Note that the color scale is different for several of the models in the right panel of Figure~\ref{fig:353_int}. This is because the models have different zero points, mainly due to uncertainty in the CIB monopole. For the generation of model {\tt d12}, a monopole of 0.09~MJy/sr is subtracted from the DR2 GNILC intensity map. In the case of model {\tt d1}, it is due to the difference in the CIB monopole between PR2 and PR3 products, as model {\tt d1} is based on PR2 data.

In Figure~\ref{fig:353_pol_int}, we compare the observations and models for polarized intensity at a resolution of $30\arcmin$, to reduce the effects of instrumental noise in the PR3 data. For both of the regions---close to the Galactic plane and in the center of the BICEP/Keck patch---all three models are in reasonable visual agreement with the observations. 

\subsection{Power spectra}
\label{sec:PS-validation}
In this section we discuss the power spectrum validation of the \texttt{PySM} dust and synchrotron models. After introducing the methodology in Section~\ref{subsubsec:methods}, we examine dust and synchrotron emission over large sky areas in Sections~\ref{sec:dust_validation} and \ref{sec:sync_validation}, respectively, and finally analyze the BICEP/Keck patch specifically in Section~\ref{sec:BK_validation}.

\subsubsection{Methodology} \label{subsubsec:methods}
For our large-area validation, we employ sky masks that leave 80\%, 60\%, 40\% and 20\% of the sky unmasked. The mask choices are different for the dust and synchrotron maps, as described further below. 

After masking, we compute the power spectra using the \texttt{anafast} function from healpy\footnote{\url{https://healpy.readthedocs.io}}~\citep{Zonca:2019} and account for the masking effects by dividing the spectra by $f_{\rm sky}$, the second moment of the mask. We do not implement further corrections for mode-mixing caused by masking, as methods for such corrections assume that the field is free from any inherent mode-coupling~\citep[e.g.,][]{Hivon:2002}, which holds for the CMB but not for the highly non-stationary Galactic emission. 

We bin the power spectra into dynamic $\ell$-intervals optimized for map noise and sky fraction. Throughout this work, we present the results as $\mathcal{D}_\ell \equiv \ell(\ell + 1) C_\ell / 2\pi$, where $C_\ell$ is the power spectrum, with all values expressed in $\mu{\rm K^2_{CMB}}$ units. All power spectrum comparisons presented here are at the native resolution of the data they are compared with.

Since polarized dust emission is the primary foreground contaminant for CMB $B$-mode observations, we conduct additional validations of dust $B$-mode polarization in small sky patches to examine the properties of the injected small scales. Our results demonstrate that the spatial modulation of the small-scale realizations in the \texttt{d9}, \texttt{d10} and \texttt{d11} models does not, on average, introduce foreground power excess in high latitude sky regions, an improvement over previous \texttt{PySM} models.

\subsubsection{Dust Emission Over the Sky} 
\label{sec:dust_validation}
To evaluate the performance of the \texttt{PySM} dust models \texttt{d1}, \texttt{d9}, \texttt{d10}, and \texttt{d12} against real data across a wide range of scales, we present their power spectra computed using both large-area and small-area masks at 353~GHz, where the dust emission is dominant. First, we generate \texttt{PySM} dust model maps at a monochromatic frequency of 353~GHz and smooth them with a $4.82\arcmin$ Gaussian beam to match the resolution of the Planck 353~GHz channel. We color-correct the Planck NPIPE 353~GHz channel maps\footnote{\url{https://portal.nersc.gov/project/cmb/planck2020/}}~\citep{PlanckCollaboration:2020} to the same single frequency using a scaling factor 1/1.098~\citep{planck2016-l11A}, and subtract the CMB dipole from the NPIPE temperature maps. We then apply identical masks (defined below) to the mean-subtracted \texttt{PySM} dust models and NPIPE A/B detector-split maps. Finally, we compute the auto-spectra for the model maps and the cross-spectra for the NPIPE data splits, as cross-spectra minimize noise bias in the observational data. 

For our large-area comparison, we use a set of $2^\circ$-apodized Galactic masks\footnote{\texttt{HFI\_Mask\_GalPlane-apo2\_2048\_R2.00.fits}} from the Planck Legacy Archive. With eight in total, these masks cover a range of sky fractions, leaving between 20\% and 99\% of the sky unmasked. For our analysis, we take four representative masks: \texttt{GAL020}, \texttt{GAL040}, \texttt{GAL060} and \texttt{GAL080}, where the number in the mask name indicates the percentage of the sky available for analysis, i.e., $100 f_{\rm sky}$.

\begin{figure}
    \centering
    \includegraphics[width=\columnwidth]{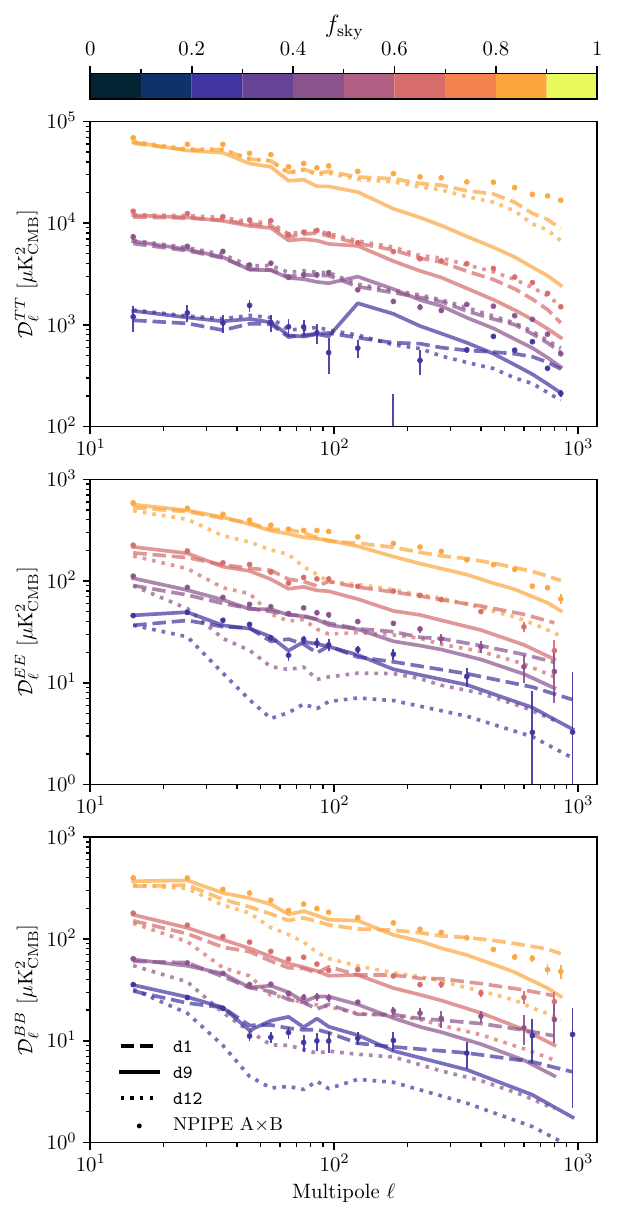}
    \caption{The 353~GHz $TT$, $EE$ and $BB$ power spectra for the \texttt{d1} (dashed lines), \texttt{d9} (solid lines) and \texttt{d12} (dotted lines) dust models, along with the NPIPE detector-split maps (circles), computed using the \texttt{GAL020}, \texttt{GAL040}, \texttt{GAL060} and \texttt{GAL080} Galactic masks. Each comparison set is colored to represent the respective sky fraction $f_{\rm sky}$. The \texttt{d10} model is not shown as it is identical to \texttt{d9} at 353~GHz.}
    \label{fig:largefield_power}
\end{figure}

In Figure~\ref{fig:largefield_power}, we compare the $TT$, $EE$, and $BB$ power spectra of the \texttt{d1}, \texttt{d9} and \texttt{d12} models against the NPIPE map. Since the \texttt{d10} model is identical to the \texttt{d9} model at 353~GHz, it is not shown separately. Error bars on the cross-spectra are derived from 200 NPIPE detector-split simulations.

We find that the \texttt{d1} and \texttt{d12} models generally match the observed $TT$ spectrum, although they underestimate the power at $\ell > 500$ when $f_{\rm sky} = 0.8$. The \texttt{d9} model shows similar agreement with observations at $\ell \lesssim 100$ across all sky fractions. However, its power declines more steeply at higher multipoles for $f_{\rm sky} = 0.6$ and $0.8$, and it exhibits a bump in the power spectrum around $\ell \sim 100$ for $f_{\rm sky} = 0.2$ and $0.4$. This artifact occurs because the parameters $l_1$, $c_1$, $c_2$ and $\gamma$ in Equations~\ref{eq:filter} and~\ref{eq:filter2}, which control the smoothness of the transition from large-scale template power to small-scale injection power, were optimized for large-sky polarization power spectra rather than total intensity power spectra. 

In polarization, \texttt{d12} significantly underestimates both $EE$ and $BB$ spectra, in particular on small scales. This may be at least partly due to the fact that independent random fluctuations are introduced in each layer on small scales, while the large-scale emission is correlated between layers. The independent fluctuations suffer line-of-sight depolarization in the sum of the layers, so that the small scale power is reduced relative to the large-scale power. On the contrary, both \texttt{d1} and \texttt{d9} align well with $EE$ data across all scales and sky fractions. \texttt{d9} demonstrates generally smooth power transitions for both $EE$ and $BB$ spectra, although some ripple artifacts emerge around $\ell \sim 100$ in smaller $f_{\rm sky}$ cases, again caused by the parameter optimization. As $f_{\rm sky}$ decreases the NPIPE $BB$ bandpowers rise at $\ell > 200$, exceeding the expected noise contribution as computed from simulations. The origin of this rise is unclear, but if it is not simply a statistical fluctuation, it may be due to unmodeled residual noise or instrumental systematic effects, which become more prominent in smaller sky patches. This behavior is not observed in $\texttt{d9}$, as its small-scale power is derived from template extrapolation.

\begin{figure}
    \centering
    \includegraphics[width=0.48\textwidth]{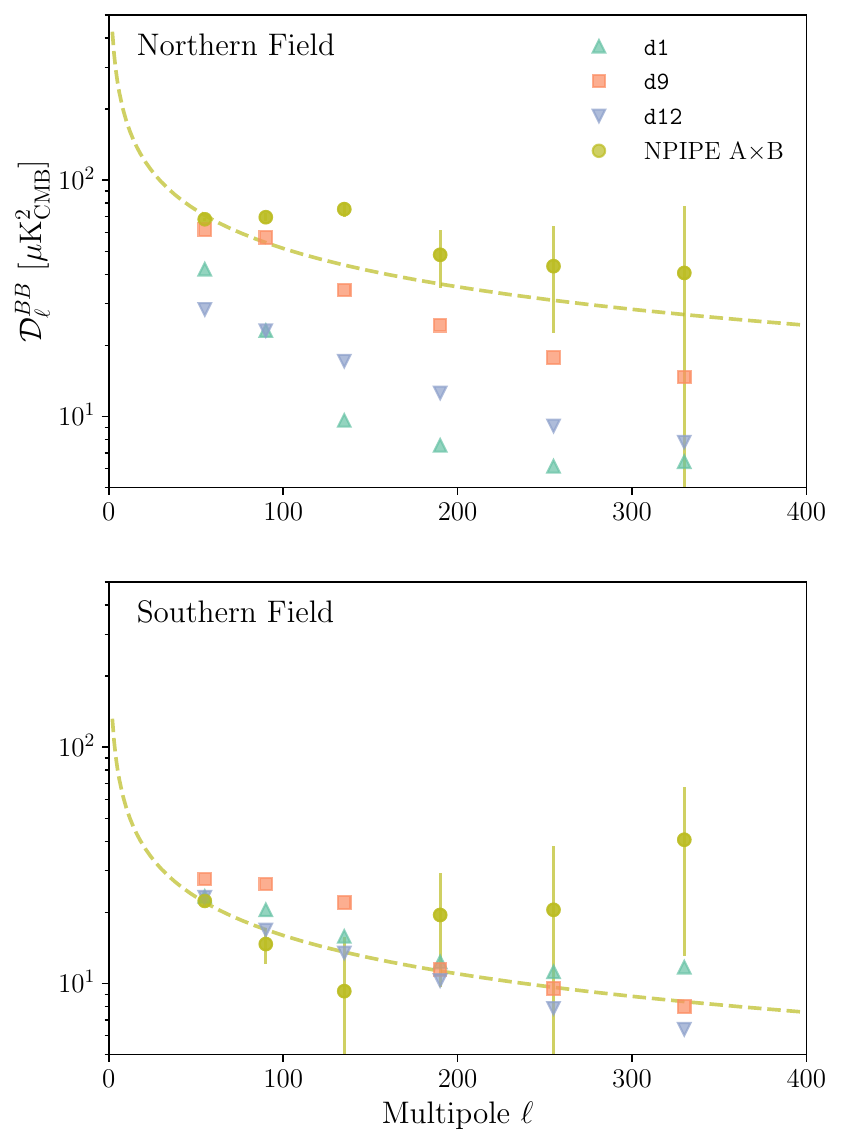}
    \caption{The binned $BB$ power spectra from the 353~GHz NPIPE detector-split maps and the dust model maps \texttt{d1}, \texttt{d9} and \texttt{d12} in two representative fields. The dashed lines indicate the best-fit of the fixed-index power law ($\mathcal{D}_\ell^{BB} = A \, \big( l/80 \big)^{\alpha}$ where $\alpha = -0.54$) to the NPIPE data points, with the fit largely driven by the first two bandpowers.}
    \label{fig:smallfield_power}
\end{figure}

\begin{figure*}
    \centering
    \includegraphics[width=2.1\columnwidth]{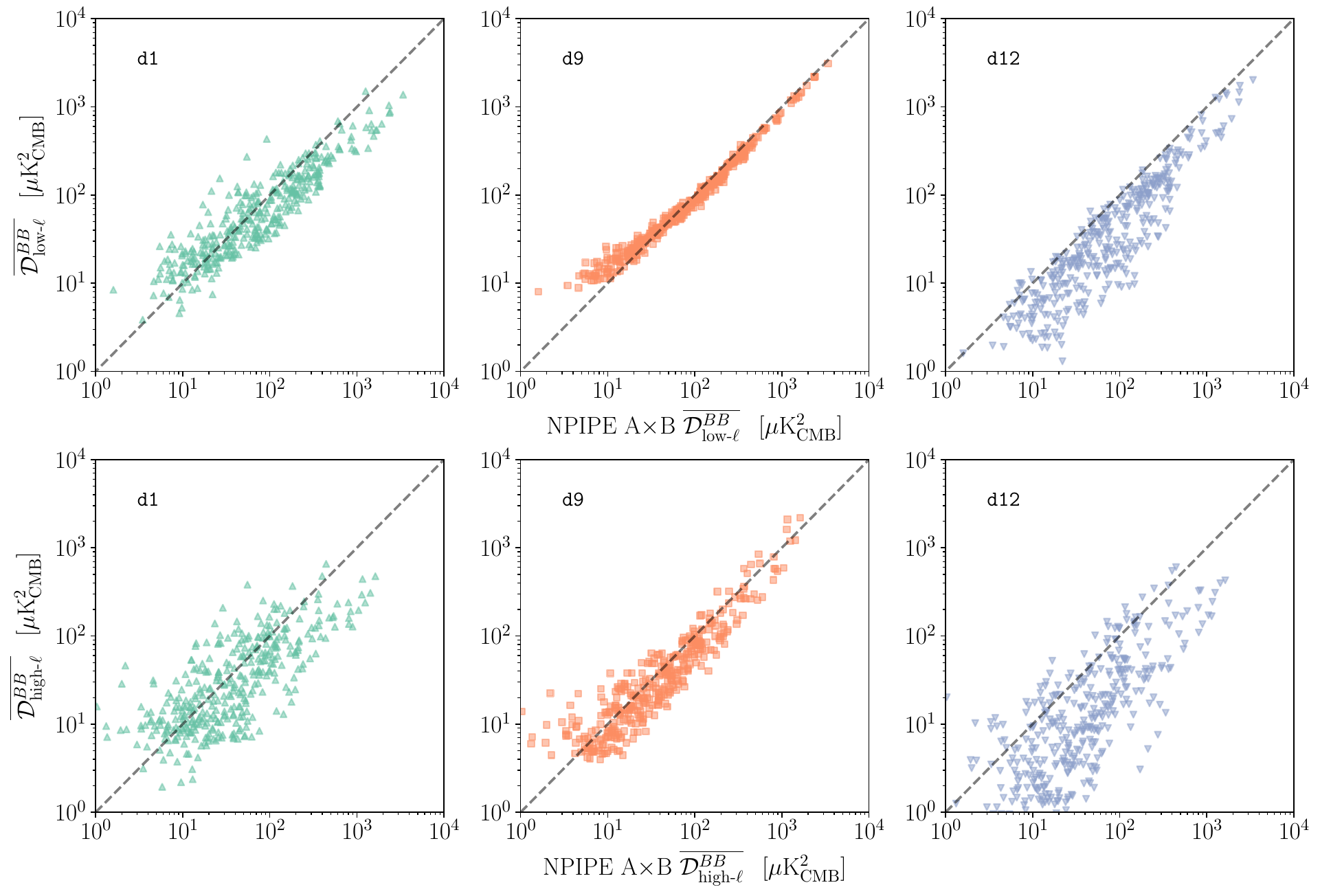}
    \caption{Scatter plots of the mean of the first two 353~GHz $BB$ bandpowers, $\overline{\mathcal{D}_{\text{low-}\ell}^{BB}}$, and the mean of the last four $BB$ bandpowers, $\overline{\mathcal{D}_{\text{high-}\ell}^{BB}}$, as illustrated in Figure~\ref{fig:smallfield_power}. The top panel compares $\overline{\mathcal{D}_{\text{low-}\ell}^{BB}}$ of \texttt{d1}, \texttt{d9}, and \texttt{d12} against that of NPIPE, while the bottom panel shows the same comparison for $\overline{\mathcal{D}_{\text{high-}\ell}^{BB}}$. Each data point represents the results from a circular sky patch with $|b| > 30^\circ$. Dashed lines indicate the 1:1 ratio.}
    \label{fig:smallfield_power_all}
\end{figure*}

For our small-area comparison, we instead follow the method described in \cite{planck2014-XXX} to define the masks for power spectrum computation. The full sky is divided into 768 patches with a HEALPix grid with $N_\text{side} = 8$. At the center of each patch, we construct a circular mask covering 400~deg$^2$, with the edges tapered by a $2^\circ$ FWHM Gaussian, yielding $f_{\rm sky} \sim 0.008$. Figure~\ref{fig:smallfield_power} presents representative results from two selected circular fields located at Galactic latitudes $|b| > 30^\circ$ in the northern and southern Galactic hemispheres. Since the computational cost of running simulations for cross-spectra error calculations in each small patch is high, the errors are estimated using analytical approximations of the cross-correlation matrix of power spectra~\citep{Tristram:2005}.

To assess whether the \texttt{PySM} models can accurately replicate the observed dust $BB$ power spectrum amplitude and scaling in $\ell$, particularly in small, high-Galactic latitude fields relevant to future CMB observations, we calculate the low-$\ell$ averaged bandpower $\overline{\mathcal{D}_{\text{low-}\ell}^{BB}}$ (over $40 \le \ell < 110$) and the high-$\ell$ averaged bandpower $\overline{\mathcal{D}_{\text{high-}\ell}^{BB}}$ (over $110 \le \ell < 370$) for both the model and NPIPE maps in each small field with $|b| > 30^\circ$. These metrics reduce the impact of noise fluctuation in the NPIPE $BB$ data in the circular sky patches, serving as proxies for comparing dust amplitudes between the model and the data at two different angular scales. The results are presented as scatter plots in Figure~\ref{fig:smallfield_power_all}.

Both the \texttt{d1} and \texttt{d9} models exhibit power spectral amplitudes that are generally consistent with NPIPE, but \texttt{d9} demonstrates a significant improvement in correlation, particularly increasing the correlation coefficient from 0.887 to 0.998 in the low-$\ell$ regime. This improvement stems from the combination of GNILC large-scale templates and small-scale modulation, which is especially evident in regions with a high signal-to-noise ratio, such as in the $\overline{\mathcal{D}_{\text{low-}\ell}^{BB}}$ comparison or in small fields with substantial dust amplitudes. 

However, in fields where $\overline{\mathcal{D}_{\text{low-}\ell}^{BB}}$ and $\overline{\mathcal{D}_{\text{high-}\ell}^{BB}}$ are smaller than 10~$\mu\rm{K^2_{CMB}}$, \texttt{d9} systematically overestimates the dust amplitude. To investigate the cause, we repeat the same analysis using GNILC template maps and find that the results closely follow the distribution of the \texttt{d9} scatter points at low-$\ell$ in Figure~\ref{fig:smallfield_power_all}. This suggests that the overestimation is likely due to a bias present in regions of the GNILC dust maps with low foregrounds, which propagates from large to small scales through the extrapolated power spectrum fit during our model construction. Conversely, $\texttt{d12}$ underestimates the dust amplitude in 94\% (85\%) of the small fields for low-$\ell$ (high-$\ell$). These opposing trends in \texttt{d9} and \texttt{d12} are also observed in individual small fields measured by ongoing $B$-mode experiments, such as the BICEP/Keck field, which will be discussed in Section~\ref{sec:BK_validation}, and in the SPIDER field independently analyzed by \cite{Ade:2025b}, although the latter results are presented at 150~GHz.

For the comparison of scaling in $\ell$, we introduce the ratio $\mathcal{R} \equiv \overline{\mathcal{D}_{\text{low-}\ell}^{BB}} \Big/ \overline{\mathcal{D}_{\text{high-}\ell}^{BB}}$ as another metric to describe changes in spatial power across the modulation scale. According to this definition, the fixed-index power law $\mathcal{D}_\ell^{BB} \propto \ell^{-0.54}$, derived from the analysis of a larger sky region with $f_{\rm sky} = 0.8$ by \cite{planck2016-l11A}, yields a value of $\mathcal{R} = 1.83$. This estimate is closely aligned with the small-field NPIPE data, which give a value of $\mathcal{R} = 1.85 \pm 0.93$. The \texttt{d1}, \texttt{d9}, and \texttt{d12} models instead produce slightly higher, though still consistent, values: $\mathcal{R} = 2.03 \pm 0.72$, $\mathcal{R} = 2.35 \pm 0.77$ and $\mathcal{R} = 2.26 \pm 0.91$ respectively. While the injected small scales in \texttt{d9} are also generated using an index of $\alpha = -0.54$ (Table~\ref{tab:smallscale_par}), this fit is performed in the $bb$ spectrum. 

During the development of the \texttt{PySM} models presented in the present study, we used the ratio $\mathcal{R}$ to evaluate the smoothness of the transition in bandpower from large to small scales in small regions. This approach ultimately guided us to adopt the improved modulation map construction method discussed in Section~\ref{subsec:methodology}, ensuring a uniform transition between scales. The remaining discrepancy in \texttt{d9} can be attributed to the non-linear transformation between $D_\ell^{bb}$ and $D_\ell^{BB}$.

\subsubsection{Synchrotron Emission Over the Sky} \label{sec:sync_validation}

In this section, we detail the validation of the new synchrotron models by comparing the power spectra with observations. For validating the \texttt{PySM} synchrotron total intensity models, we use the synchrotron map\footnote{\url{https://beyondplanck.science/products/files\_v1}} from the BeyondPlanck re-analysis of Planck LFI data \citep{Andersen:2023}. This map is at a reference frequency of 30\,GHz and has an angular resolution of $2^\circ$. We produce single-frequency total intensity maps for the different synchrotron models at 30\,GHz and smoothed to a FWHM of $2^\circ$.

We have purposely chosen an earlier release of BeyondPlanck for our analysis. Later data release versions of BeyondPlanck, or its successor CosmoGlobe, produce synchrotron intensity maps at 408~MHz~\citep{Watts:2023b}. The \texttt{PySM} synchrotron models \texttt{s5} and \texttt{s7} are not suitable for producing synchrotron simulations at 408\,MHz. This is a consequence of scaling the Haslam map from 408\,MHz to 30\,GHz with a constant $\beta_s$ for constructing the template for synchrotron intensity, and then applying a spatially variable $\beta_s$ to evaluate the model (see Section~\ref{subsec:spec_params_overview}). 

For synchrotron polarization validation, we compare the models with Planck Revisited synchrotron polarization maps\footnote{\url{https://portal.nersc.gov/project/cmb/Planck\_Revisited}} \citep{Delabrouille:2024}. These are the lowest-noise full sky maps of polarized synchrotron at 30\,GHz at $1^\circ$ resolution at the time of analysis. 

\begin{figure}
    \centering
    \includegraphics[width=0.46\textwidth]{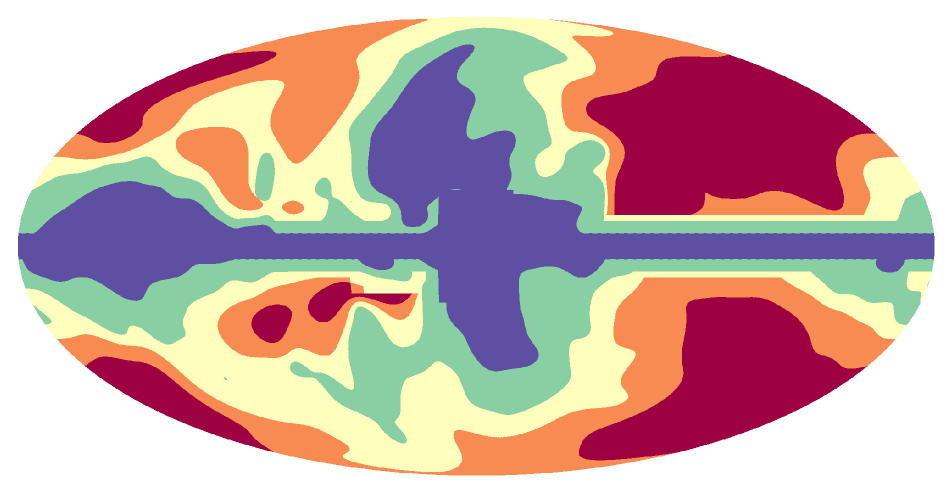}
    \caption{The four different Galactic masks used for the synchrotron validation. The red region is the 20\% mask; the orange region shows the additional sky coverage for 40\% mask; the yellow region shows the added coverage for 60\%; the green region is the added sky patch of 80\% mask. The purple region is excluded in all masks.}
    \label{fig:sync_masks}
\end{figure}

We do not use the Planck Galactic masks for the synchrotron power spectra validation. The Planck Galactic masks capture the shape of the Galactic dust signal, as it is the brightest foreground at CMB frequencies. The shape of the Galactic synchrotron signal differs significantly from the shape of the Galactic masks. Therefore, we construct masks for the synchrotron by thresholding the synchrotron polarized intensity smoothed with a $8.5^\circ$ beam. We additionally apply a Galactic latitude cut of $|b| < 4^\circ$ for the 80 \% sky coverage mask, $|b| < 8^\circ$ for 60 \% and $|b| < 10^\circ$ for both 40 \% and 20 \% sky coverage masks. This ensures that we are always excluding the Galactic plane. In Figure~\ref{fig:sync_masks} we show the coverages of these masks. 

\begin{figure}
   \centering
   \includegraphics[width=0.48\textwidth]{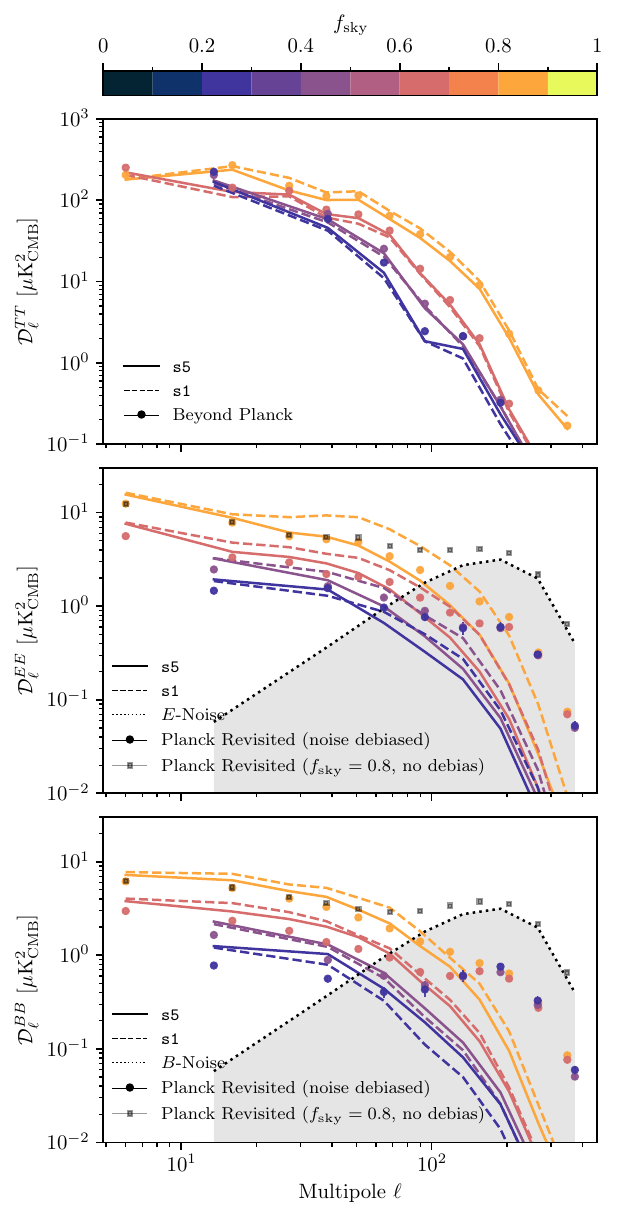}
    \caption{Comparison of the \texttt{PySM} \texttt{s1} and \texttt{s5} power spectra with observations over different sky fractions. The observed $TT$ power spectra are computed from the synchrotron intensity map from  BeyondPlanck Release 1~\citep{Andersen:2023} at 30~GHz with a $2^\circ$ beam. The observed $EE$ and $BB$ power spectra are derived from the Planck Revisited~\citep{Delabrouille:2024} polarized synchrotron map at 30~GHz with a $1^\circ$ beam. We show the $EE$ and $BB$ spectra with (colored markers) and without noise debiasing (gray markers). The noise power spectra, computed on 200 simulations, is showed by the gray shaded region.}
   \label{fig:Dl_sync_galmask}
\end{figure}

For the polarization analysis, we further apply the Planck Revisited point source mask to exclude the brightest point sources. The combined mask is apodized with a $2^\circ$--$6^\circ$ cosine taper, with the apodization length increasing as the sky fraction decreases. Using 200 noise realizations, we compute the mean and standard deviation of the noise spectra. The polarization power spectra are then noise-debiased, with error bars reflecting the noise standard deviation.

In Figure~\ref{fig:Dl_sync_galmask} we present $TT$, $EE$ and $BB$ power spectrum comparisons for models \texttt{s1} and \texttt{s5}. Since models \texttt{s4}, \texttt{s6} and \texttt{s7} yield nearly identical results, we display only the results for model \texttt{s5}. The $TT$ power spectra across all sky fractions for both models show excellent agreement with the BeyondPlanck synchrotron total intensity power spectra. For the $EE$ power spectra, model \texttt{s5} demonstrates good agreement in the multipole range where the signal-to-noise is $\gtrsim 1$ for the Planck Revisited polarized synchrotron power spectra. In contrast, model \texttt{s1} exhibits higher power for most multipoles compared to the Planck Revisited $EE$-synchrotron spectrum for the 80\% and 60\% sky fractions, although they align well for the 40\% and 20\% sky fractions. We also observe a distinct change in the shape of the $EE$ power spectrum for the \texttt{s1} model around the injection scale, $\ell_*\sim 36$ \citep{Thorne:2017}. The improved performance for model \texttt{s5} comes from the empirical optimization of the filters in Equation~\ref{eq:filter}. Both models have higher $BB$ power for 80\% and 60\% sky fractions.

For $\ell \sim 200$, we find a bump in the power spectrum of the Planck Revisited observations. This is likely caused by our inability to perform noise debiasing where the signal-to-noise ratio is low. However, the combined effect of residual point sources and suboptimal masking choices may also contribute to the residual bias. Based on the current results, we conclude that while our synchrotron models are validated for $\ell \lesssim 300$ for intensity, they are only consistent with observations up to $\ell \lesssim 100$ in polarization, due to the limited signal-to-noise of the polarized synchrotron data at smaller scales.

\subsubsection{Dust and Synchrotron Emission in the BICEP/Keck Patch}
\label{sec:BK_validation}
\begin{figure*}
    \centering
    \includegraphics[width=2.\columnwidth]{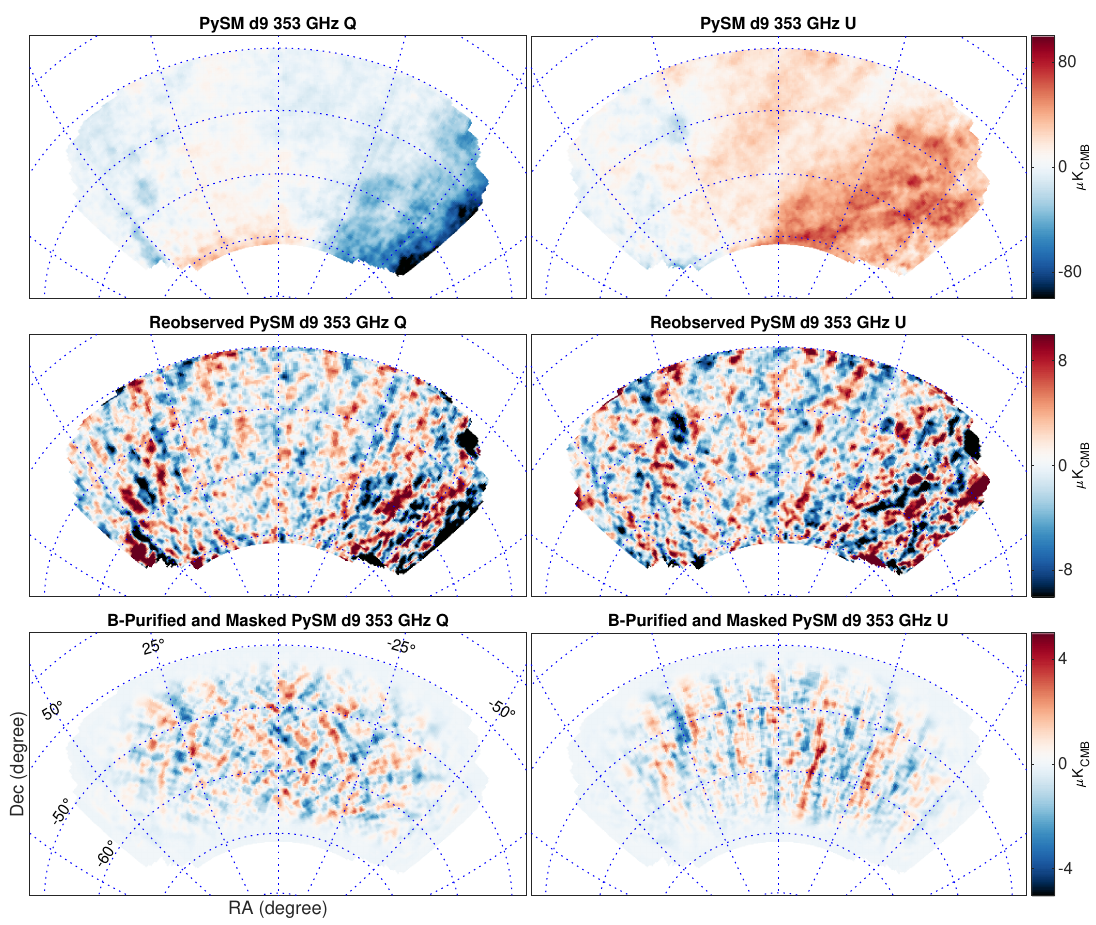}
    \caption{The \texttt{PySM} \texttt{d9} maps, reobserved maps and purified \& apodized maps in the BICEP/Keck sky patch at 353\,GHz. Each row shows the
    \texttt{PySM} $Q$ and $U$ maps reprocessed by the BICEP3 beam profile, observation matrix, and purification matrix \& apodization mask successively.}
    \label{fig:psym_BKmatrix}
\end{figure*}

The cleanest regions of sky in the southern hemisphere are particularly crucial for ongoing and future CMB experiments aimed at measuring primordial $B$-mode polarization. The most powerful current dataset for constraining the tensor-to-scalar ratio $r$ to date comes from the BICEP/Keck (BK) experiment~\citep[][henceforth ``BK18'']{Ade:2021}. This analysis includes observations up to and including the 2018 observation season, covering a $\sim 600$ square degree sky patch centered at RA 0h, Dec. $-57.5^{\circ}$, and incorporates NPIPE and WMAP data across the 23--353~GHz range from the same region. In the coming years, a collaborative effort between BICEP/Keck and the South Pole Telescope (SPT) will combine the BK maps with overlapping SPT-3G maps being used to ``delens'' the observed CMB $B$-modes, further tightening constraints on $r$~\citep{TheBICEP/KeckCollaboration:2024}. We hence provide a dedicated analysis of our models in this well-studied patch of sky.

\begin{figure}
    \centering
    \includegraphics[width=0.48\textwidth]{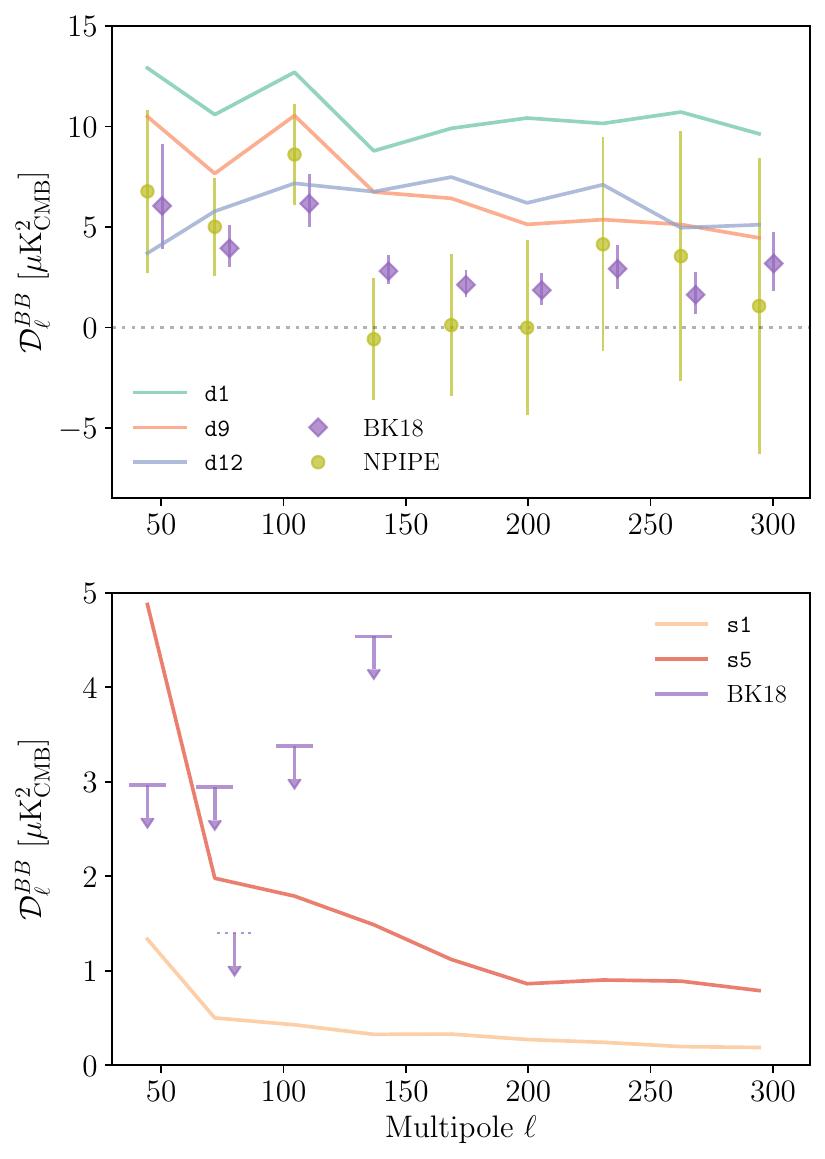}
    \caption{Comparison of the \texttt{PySM} and NPIPE $BB$ power spectra for the BK patch against the BK18 foreground measurements. The top panel shows the dust results at 353~GHz, while the bottom panel presents the synchrotron results at 23~GHz. These nine bandpowers correspond to the BK science bins used for delivering constraints. For dust, the BK18 data points (offset horizontally for clarity) represent the most probable values from the spectral decomposition analysis, with error bars indicating 68\% confidence intervals. For synchrotron, 95\% upper limits are shown instead, omitting those exceeding 5~$\mu\rm{K^2_{CMB}}$. The BK18 95\% upper limit $A_{s,\ell=80} < 1.4$~$\mu\rm{K^2_{CMB}}$, inferred from all nine bandpowers, is indicated with a dotted line.}
    \label{fig:BKfield_power}
\end{figure}

We utilize the analysis method illustrated in Figure~\ref{fig:psym_BKmatrix}, which shows \texttt{PySM} dust model \texttt{d9} maps that have been ``reobserved'' to replicate the impact of the BK timestream processing and map-making pipeline. In the first panel, the \texttt{d9} $Q$ and $U$ maps at 353~GHz (delta function bandpass) are masked to the BICEP3 observation region and convolved with the BICEP3 beam. The second panel displays the results after these smoothed maps are multiplied by the BICEP3 observation matrix, simulating the data reduction steps, such as filtering applied along RA scans and beam deprojection in the linear map-making process~\citep{BICEP2Collaboration:2016}. This process results in output maps that appear ``as observed'' through the BICEP3 pipeline, with large scales filtered out. In the third panel, the intermediate maps are multiplied by the corresponding purification matrix to remove $E$- and ambiguous-modes, leaving pure $B$-mode dominated $Q$ and $U$ maps with distinct ``cross-like'' and ``plus-like'' patterns,  respectively. Finally, the inverse noise-variance apodization mask is applied. 

By applying the same process to other \texttt{PySM} models at frequencies of interest (\texttt{s1}, \texttt{s4}, \texttt{s5}, \texttt{s6}, \texttt{s7} at 23, 30, 40~GHz and \texttt{d1}, \texttt{d9}, \texttt{d10}, \texttt{d11}, \texttt{d12} at 85, 150, 220, 270, 353~GHz), we generate a set of BK pipeline-propagated, pure $B$-mode model foreground maps that can be compared to BK measurements. We then calculate their $BB$ auto power spectra using the BK binning and present selected results. The color-corrected and noise-biased NPIPE $BB$ power spectrum at 353~GHz, derived from the NPIPE real and simulation maps processed using the same BK analysis method\footnote{\url{http://bicepkeck.org/bk18_2021_release.html}}, are also shown in this section. 

In Figure~\ref{fig:BKfield_power}, we present the $BB$ bandpower at 353~GHz for the \texttt{d1}, \texttt{d9}, and \texttt{d12} models, along with the bandpower at 23~GHz for the \texttt{s1} and \texttt{s5} models---these are the reference frequencies adopted in the BK analyses as well as the \texttt{PySM} models. The \texttt{PySM} power spectra are compared against the NPIPE power spectrum and BK18 foreground measurements. For dust, we show the 353~GHz dust component of the BK $BB$ spectra, derived from a reanalysis of the per-bandpower spectral decomposition at 95~GHz and 150~GHz, as shown in Figure~16 of~\cite{Ade:2021}. For synchrotron, since the BK18 analysis does not detect a synchrotron foreground, we conduct a similar analysis at 23~GHz and instead plot the 95\% upper limits on the synchrotron bandpowers. To enable a direct comparison, the \texttt{PySM} model and NPIPE values have been corrected for the suppression effects and partial-sky coverage which are applied by the beam, matrices and apodization mask. The figure therefore presents the full-sky $D_\ell$ values evaluated at delta function bandpasses.

All models exhibit excess dust $BB$ power relative to the measured values in this sky patch at 353~GHz. Among them, \texttt{d1} overestimates the amplitude by about a factor of three---the largest discrepancy---and it does not follow a power law in $\ell$, leading to high-$\ell$ values approximately twice as high as the other models. In contrast, \texttt{d9} and \texttt{d12} power show smaller deviation ratios, and generally reproduce the spatial variations observed in the BK data, although \texttt{d12} shows a drop in power at low $\ell$.

At large angular scales $\ell \lesssim 100$, the new \texttt{PySM} models are largely influenced by their underlying GNILC templates, while the BK18 constraints on dust $BB$ power come primarily from BK measurements at 220~GHz. The NPIPE power spectrum in the same plot reveals that the variations seen in the first three bandpowers of the \texttt{d9} model are directly inherited from its large-scale templates. The remaining amplitude discrepancy is likely caused by the bias present in the GNILC maps.

\begin{deluxetable}{lccccc}
    \caption{Comparison to BK18}
     \tablehead{& \colhead{85~GHz} & \colhead{150~GHz} & \colhead{220~GHz} & \colhead{270~GHz} & \colhead{353~GHz}}
    \startdata
    \texttt{d1}  & 2.42	& 2.63 & 2.81 & 2.94 & 3.13 \\
    \texttt{d9}  & 2.17 & 2.12 & 2.08 & 2.06 & 2.03 \\
    \texttt{d10} & 0.96 & 1.30 & 1.59 & 1.77 & 2.03 \\
    \texttt{d11} & 0.97 \(\pm\) 0.03 & 1.31 \(\pm\) 0.04 & 1.61 \(\pm\) 0.05 & 1.80 \(\pm\) 0.06 & 2.07 \(\pm\) 0.07 \\
    \texttt{d12} & 2.76	& 2.27 & 2.08 & 2.02 & 1.96 \\
    \enddata
    \tablecomments{Deviation ratios of the reobserved \texttt{PySM} dust $BB$ spectra to the dust component of the BK18 maximum likelihood model at each frequency. The \texttt{d11} values show the mean and standard deviation derived from 100 realizations, demonstrating that \texttt{d10} is indeed a representative realization from this stochastic model.}
    \label{tab:BB_dustratio}
\end{deluxetable}

Table~\ref{tab:BB_dustratio} further explores how the model amplitudes vary across frequencies, presenting the results of a single-amplitude-parameter fit between the \texttt{PySM} models and BK measurements down to 85~GHz. These deviation ratios are determined by fitting the nine bandpowers of the \texttt{PySM} models to the dust component of the BK18 maximum likelihood foreground model, which follows a modified blackbody power spectrum (with $T_d$ fixed at $19.6~\text{K}$) and is characterized by parameters $A_{d,\ell=80} = 4.4 \; \mu\rm{K^2_{CMB}}$, $\beta_d = 1.5$ and $\alpha_d = -0.66$. \texttt{d9} consistently maintains a deviation factor of around two, while \texttt{d1} and \texttt{d10} deviate less from the measurements at lower frequencies. This suggests that these two models, constructed using the $\beta_d$ maps from \texttt{Commander} \citep[$\beta_d = 1.51 \pm 0.06$,][]{planck2014-a12} and GNILC \citep[$\beta_d = 1.6 \pm 0.1$,][]{planck2016-XLVIII} as their respective spectral parameter templates, have too large a value of $\beta_d$ relative to the BK measured value of $1.49^{+0.13}_{-0.12}$ in this patch. The frequency-dependent trend observed in \texttt{d12} is incompatible with a simple change of $\beta_d$ and likely reflects the multiple-layer behavior inherent to this model. 

Overall, the \texttt{d9} and \texttt{d10} models demonstrate substantial improvements over \texttt{d1}, particularly in their ability to extrapolate the large-scale templates and capture the preferred small-scale power decay that the NPIPE data are unable to reveal in this specific field. However, while the discrepancies in amplitude are reduced, notable differences still persist when compared to the BK18 measurements. Given the way in which the models are constructed, it is not to be expected that they will track reality perfectly in small clean regions, especially in the presence of the GNILC template bias we have identified. Current power spectrum-based modeling techniques require a certain degree of global averaging of power spectrum parameters, which are known to vary across the sky in small patches~\citep{planck2016-l04, CordovaRosado:2024}. For instance, while dust amplitude can be modulated using large-scale templates, the power-spectral tilt $\alpha_d$ has to be fixed to a single value for the entire sky, which is inherently unrealistic. Further refinement of the new models, which could bring this sky patch into better agreement with the BK data, will be left for future studies. 

Lastly, we compare the synchrotron models with the BK results, focusing on \texttt{s1} and \texttt{s5}, as other new models yield similar outcomes. As illustrated in the lower panel of Figure~\ref{fig:BKfield_power}, at 23~GHz, \texttt{s5} exhibits excess power at $\ell \lesssim 50$ (or $\ell \lesssim 70$ when compared to the broad-band 95\% upper limit $A_{s,\ell=80} < 1.4$~$\mu\rm{K^2_{CMB}}$) and a robustly nonzero $\alpha_s$, while \texttt{s1} maintains relatively constant spatial power within the 95\% upper limits. This trend extends to 30~GHz and 40~GHz as well. The low-$\ell$ power spectrum discrepancies between models and data are likely because the models rely on the WMAP 23~GHz map as a template only up to $\ell = 38$, whereas the NPIPE 30/40~GHz maps, which favor a smaller $A_s$, also contribute to the BK constraints~\citep[][Figure~21]{Ade:2021}. While the BK18 data provide little constraining power on the parameters $\alpha_s$ and $\beta_s$, this comparison will be worth revisiting when new measurements from the BICEP Array telescope, especially its dedicated 30/40~GHz receiver, become available~\citep{Moncelsi:2020}.

\subsection{Decorrelation} \label{subsec:decorrelation}

One of the most challenging aspects of dust emission for CMB analyses is that its frequency dependence varies across the sky. If dust had the same SED everywhere, it would be sufficient to measure it at a frequency where it dominates the submillimeter emission and then subtract off all emission in lower frequency maps that is correlated with that template. The extent to which a map of dust emission at one frequency differs from a map of dust emission at a different frequency (aside from an overall normalization) is referred to as ``frequency decorrelation''. Frequency decorrelation has been identified as a major uncertainty for ongoing and upcoming analyses \citep{Ade:2021}.

The level of decorrelation between two frequencies $\nu_1$ and $\nu_2$ can be quantified by the spectral correlation parameter $\mathcal{R}_\ell$, defined as

\begin{equation} \label{eq:R_ell}
    \mathcal{R}^{XY}_\ell(\nu_1\times\nu_2) \equiv \frac{\mathcal{D}_\ell^{XY}(\nu_1\times\nu_2)}{\sqrt{\mathcal{D}_\ell^{XY}(\nu_1\times\nu_1)\mathcal{D}_\ell^{XY}(\nu_2\times\nu_2)}}
    ~~~,
\end{equation}
where $X$ and $Y$ can be any of $T$, $E$, or $B$ \citep{planck2016-L}. Here we focus on $\mathcal{R}_\ell^{BB}$.

The 353 and 217\,GHz Planck channels have the highest sensitivity to polarized dust emission and so furnish the current best constraints on the level of dust decorrelation. Analyzing 71\% of the sky over multipoles $50 \leq \ell \leq 160$, \citet{planck2016-l11A} found $\mathcal{R}_\ell^{BB} > 0.991$ (97.5\% confidence) between these two frequencies.

\begin{figure}
    \centering
    \includegraphics[width=\columnwidth]{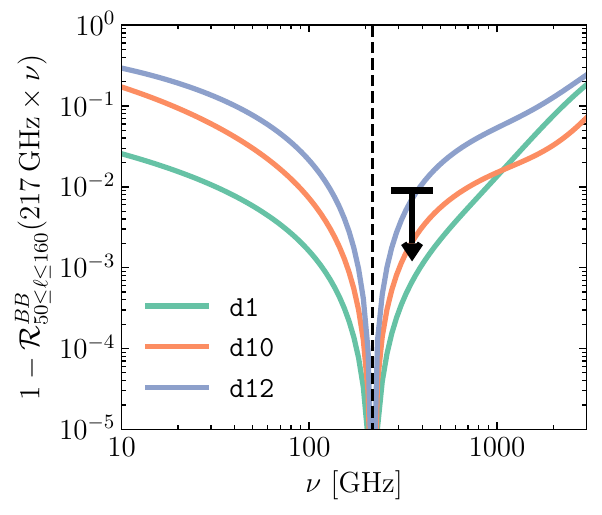}
    \caption{The decorrelation parameter $1-\mathcal{R}_\ell^{BB}$, where $\mathcal{R}_\ell^{BB}$ is the correlation parameter defined in Equation~\eqref{eq:R_ell}, for the \texttt{d1}, \texttt{d10}, and \texttt{d12} dust models between a reference frequency of 217\,GHz and variable $\nu$. $\mathcal{R}_\ell^{BB}$ is evaluated between $50 \leq \ell \leq 160$ over the Planck 70\% sky mask. The 97.5\% upper limit on decorrelation derived by \citet{planck2016-l11A} over 71\% of the sky ($\mathcal{R}_\ell^{BB}$ > 0.991) is indicated by the black arrow at 353\,GHz. The models span a large range in decorrelation level and each has a distinctive dependence on frequency.}
    \label{fig:decorrelation}
\end{figure}

In Figure~\ref{fig:decorrelation}, we compare the \texttt{d1}, \texttt{d10}, and \texttt{d12} models to this upper limit and analyze more generally how the level of decorrelation with respect to the 217\,GHz map changes with frequency. For each model, we compute $\mathcal{R}_\ell^{BB}\left(217\times\nu\right)$ over the multipole range $50 \leq \ell \leq 160$ over the Planck \texttt{GAL070} mask as a function of $\nu$. We adopt a uniform weighting to average the spectra over the broad multipole bin, since $\mathcal{D}_\ell^{BB}$ for dust scales roughly as $\ell^{-0.5}$ (see Table~\ref{tab:smallscale_par}). However, nearly identical results are obtained with uniform weighting in $C_\ell$ instead.

We find that all three models respect the upper limit set by \citet{planck2016-l11A}, with \texttt{d12} coming closest to saturating it ($\mathcal{R}_\ell^{BB}(217\times353) = 0.9930$), then \texttt{d10} (0.9979), then \texttt{d1} (0.9993). The models thus span a range of viable levels of decorrelation. Because the spectral parameters ($T_d$ and $\beta_d$) of the \texttt{d9} model do not vary across the sky, that model has no decorrelation by construction (i.e., $\mathcal{R}_\ell = 1$).

It is noteworthy that the frequency dependence of $\mathcal{R}_\ell^{BB}(217\times\nu)$ is different for each model. For instance, the \texttt{d1} model has less decorrelation than \texttt{d10} at frequencies near 217\,GHz, but more decorrelation at $\nu \gtrsim 1$\,THz. At frequencies much lower than the peak of the dust SED, the dust emission is in the Rayleigh-Jeans tail of the Planck function and is thus linearly proportional to $T_d$. In this limit, $T_d$ cannot contribute to frequency decorrelation, as changes in $T_d$ do not affect the ratio of the emission in two bands. At low frequencies, then, decorrelation is sensitive only to variations in $\beta_d$. In contrast, dust emission near the peak is a non-linear function of $T_d$, rendering changes in the dust temperature much more important to decorrelation. 

The \texttt{d1} model is based on component separation with \texttt{Commander} that placed a Gaussian prior on $\beta_d$ with $\sigma_{\beta_d} = 0.1$ \citep{planck2014-a12}. Therefore, most of the observed variability of the dust SED is explained in this model via fluctuations in $T_d$. Further, the \texttt{Commander} data model did not account for CIB fluctuations, and so the resulting maps of dust parameters have enhanced small-scale fluctuations from CIB contamination (see Section~\ref{sec:CIBcontamination}). In contrast, the component separation based on GNILC that led to the $T_d$ and $\beta_d$ maps used in the \texttt{d10} model \citep{planck2016-XLVIII} permitted large variations in $\beta_d$ and largely removed CIB fluctuations. Indeed, the \texttt{Commander}-based parameter maps have less variance in $\beta_d$ and more variance in $T_d$ than the parameter maps from the GNILC-based analysis \citep[see][Table~1]{planck2016-XLVIII}. The result is that \texttt{d10} predicts much larger values of decorrelation than \texttt{d1} at low frequencies where $\beta_d$ is the dominant driver of decorrelation, but somewhat smaller values at high frequencies where $T_d$ is the dominant driver.

At present, it is unclear whether \texttt{d1}, \texttt{d9}, \texttt{d10}, or \texttt{d12} is a more accurate description of the spatial variability of $T_d$ and $\beta_d$ and thus of frequency decorrelation. Constraints on decorrelation at higher frequencies where the models diverge most sharply---such as from CCAT \citep{CCAT-PrimeCollaboration:2023}, SPIDER-2 \citep{Shaw:2024}, Taurus \citep{May:2024}, and potential balloon-borne or space-based far-infrared/submillimeter platforms like BLAST Observatory \citep{Coppi:2024}, LiteBIRD \citep{LiteBIRDCollaboration:2023}, PICO \citep{Hanany:2019}, PIXIE \citep{Kogut:2025}, and PRIMA \citep{Dowell:2024}---would enable more accurate predictions for the level of decorrelation expected at CMB frequencies. As component separation techniques continue to improve by incorporating information in both the pixel domain and the harmonic domain, and as new data sets are becoming available, it is also timely to revisit the component separation analysis with the aim of making improved $T_d$ and $\beta_d$ maps.

\subsection{Extragalactic contamination} \label{sec:CIBcontamination}

\begin{figure}[ht!]
    \centering
    \includegraphics[width=\columnwidth]{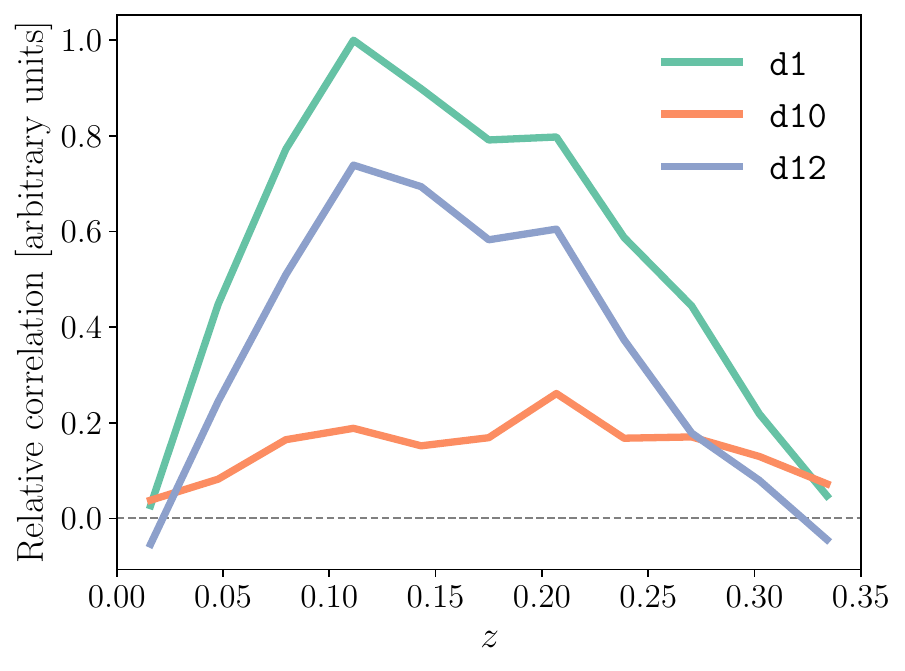}
    \caption{Relative extragalactic contamination (arbitrary units) in the new \texttt{d10} and \texttt{d12} dust total intensity templates, compared to the \texttt{d1} model. Contamination is quantified as the excess 857\,GHz emission within $11\arcmin$ of galaxies from the GLADE+ catalog, normalized to the maximum excess across all models, and plotted as a function of redshift $z$. The new dust templates contain less extragalactic contamination than older dust models because they are based on GNILC-processed Planck data. The improvement is most significant for \texttt{d10}.}
    \label{fig:extragal_contamination}
\end{figure}

We quantify the extragalactic contamination present in our dust models using a tomographic redshift-clustering technique \citep{Schmidt:2015, Chiang:2019}. Our intensity-based Galactic dust templates inevitably contain emission from both Galactic dust and external galaxies. As described in Section \ref{sec:dustamplitude}, the new \texttt{d9} and \texttt{d10} dust templates are derived from GNILC-processed Planck data, while older \texttt{PySM} dust templates used \texttt{Commander} data products. We thus expect that the new Galactic dust models are significantly less affected by CIB contamination than previous models. Here we quantify this contamination by measuring the cross-correlation between our dust models and the clustering of galaxies as a function of redshift in spectroscopic survey data. A perfect Galactic dust template would be uncorrelated with such clustering; the signature of CIB contamination is excess template emission correlated with galaxy clustering. 

Following a procedure similar to \citet{Chiang:2019}, we compute the cross-correlation between local fluctuations in the Galactic emission maps and galaxy density maps. The latter are constructed by stacking the number of galaxies from the GLADE+ catalog~\citep{Dalya:2022} in $N_{\rm side} = 2048$ HEALPix pixels within redshift bins of $\Delta z \sim 0.03$ over the range $ 0 < z < 0.35$, and then smoothing the resulting maps to $22\arcmin$. We compute this cross-correlation for each of the \texttt{d1, d10}, and \texttt{d12} dust emission templates at 857 GHz, at high Galactic latitudes (in the \texttt{GAL70} mask) within the catalog footprint. Figure \ref{fig:extragal_contamination} shows that while each of the new GNILC-based maps contain less extragalactic contamination than \texttt{d1}, the decreased contamination is more marked in \texttt{d10} than in \texttt{d12}. This is expected, as \texttt{d10} and similar templates have random small-scale structure, while \texttt{d12} retains more of the structure of the data at small scales. Although the shape of the correlation curves is affected by the galaxy distribution and catalog completeness as a function of redshift, the GLADE+ catalog is up to 90\% complete at $\sim500~\mathrm{Mpc}$~\citep[$z\sim0.1$;][]{Dalya:2022}.

We perform the same cross-correlation analysis to quantify the extragalactic contamination in the $\beta_d$ maps. We find, as expected, higher extragalactic contamination in the \texttt{d1} $\beta_d$ map than in the \texttt{d10} $\beta_d$ map. We emphasize that the mitigation of CIB contamination in the GNILC-processed Planck data affects both the spatial structure of the template frequency maps \textit{and} their extrapolation to other frequencies via the spectral parameter maps. CIB contamination in dust models can create spurious small-scale map structure, depress the map polarization fraction, and imply specious dust spectral parameters that affect the map frequency scaling.  

\subsection{Non-Gaussianity} \label{sec:nongaussianity}

\begin{figure*}
    \centering
    \includegraphics[width=\textwidth]{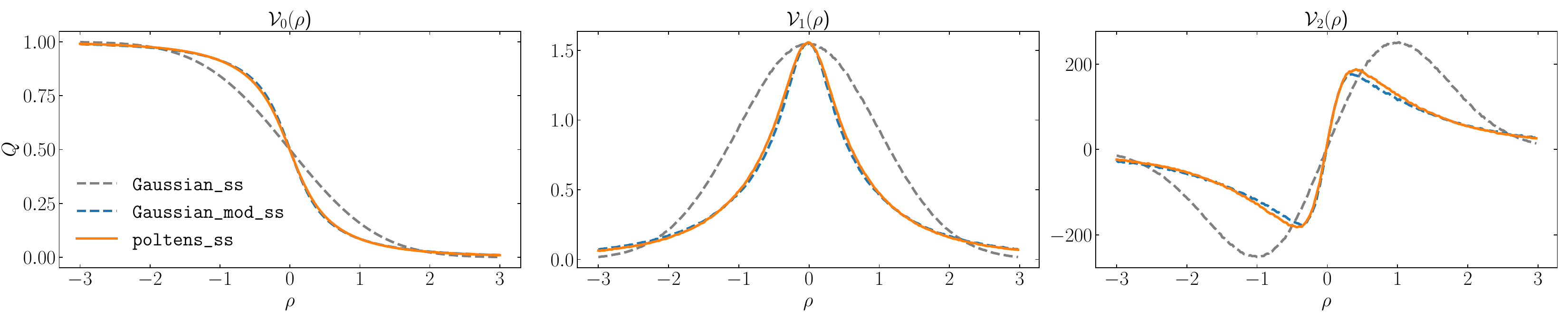}
    \caption{MFs for the small scales of three sets of $Q$ maps on the sphere with \texttt{GAL080} mask. The large scales are filtered out by excluding multipoles $\ell < 200$ in the maps, and we choose $\ell_\text{max} = 2048$ when obtaining the small scales. We show the MFs as a function of threshold $\rho$, for  a map employing  unmodulated (\texttt{Gaussian\_ss}, dashed gray) and modulated (\texttt{Gaussian\_mod\_ss}, dashed blue) Gaussian small scales. The \texttt{poltens\_ss} (solid orange) MFs show the case for modulated non-Gaussian small scales obtained by high-pass filtering the \texttt{d10} $Q$ map.}
    \label{fig:MF:sphere}
\end{figure*}

We quantify the level of non-Gaussianity in the small-scale dust emission generated through the polarization fraction tensor transformation used in dust model \texttt{d10} at 353 GHz (which is identical to \texttt{d9} at that frequency). To measure the non-Gaussianity in the maps we consider the Minkowski Functionals \citep[MFs,][]{Minkowski1903}, which are a common tool to quantify map-space, higher-order statistical properties \citep[e.g.,][]{Rahman:2021, Martire:2023, Carones:2024}. Hadwiger’s theorem implies that, for any $n$-dimensional excursion set defined with a threshold value $\rho$, there exist $n+1$ MFs that geometrically and topologically describe the morphology of the set \citep{hadwigerVorlesungenUeberInhalt1957}. In our case, for two-dimensional maps, we have three kinds of MFs. These are $\mathcal{V}_0$, $\mathcal{V}_1$, and $\mathcal{V}_2$, which correspond to the area, the perimeter and the connectivity of the excursion set (iso-intensity contour), respectively.

We use these MFs to compare the small-scale structure in \texttt{d10} to those of two other sets of maps: (i) maps where the small-scale structures are fully Gaussian and isotropic and (ii) maps where the small-scale structures are Gaussian but anisotropic across the sky. As described Section~\ref{subsec:methodology}, the small scale structures in the \texttt{d10} model are generated as a Gaussian random field in $i$, $q$, and $u$, and are multiplied by the $m_i$ and $m_p$ modulation maps before they are coadded to the large-scale maps and transformed back into $I$, $Q$ and $U$. We want to understand the impact of this effective modulation on the MFs, and therefore construct a set of maps that are amplitude-modulated versions of Gaussian-random-field maps. This allows us to disentangle any non-Gaussianity generated through the modulation from the potential non-Gaussianity introduced due to the polarization fraction tensor transformation.

The first set of maps contains isotropic small-scale structure. We construct the small-scale structure by generating a Gaussian random field with power-law power spectra in $TT$, $EE$, and $BB$, using power laws fit to the power spectrum of the \texttt{d10} maps on the \texttt{GAL097} mask in the multipole range $[800, 2000]$. This Gaussian random field is then high-pass filtered to remove power below $\ell_{cut} = 200$ in both total intensity and polarization, and co-added to the large-scale dust template.

For the second set of maps, we introduce a modulation in $I$, $Q$, and $U$ that mimics the effective modulation applied to $i$, $q$, and $u$ in the construction of the \texttt{d10} maps. We generate modulation maps $m_I$ and $m_P$ from $IQU$, following the same procedure as done in $iqu$ space (Equations~\ref{eq:mod_maps}--\ref{eq:mod_maps2}). We multiply the Gaussian isotropic small-scale map described above with these modulation maps, and then co-add them with the large-scale dust template. However, we want to ensure that the power spectra of modulated Gaussian small scales computed on the individual sky-fraction masks are as close as possible to the \texttt{d10} map. To achieve this, we adjust the modulation maps $m_I$ and $m_P$ by applying different multiplicative factors to non-overlapping regions of the sky. We multiply $m_I$ and $m_P$ on the \texttt{GAL40} mask by scalar factors that adjust the mean power spectrum of the modulated small-scale structure to be equal to the power spectrum of \texttt{d10} on the same mask. We repeat this process for each successively larger sky-fraction mask, applying the approximately order-unity multiplication factor to the non-overlapping sky region at each iteration. 

We thus consider three sets of maps with different small scales co-added: model \texttt{d10}, a map with purely Gaussian small-scale structure, and a map with modulated Gaussian small-scale structure. We refer to these three maps as \texttt{poltens\_ss}, \texttt{Gaussian\_ss}, and \texttt{Gaussian\_mod\_ss}, respectively.

We apply a high-pass filter with $\ell_{min} = 200$, using a smooth function similar to Equation~\ref{eq:filter2}, to retain only the small scales of these maps. We calculate the MFs both on the sphere and in several selected regions of the sky projected into a Cartesian projection. 

\subsubsection{Minkowski Functionals on the sphere}
Following the algorithm in \cite{Grewal:2022}, we calculate the MFs for the three $Q$ maps on the sphere, i.e., in HEALPix format, on the \texttt{GAL080} mask. We first normalize the maps by dividing each map by its standard deviation, and compute the MFs for iso-intensity contours in the range $[-3, 3]$ (Figure~\ref{fig:MF:sphere}). The MFs of the corresponding $U$ maps look very similar to $Q$ maps and are not shown here.

Figure~\ref{fig:MF:sphere} shows that when averaging over a large sky area, the MFs of \texttt{Gaussian\_mod\_ss} and \texttt{poltens\_ss} are almost identical, while the MFs of \texttt{Gaussian\_ss} differ substantially. The difference in MFs between \texttt{poltens\_ss} and \texttt{Gaussian\_ss} indicates the existence of non-Gaussianity in \texttt{poltens\_ss}, but the similarity between \texttt{poltens\_ss} and \texttt{Gaussian\_mod\_ss} demonstrates that the non-Gaussianity in \texttt{poltens\_ss} comes from the anisotropy in the maps, which originates from the modulation, rather than from the polarization fraction tensor transformation.

\begin{figure*}
    \centering
    \includegraphics[trim={0, 2.5cm, 0, 0},clip,width=\textwidth]{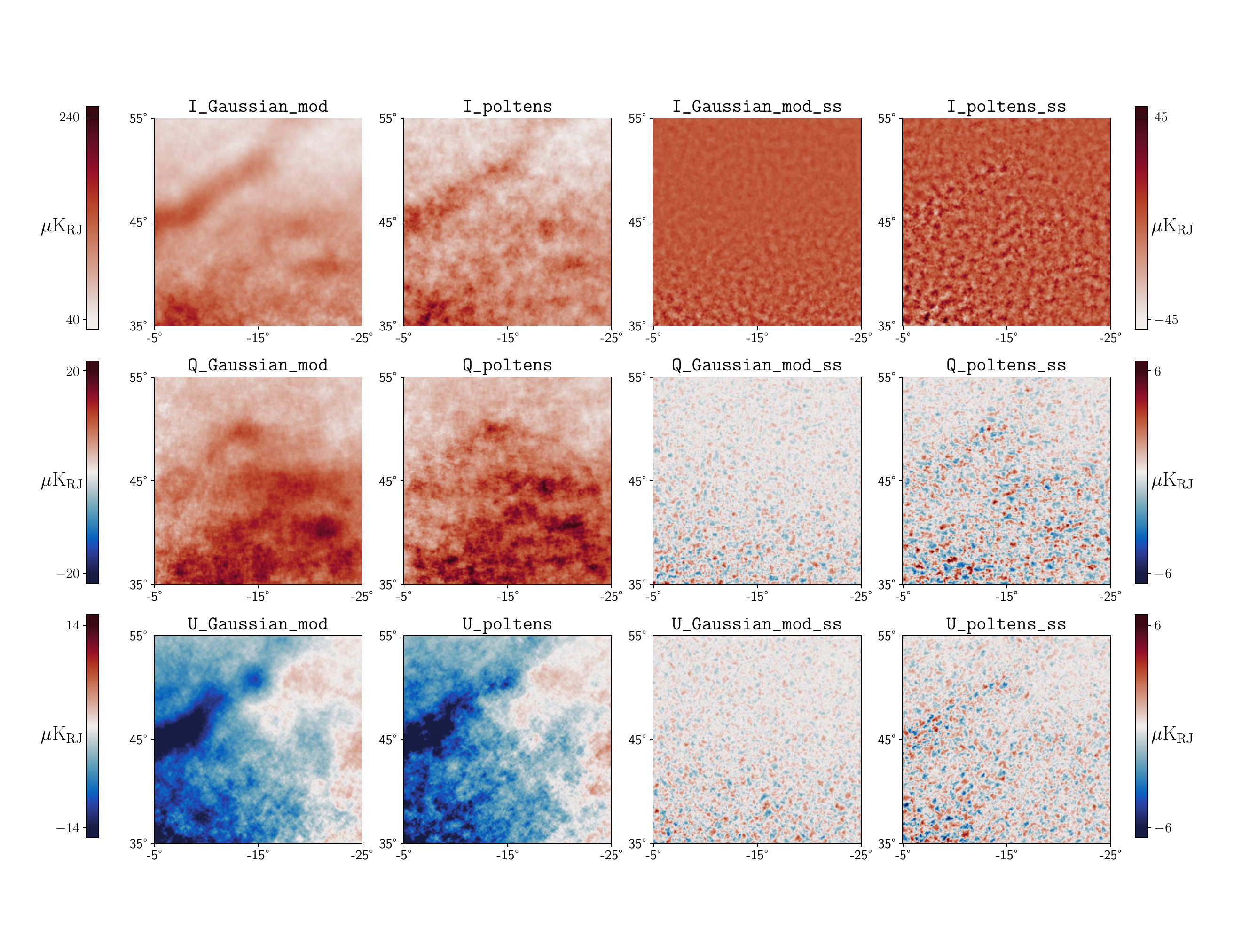}
    \caption{Zoom-in plots of an intermediate Galactic latitude patch centered at $(l, b) = (-15^{\circ}$, $45^{\circ})$. The $I$, $Q$, and $U$ maps (top to bottom) constructed using \texttt{Gaussian\_mod\_ss} and \texttt{poltens\_ss} are shown in the first two columns, respectively. The latter is the $\texttt{d10}$ map. The two rightmost columns show these maps high-pass filtered to preserve only scales with $\ell > 200$.
    The colorbar on the left indicates the pixel values in the left-most two columns in the units of $\mu \rm K_{RJ}$ and the colorbar on the right is for the last two columns.  }
    \label{fig:maps:patch2}
\end{figure*}

\begin{figure*}
    \centering
    \includegraphics[width=\textwidth]{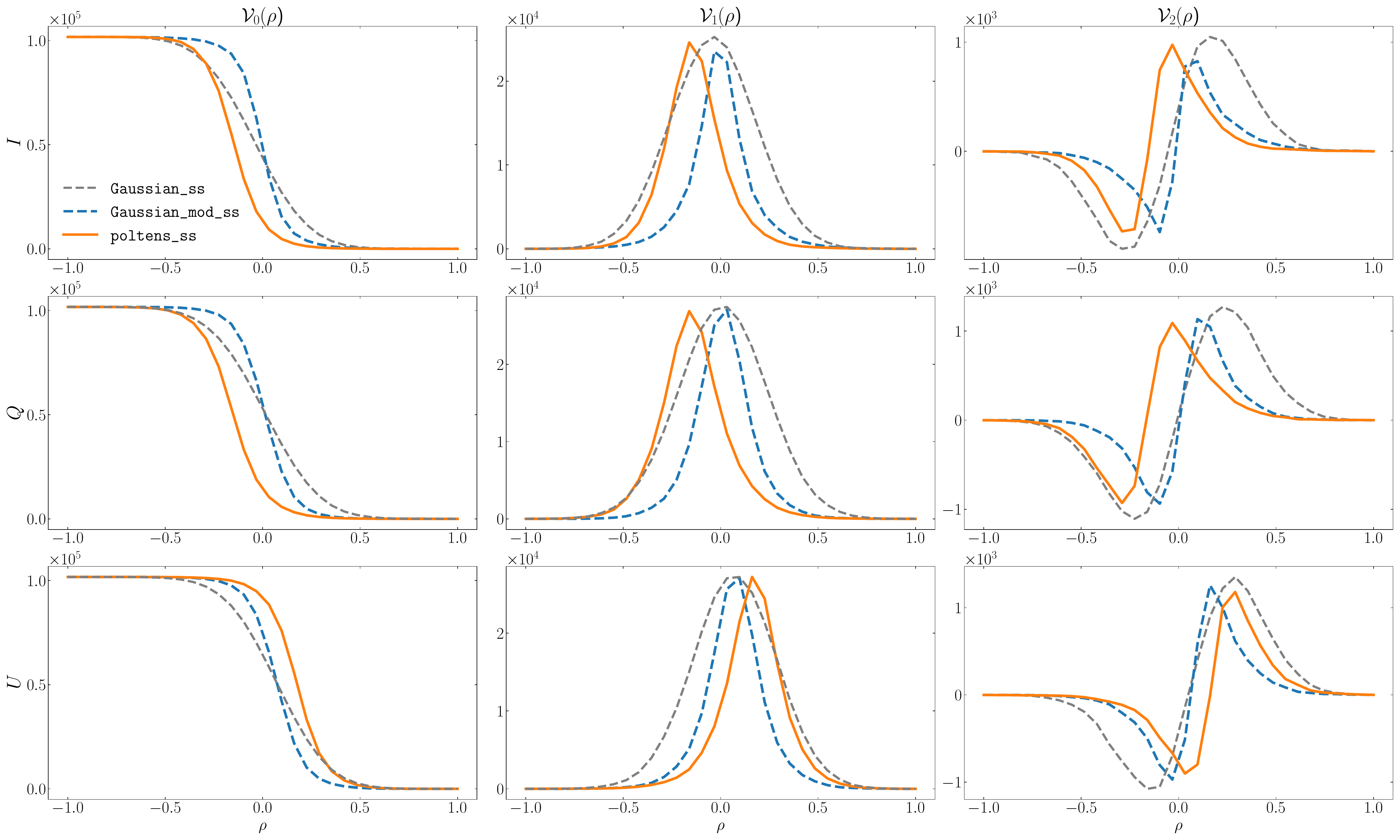}
    \caption{Minkowski Functionals as a function of the threshold $\rho$ for one realization of $I$, $Q$ and $U$  small scales ($\ell > 200$) in the patch centered on $(l, b) = (-15^{\circ}$, $45^{\circ})$ in Galactic coordinates. Each row shows three kinds of Minkowski Functionals. The blue dashed one is for \texttt{Gaussian\_mod\_ss} while the orange solid one is from \texttt{poltens\_ss}. We also show the \texttt{Gaussian\_ss} in dashed gray line as a comparison. }
    \label{fig:MF:patch2}
\end{figure*}

\subsubsection{Minkowski Functionals on small regions}
We consider an intermediate-latitude $20^{\circ}\times20^{\circ}$ region centered at $(l, b) = (-15^{\circ}$, $45^{\circ})$ to determine whether significant differences in the MFs between \texttt{Gaussian\_mod\_ss} and \texttt{poltens\_ss} sets of maps exist in small regions of sky. Those maps are shown in Figure~\ref{fig:maps:patch2}. We can see by eye that \texttt{poltens\_ss} contains structure that is not present in the \texttt{Gaussian\_mod\_ss} maps. We calculate the MFs of these small-scale maps, following \cite{Mantz:2008} for the calculation of MFs for a square patch. Before the calculation, we also rescale all the small scales linearly to be in the range $[-1, 1]$.

Figure~\ref{fig:MF:patch2} shows the MFs of the \texttt{Gaussian\_mod\_ss} and \texttt{poltens\_ss} maps presented in Figure~\ref{fig:maps:patch2}. In contrast with the large-area results presented in Figure~\ref{fig:MF:sphere}, in this case we do measure a departure of the \texttt{poltens\_ss} MFs from the \texttt{Gaussian\_mod\_ss} ones. This means that the polarization fraction tensor transformation introduced non-Gaussian small-scale structure, distinct from pure modulation effects, that is detectable over small sky regions. We conclude that the level of induced non-Gaussianity differs from region to region and does not significantly impact the statistical properties when averaged over large sky fractions.

\section{Future Outlook} \label{sec:discussion}

The Galactic emission models presented in this work are created from the latest data from large-area surveys like Planck and are informed by the latest literature constraints on the spectral behavior of Galactic emission components. These models incorporate some of the expected complexity of polarized emission at scales that are not yet well-constrained by data---in particular, the non-Gaussianity of interstellar emission structure. While this represents a step forward in the realism of these models over previous all-sky Galactic emission models, there are a number of idealizations that could be improved in future work.

The Galactic emission templates employed in this work are derived from component separation of the microwave sky and thus are subject to both noise and fitting degeneracies. This is particularly evident in the spectral parameter maps. The $T_d$ and $\beta_d$ template maps have a strong anti-correlation that can arise from fitting degeneracies, particularly at low signal-to-noise ratios \citep{Shetty:2009}. These maps also exhibit correlations with the dust intensity maps, particularly in diffuse, low-signal regions. While all dust models developed in this work respect current observational constraints on frequency decorrelation (see Section~\ref{subsec:decorrelation}), it is likely that some variation in the spectral parameter maps is attributable to noise rather than true astrophysical variation. Construction of improved spectral parameter maps is feasible by employing improved algorithms on existing data \citep[e.g.,][]{Watts:2024} and should be a high priority for future work.

By imposing a uniform cut-off in $\ell$ over which data-driven templates are employed, the models developed here are discarding high signal-to-noise information at small scales over portions of the sky where emission is bright. Future methodologies would ideally retain information from the templates on as small a scale as possible, with the cut-off scale varying across the sky. Artifacts from the splicing of large-scale templates with small-scale synthetic structures, including the ratio of spatial power across the modulation scale discussed in Section \ref{sec:dust_validation}, should be further mitigated. We note that the current methodology is fundamentally not designed to be able to reproduce the observed power spectra of all quantities over all sky areas, but future approaches could improve the behavior of the power spectrum beyond the modulation scale and for smaller sky masks. Further, particularly as data are being collected over partial sky areas by sensitive ground-based experiments, methods to incorporate partial sky templates should be developed.

The current models do not explicitly impose any non-zero parity-odd correlations in the polarized dust emission, i.e., the $TB$ and $EB$ correlations are zero in the power spectra that are used to extrapolate the small-scale dust emission structure. However, analysis of the Planck polarized dust emission finds a significant nonzero $TB$ correlation at $\ell \lesssim 600$ \citep{planck2016-l11A, Weiland:2020}. A proposed physical mechanism for the origin of the nonzero $TB$ signal is misalignment between dusty filaments and the projected magnetic field orientation \citep{Huffenberger:2020, Clark:2021, Cukierman:2023}: this picture also predicts the sign and amplitude of an expected nonzero $EB$ correlation. Future work could incorporate this parity-odd contribution to the polarized dust emission \citep[e.g.,][]{Hervias-Caimapo:2025}. Such models would be of particular use for the development of analysis techniques that seek to constrain signatures of cosmic birefringence or other parity-violating physics in the CMB \citep[e.g.,][]{Choi:2020, Minami:2020, Eskilt:2022, Ade:2025a}.

Future work could also improve the physical realism of the small-scale emission structure. The structure of polarized dust emission on small scales is highly filamentary \citep[e.g.,][]{Clark:2015, Halal:2024b}. Realizations of small-scale structure that have the particular character of non-Gaussianity that is realized in the sky could be generated using, for example, generative adversarial neural networks \citep{Krachmalnicoff:2021, Yao:2024}, a transformer-based super-resolution approach \citep{Halal:2024thesis}, or scattering-transform-based techniques \citep{Regaldo-SaintBlancard:2020, Delouis:2022, Mousset:2024}, which could be informed by the non-Gaussian statistics of diffuse ISM gas tracers \citep{Lei:2023, Lei:2025}. Forthcoming high-angular resolution dust maps from, e.g., SO \citep{Ade:2019}, the South Pole Observatory~\citep{TheBICEP/KeckCollaboration:2024}, or CCAT \citep{CCAT-PrimeCollaboration:2023} could also be used to inform a statistical description of small-scale dust structure. Another approach is to explicitly model the observed emission as arising from a superposition of many interstellar filaments, perhaps informed by the structure of interstellar neutral hydrogen emission \citep{Clark:2019}, or as a flexible parametric model \citep{Hervias-Caimapo:2022, Hervias-Caimapo:2025}.

As our knowledge of diffuse ISM physics improves, there are further opportunities for more sophisticated modeling of the Galactic emission. A recently discovered link between the mass fraction of neutral hydrogen in the cold neutral medium and the 353\,GHz dust polarization fraction implies that dust polarization models could be further improved by incorporating information on the phase structure of the neutral ISM \citep{Lei:2024}. It was also recently found that the three-dimensional structure of nearby interstellar dust has a measurable imprint on the dust polarization structure at CMB-relevant frequencies \citep{Halal:2024a}. Thus, as models of the 3D dust distribution \citep[e.g.,][]{Edenhofer:2024, Zhang:2025} improve, it may be possible to incorporate these data products into models of the dust polarization field. Starlight polarization data can further constrain the plane-of-sky magnetic field orientation of nearby dust structures \citep[e.g.,][]{Pelgrims:2024, Panopoulou:2025}. Although our focus is on data-driven sky modeling, numerical simulations \citep[e.g.,][]{Kim:2019, Stalpes:2024} could plausibly inform modeling choices for aspects of the models that are poorly constrained by data, e.g., higher-order statistics of the dust polarization \citep[e.g.,][]{Portillo:2018, Williamson:2024}. 

In addition to the small-scale structure of the diffuse dust and synchrotron emission, the real sky at these frequencies contains emission from discrete sources that are not explicitly included in our models. These include cold clumps \citep{Clancy:2023}, as well as planetary nebulae, supernova remnants, pulsar wind nebulae, and so forth \citep{Naess:2020, Guan:2021}. Nearby galaxies, while not technically within the scope of Galactic emission modeling, should also be incorporated as they are not included in existing simulations of the extragalactic sky.

\section{Recommended Model Suite}\label{sec:modelsuite}

{\small 
\begin{deluxetable*}{l l l}
    \tablecaption{Recommended Model Suite\label{tab:modelsuite}}
    \tablehead{\colhead{Complexity} & \colhead{Model set} & \colhead{Short description}}
    \startdata
    Low  & \texttt{d9, s4, f1, a1, co1} & \begin{tabular}{@{}l@{}}Small-scale emission fluctuations in amplitude only; \\ no frequency decorrelation in dust or synchrotron. \\ Unpolarized CO emission.\end{tabular} \\
    \hline
    Medium  & \texttt{d10, s5, f1, a1, co3} & \begin{tabular}{@{}l@{}}Extrapolation to small scales for both amplitude and \\ spectral parameters in dust and synchrotron. \\ CO polarization at the 1\% level.\end{tabular} \\
    \hline
    High  & \texttt{d12, s7, f1, a2, co3} & \begin{tabular}{@{}l@{}}Dust layer model, spatially varying synchrotron \\ curvature, polarized AME and CO.\end{tabular} \\
   \enddata
    \tablecomments{Summary of the suite of model sets described in Section~\ref{sec:modelsuite}. These are recommended combinations of models at three levels of complexity (low, medium, and high).}
\end{deluxetable*}
}

The models available for each emission component can be used in various combinations to form a number of unique Galactic sky models. While every user has this combinatoric freedom, we also prescribe a suite of recommended model sets. Our goal is to facilitate analyses that use a common set of assumptions. Community-wide use of this suite of Galactic emission model sets will enable easier comparisons between scientific forecasts for various experimental designs. Further, this will enable exploration of synergies between multiple experiments---for example, optimizing joint analyses of data from multiple telescopes.

Table~\ref{tab:modelsuite} details three model sets, representing increasing levels of complexity. The low complexity model set is highly idealized. Because this model implements synchrotron and dust variability in amplitude alone and not in their spectral parameters, these components exhibit no frequency decorrelation. The medium complexity model set includes Galactic emission properties that are expected physically, like sky-variable spectral parameters for the dust and synchrotron, extrapolated to small scales. The high complexity model set models Galactic emission properties that are physically realistic but as-yet undetected, like polarized AME and spatially varying synchrotron curvature. The model set with high complexity is based on the layer model for Galactic dust emission as detailed in Section~\ref{sec:layers}, and consequently includes line-of-sight frequency decorrelation. Both effects  result in higher levels of foreground residuals when compared to using medium complexity models, especially when traditional foreground cleaning techniques are applied.
The degree-scale decorrelation at frequencies dominated by the dust emission is near the maximum allowed by current constraints (Figure~\ref{fig:decorrelation}).

These model sets have already been constructed for CMB-S4, LiteBIRD, Simons Observatory, and South Pole Telescope 3G using the most recent bandpasses and beam models for each experiment, with model sets for other experiments in preparation. The results and documentation are publicly available in the CMB-S4 Data Portal at \url{https://data.cmb-s4.org}. The data products can be downloaded individually via direct HTTPS access as FITS files with no need of authentication, or they can be downloaded in batch mode using Globus to a Supercomputer or another Globus Endpoint.

All component models implemented in these suites are intended to be realistic over the range 1--1000\,GHz. The \texttt{d9}, \texttt{d10}, and \texttt{d11} models have dust spectral parameter maps constrained by far-infrared data and thus could be used up to $\sim$3000\,GHz. Caution is warranted in extrapolating any model beyond this range. Most component models are valid up to $N_{\rm side}=8192$, though \texttt{d12} and the CO models have a maximum $N_{\rm side}=2048$. The legacy \texttt{f1} free-free model has known high-$\ell$ artifacts even at $N_{\rm side} = 2048$: refinement of the free-free model is planned for future work.

\section{Summary and Conclusions} \label{sec:summary}

This work presents new models of Galactic emission in total intensity and polarization at frequencies relevant to CMB experiments ($\sim$1--1000\,GHz). The key conclusions of this work are as follows:

\begin{enumerate}
    \item We develop new models of small-scale, non-Gaussian Galactic dust and synchrotron emission based on the polarization fraction tensor framework. We combine realizations of this small-scale synthetic emission with well-measured large scales to produce maps with plausible levels of fluctuations at all angular scales, including in maps of spectral parameters (see Figure~\ref{fig:flowchart} for a schematic). The result is a set of all-sky models of dust and synchrotron emission that agree with observational data at large scales and that are consistent but stochastic at small scales.
    \item We implement the dust layer model of \citet{Martinez-Solaeche:2018} into \texttt{PySM}, providing an alternative dust model that has more realistic line-of-sight integration, producing line-of-sight frequency decorrelation.
    \item We implement three models of CO line emission and polarization based on Planck CO maps \citep{planck2013-p03a} and on models of high-Galactic latitude CO emission from the literature \citep{Puglisi:2017}.
    \item The dust and synchrotron models implemented in this work show improved overall agreement with observational data compared to previous \texttt{PySM} models. In particular, the small-area dust $BB$ amplitude and multipole dependence in the power spectra of the \texttt{d9}, \texttt{d10}, and \texttt{d11} models align more closely with those of the 353~GHz Planck NPIPE maps across the sky (Figure~\ref{fig:smallfield_power_all}).
    \item We quantify the frequency decorrelation with the decorrelation parameter $\mathcal{R}_\ell$ (Equation~\ref{eq:R_ell}). While all models are consistent with Planck constraints on decorrelation between 217 and 353\,GHz, they span a large range of decorrelation level and have distinct frequency dependence of $\mathcal{R}_\ell$ (Figure~\ref{fig:decorrelation}).
    \item We demonstrate that by using dust amplitude and spectral templates derived from the GNILC algorithm, the new \texttt{PySM} models contain significantly less contamination from the CIB than previous models (Figure~\ref{fig:extragal_contamination}).
    \item We find that the polarization fraction tensor framework yields maps with appreciable non-Gaussian small-scale structure in small patches (Figure~\ref{fig:MF:patch2}). However, we find that the non-Gaussianity greatly diminishes when averaging over large sky areas (Figure~\ref{fig:MF:sphere}).
    \item We define a set of three model suites denoted low, medium, and high complexity for use in CMB analyses (Table~\ref{tab:modelsuite}). The suites bracket the range of allowed dust frequency decorrelation (from none in the low complexity model to near maximal in the high complexity model) and in general progress from the simplest allowed by current constraints to featuring emission components that are plausible but that have not been detected (e.g., polarized AME). 
    \item All models presented here are available in the current \texttt{PySM}~3 release. Synthetically observed sky maps using these models are available at the CMB-S4 Data Portal \url{https://data.cmb-s4.org} using the beams and bandpasses of CMB-S4, LiteBIRD, Simons Observatory, and South Pole Telescope 3G.
\end{enumerate}

The construction of Galactic emission models is necessarily an iterative process: as more data at higher sensitivity, higher angular resolution, and different frequencies become available, more well-constrained and thus more realistic models can be constructed. While significant uncertainties remain, this work represents a new iteration of this ongoing effort to make CMB polarization science robust to the complexities of Galactic emission in both experiment design and data analysis methods.

\section*{Acknowledgments}
We thank the Atacama Cosmology Telescope (ACT), BICEP/Keck, BLAST, CCAT, CMB-S4, LiteBIRD, Simons Observatory (SO), and South Pole Telescope (SPT) Collaborations for their support of the Pan-Experiment Galactic Science Group, which coordinated this community-wide effort. We thank the BICEP/Keck Collaboration for sharing the BK18 matrix products used in this study. We are grateful to Irene Abril-Cabezas, Carlo Baccigalupi, Colin Bischoff, Dick Bond, Fran\c{c}ois Boulanger, Yi-Kuan Chiang, Tuhin Ghosh, Bill Jones, Fran\c{c}ois Levrier, Fazlu Rahman, and many other members of the Pan-Experiment Galactic Science Group for stimulating discussions that motivated and improved this work. We thank the anonymous referee for helpful feedback.

This work was supported by the National Science Foundation under grants No. AST-2106607 and AST-2441452 (PI S.E.C.). S.E.C. additionally acknowledges support from an Alfred P. Sloan Research Fellowship. This research used resources of the National Energy Research Scientific Computing Center, a DOE Office of Science User Facility supported by the Office of Science of the U.S. Department of Energy under Contract No. DE-AC02-05CH11231 using NERSC award HEP-ERCAP0032657. N.K. and M.R. acknowledge support by the RadioForegroundsPlus Project HORIZON-CL4-2023-SPACE-01, GA 101135036. M.M.N. acknowledges support from the Princeton Undergraduate Summer Research Program. M.D.H. and S.E.C. acknowledge support from the Stanford Physics Undergraduate Summer Research Program. M.R. acknowledges the support of the Spanish Ministry of Science and Innovation through grants PID2022-139223OB-C21 and PID2022-140670NA-I00. This work was carried out in part at the Jet Propulsion Laboratory, California Institute of Technology, under a contract with the National Aeronautics and Space Administration.

Some of the results in this paper have been derived using the \texttt{healpy} and \texttt{HEALPix} packages.

\software{Astropy \citep{AstropyCollaboration:2013, AstropyCollaboration:2018, AstropyCollaboration:2022}, cmocean \citep{Thyng:2016}, HEALPix \citep{Gorski:2005}, healpy \citep{Zonca:2019}, Matplotlib \citep{Hunter:2007}, NaMaster \citep{Alonso:2019}, NumPy \citep{vanderWalt:2011, Harris:2020}, Pynkowski\footnote{\url{https://javicarron.github.io/pynkowski/pynkowski.html}}, PySM \citep{Thorne:2017, Zonca:2021}, SciPy \citep{Virtanen:2020}}

\bibliography{refs, refsADS, refsPlanck}

@article{Thyng:2016,
	  author = {{Thyng}, Kristen~M. and {Greene}, Chad~A. and {Hetland}, Robert~D. and {Zimmerle}, Heather~M. and {DiMarco}, Steven~F. },
	  title = {True Colors of Oceanography: Guidelines for Effective and Accurate Colormap Selection},
	  journal = {Oceanography},
	  year = {2016},
	  month = {September},
	  note = {},
	  volume = {29},
        pages = {9-13},
	  url = {https://doi.org/10.5670/oceanog.2016.66},
        doi = {10.5670/oceanog.2016.66},
}

@article{Minkowski1903,
author = {Minkowski, Hermann},
journal = {Mathematische Annalen},
pages = {447-495},
title = {Volumen und Oberfläche},
url = {http://eudml.org/doc/158108},
volume = {57},
year = {1903},
}

@book{hadwigerVorlesungenUeberInhalt1957,
	series = {Die {Grundlehren} der mathematischen {Wissenschaften}},
	title = {Vorlesungen ueber {Inhalt}, {Oberflache} und {Isoperimetrie}},
	url = {https://books.google.it/books?id=YiA4tAEACAAJ},
	publisher = {Springer},
	author = {Hadwiger, H.},
	year = {1957},
}

@phdthesis{Halal:2024thesis,
    author = {George Halal},
    title = {Polarized dust emission and the morphology of the interstellar medium},
    school = {Stanford University},
    year = {2024},
    url = {https://purl.stanford.edu/rn650nf7483}
}

@ARTICLE{Lei:2025,
       author = {{Lei}, Minjie and {Clark}, S.~E. and {Morel}, Rudy and {Allys}, E. and {Butsky}, Iryna S. and {Redshaw}, Caleb and {Fielding}, Drummond B.},
        title = "{Neutral Gas Phase Distribution from H I Morphology: Phase Separation with Scattering Spectra and Variational Autoencoders}",
      journal = {\apj},
         year = 2025,
        month = nov,
       volume = {993},
       number = {1},
          eid = {4},
        pages = {4},
          doi = {10.3847/1538-4357/ae044a},
archivePrefix = {arXiv},
       eprint = {2505.20407},
 primaryClass = {astro-ph.GA},
       adsurl = {https://ui.adsabs.harvard.edu/abs/2025ApJ...993....4L}
}

@ARTICLE{Kogut:2025,
       author = {{Kogut}, Alan and {Aghanim}, Nabila and {Chluba}, Jens and {Chuss}, David T. and {Delabrouille}, Jacques and {Dvorkin}, Cora and {Fixsen}, Dale and {Ghosh}, Shamik and {Hensley}, Brandon S. and {Hill}, J. Colin and {Maffei}, Bruno and {Pullen}, Anthony R. and {Rotti}, Aditya and {Sabyr}, Alina and {Switzer}, Eric R. and {Thiele}, Leander and {Wollack}, Edward J. and {Zelko}, Ioana},
        title = "{The Primordial Inflation Explorer (PIXIE): mission design and science goals}",
      journal = {\jcap},
         year = 2025,
        month = apr,
       volume = {2025},
       number = {4},
          eid = {020},
        pages = {020},
          doi = {10.1088/1475-7516/2025/04/020},
archivePrefix = {arXiv},
       eprint = {2405.20403},
 primaryClass = {astro-ph.CO},
       adsurl = {https://ui.adsabs.harvard.edu/abs/2025JCAP...04..020K}
}

@ARTICLE{Hervias-Caimapo:2025,
       author = {{Herv{\'\i}as-Caimapo}, Carlos and {Cukierman}, Ari J. and {Diego-Palazuelos}, Patricia and {Huffenberger}, Kevin M. and {Clark}, Susan E.},
        title = "{Modeling parity-violating spectra in Galactic dust polarization with filaments and its applications to cosmic birefringence searches}",
      journal = {\prd},
         year = 2025,
        month = apr,
       volume = {111},
       number = {8},
          eid = {083532},
        pages = {083532},
          doi = {10.1103/PhysRevD.111.083532},
archivePrefix = {arXiv},
       eprint = {2408.06214},
 primaryClass = {astro-ph.CO},
       adsurl = {https://ui.adsabs.harvard.edu/abs/2025PhRvD.111h3532H}
}

@ARTICLE{Ade:2025a,
       author = {{Ade}, P.~A.~R. and {Ahmed}, Z. and {Amiri}, M. and {Barkats}, D. and {Basu Thakur}, R. and {Bischoff}, C.~A. and {Beck}, D. and {Bock}, J.~J. and {Boenish}, H. and {Buza}, V. and {Cheshire}, J.~R. and {Connors}, J. and {Cornelison}, J. and {Crumrine}, M. and {Cukierman}, A.~J. and {Denison}, E. and {Duband}, L. and {Eiben}, M. and {Elwood}, B.~D. and {Fatigoni}, S. and {Filippini}, J.~P. and {Fortes}, A. and {Gao}, M. and {Giannakopoulos}, C. and {Goeckner-Wald}, N. and {Goldfinger}, D.~C. and {Grayson}, J.~A. and {Greathouse}, A. and {Grimes}, P.~K. and {Hall}, G. and {Halal}, G. and {Halpern}, M. and {Hand}, E. and {Harrison}, S.~A. and {Henderson}, S. and {Hubmayr}, J. and {Hui}, H. and {Irwin}, K.~D. and {Kang}, J.~H. and {Karkare}, K.~S. and {Kefeli}, S. and {Kovac}, J.~M. and {Kuo}, C. and {Lau}, K. and {Lautzenhiser}, M. and {Lennox}, A. and {Liu}, T. and {Megerian}, K.~G. and {Minutolo}, L. and {Moncelsi}, L. and {Nakato}, Y. and {Nguyen}, H.~T. and {O'Brient}, R. and {Patel}, A. and {Petroff}, M.~A. and {Polish}, A.~R. and {Prouve}, T. and {Pryke}, C. and {Reintsema}, C.~D. and {Romand}, T. and {Salatino}, M. and {Schillaci}, A. and {Schmitt}, B. and {Singari}, B. and {Sjoberg}, K. and {Soliman}, A. and {St Germaine}, T. and {Steiger}, A. and {Steinbach}, B. and {Sudiwala}, R. and {Thompson}, K.~L. and {Tsai}, C. and {Tucker}, C. and {Turner}, A.~D. and {Verg{\`e}s}, C. and {Vieregg}, A.~G. and {Wandui}, A. and {Weber}, A.~C. and {Willmert}, J. and {Wu}, W.~L.~K. and {Yang}, H. and {Yu}, C. and {Zeng}, L. and {Zhang}, C. and {Zhang}, S. and {(Bicep/}},
        title = "{BICEP/Keck XVIII: Measurement of BICEP3 polarization angles and consequences for constraining cosmic birefringence and inflation}",
      journal = {\prd},
         year = 2025,
        month = mar,
       volume = {111},
       number = {6},
          eid = {063505},
        pages = {063505},
          doi = {10.1103/PhysRevD.111.063505},
archivePrefix = {arXiv},
       eprint = {2410.12089},
 primaryClass = {astro-ph.CO},
       adsurl = {https://ui.adsabs.harvard.edu/abs/2025PhRvD.111f3505A}
}

@ARTICLE{Zhang:2025,
       author = {{Zhang}, Xiangyu and {Green}, Gregory M.},
        title = "{Three-dimensional maps of the interstellar dust extinction curve within the Milky Way galaxy}",
      journal = {Science},
         year = 2025,
        month = mar,
       volume = {387},
       number = {6739},
        pages = {1209-1214},
          doi = {10.1126/science.ado9787},
archivePrefix = {arXiv},
       eprint = {2407.14594},
 primaryClass = {astro-ph.GA},
       adsurl = {https://ui.adsabs.harvard.edu/abs/2025Sci...387.1209Z}
}

@ARTICLE{Panopoulou:2025,
       author = {{Panopoulou}, G.~V. and {Zucker}, C. and {Clemens}, D. and {Pelgrims}, V. and {Soler}, J.~D. and {Clark}, S.~E. and {Alves}, J. and {Goodman}, A. and {Becker Tjus}, J.},
        title = "{The magnetic field of the Radcliffe wave: Starlight polarization at the nearest approach to the Sun}",
      journal = {\aap},
         year = 2025,
        month = feb,
       volume = {694},
          eid = {A97},
        pages = {A97},
          doi = {10.1051/0004-6361/202450991},
archivePrefix = {arXiv},
       eprint = {2406.03765},
 primaryClass = {astro-ph.GA},
       adsurl = {https://ui.adsabs.harvard.edu/abs/2025A&A...694A..97P}
}

@ARTICLE{Ade:2025b,
       author = {{Ade}, P.~A.~R. and {Amiri}, M. and {Benton}, S.~J. and {Bergman}, A.~S. and {Bihary}, R. and {Bock}, J.~J. and {Bond}, J.~R. and {Bonetti}, J.~A. and {Bryan}, S.~A. and {Chiang}, H.~C. and {Contaldi}, C.~R. and {Dor{\'e}}, O. and {Duivenvoorden}, A.~J. and {Eriksen}, H.~K. and {Filippini}, J.~P. and {Fraisse}, A.~A. and {Freese}, K. and {Galloway}, M. and {Gambrel}, A.~E. and {Gandilo}, N.~N. and {Ganga}, K. and {Gourapura}, S. and {Gualtieri}, R. and {Gudmundsson}, J.~E. and {Halpern}, M. and {Hartley}, J. and {Hasselfield}, M. and {Hilton}, G. and {Holmes}, W. and {Hristov}, V.~V. and {Huang}, Z. and {Irwin}, K.~D. and {Jones}, W.~C. and {Karakci}, A. and {Kuo}, C.~L. and {Kermish}, Z.~D. and {Leung}, J.~S. -Y. and {Li}, S. and {Mak}, D.~S.~Y. and {Mason}, P.~V. and {Megerian}, K. and {Moncelsi}, L. and {Morford}, T.~A. and {Nagy}, J.~M. and {Netterfield}, C.~B. and {Nolta}, M. and {O'Brient}, R. and {Osherson}, B. and {Padilla}, I.~L. and {Racine}, B. and {Rahlin}, A.~S. and {Reintsema}, C. and {Ruhl}, J.~E. and {Runyan}, M.~C. and {Ruud}, T.~M. and {Shariff}, J.~A. and {Shaw}, E.~C. and {Shiu}, C. and {Soler}, J.~D. and {Song}, X. and {Trangsrud}, A. and {Tucker}, C. and {Tucker}, R.~S. and {Turner}, A.~D. and {van der List}, J.~F. and {Weber}, A.~C. and {Wehus}, I.~K. and {Wiebe}, D.~V. and {Young}, E.~Y. and {S}},
        title = "{Analysis of Polarized Dust Emission Using Data from the First Flight of SPIDER}",
      journal = {\apj},
         year = 2025,
        month = jan,
       volume = {978},
       number = {2},
          eid = {130},
        pages = {130},
          doi = {10.3847/1538-4357/ad900c},
archivePrefix = {arXiv},
       eprint = {2407.20982},
 primaryClass = {astro-ph.GA},
       adsurl = {https://ui.adsabs.harvard.edu/abs/2025ApJ...978..130A}
}

@ARTICLE{Williamson:2024,
       author = {{Williamson}, Victoria and {Sunseri}, James and {Slepian}, Zachary and {Hou}, Jiamin and {Greco}, Alessandro},
        title = "{First Measurements of the 4-Point Correlation Function of Magnetohydrodynamic Turbulence as a Novel Probe of the Interstellar Medium}",
      journal = {arXiv e-prints},
         year = 2024,
        month = dec,
          eid = {arXiv:2412.03967},
        pages = {arXiv:2412.03967},
          doi = {10.48550/arXiv.2412.03967},
archivePrefix = {arXiv},
       eprint = {2412.03967},
 primaryClass = {astro-ph.GA},
       adsurl = {https://ui.adsabs.harvard.edu/abs/2024arXiv241203967W}
}

@ARTICLE{Mousset:2024,
       author = {{Mousset}, L. and {Allys}, E. and {Price}, M.~A. and {Aumont}, J. and {Delouis}, J. -M. and {Montier}, L. and {McEwen}, J.~D.},
        title = "{Generative models of astrophysical fields with scattering transforms on the sphere}",
      journal = {\aap},
         year = 2024,
        month = nov,
       volume = {691},
          eid = {A269},
        pages = {A269},
          doi = {10.1051/0004-6361/202451396},
archivePrefix = {arXiv},
       eprint = {2407.07007},
 primaryClass = {astro-ph.IM},
       adsurl = {https://ui.adsabs.harvard.edu/abs/2024A&A...691A.269M}
}

@ARTICLE{Shaw:2024,
       author = {{Shaw}, Elle C. and {Ade}, Peter A.~R. and {Akers}, Scott and {Amiri}, Mandana and {Austermann}, Jason and {Beall}, James and {Becker}, Daniel T. and {Benton}, Steven J. and {Bergman}, Amanda S. and {Bock}, James J. and {Bond}, John R. and {Bryan}, Sean A. and {Chiang}, H. Cynthia and {Contaldi}, Carlo R. and {Domagalski}, Rachel S. and {Dor{\'e}}, Olivier and {Duff}, Shannon M. and {Duivenvoorden}, Adri J. and {Eriksen}, Hans K. and {Farhang}, Marzeih and {Filippini}, Jeffrey P. and {Fissel}, Laura M. and {Fraisse}, Aurelien A. and {Freese}, Katherine and {Galloway}, Mathew and {Gambrel}, Anne E. and {Gandilo}, Natalie N. and {Ganga}, Kenneth and {Gibbs}, Sho M. and {Gourapura}, Suren and {Grigorian}, Arpi and {Gualtieri}, Ricardo and {Gudmundsson}, J{\'o}n E. and {Halpern}, Mark and {Hartley}, John and {Hasselfield}, Matthew and {Hilton}, Gene and {Holmes}, Warren and {Hristov}, Viktor V. and {Huang}, Zhiqi and {Hubmayr}, Johannes and {Irwin}, Kent D. and {Jones}, William C. and {Kahn}, Asad and {Kermish}, Zigmund D. and {King}, Cesiley and {Kuo}, Calvin L. and {Lennox}, Amber R. and {Leung}, Jason S. -Y. and {Li}, Steven and {Luu}, Thuy Vy T. and {Mason}, Peter V. and {May}, Jared and {Megerian}, Krikor and {Moncelsi}, Lorenzo and {Morford}, Thomas A. and {Nagy}, Johanna M. and {Nie}, Rong and {Netterfield}, Calvin B. and {Nolta}, Michael and {Osherson}, Benjamin and {Padilla}, Ivan L. and {Rahlin}, Alexandra S. and {Redmond}, Susan and {Reintsema}, Carl and {Romualdez}, L. Javier and {Ruhl}, John E. and {Runyan}, Marcus C. and {Shariff}, Jamil A. and {Shiu}, Corwin and {Soler}, Juan D. and {Song}, Xue and {Tartakovsky}, Simon and {Thommesen}, Harald and {Trangsrud}, Amy and {Tucker}, Carole and {Turner}, Anthony D. and {Ullom}, Joel and {van der List}, Joseph F. and {Van Lanen}, Jeff and {Vissers}, Michael R. and {Weber}, Alexis C. and {Wehus}, Ingunn K. and {Wen}, Shyang and {Wiebe}, Donald V. and {Young}, Edward Y.},
        title = "{In-flight performance of SPIDER's 280-GHz receivers}",
      journal = {Journal of Astronomical Telescopes, Instruments, and Systems},
         year = 2024,
        month = oct,
       volume = {10},
          eid = {044012},
        pages = {044012},
          doi = {10.1117/1.JATIS.10.4.044012},
archivePrefix = {arXiv},
       eprint = {2408.10444},
 primaryClass = {astro-ph.IM},
       adsurl = {https://ui.adsabs.harvard.edu/abs/2024JATIS..10d4012S}
}

@ARTICLE{Stalpes:2024,
       author = {{Stalpes}, Kye A. and {Collins}, David C. and {Huffenberger}, Kevin M.},
        title = "{Planck Dust Polarization Power Spectra Are Consistent with Strongly Supersonic Turbulence}",
      journal = {\apj},
         year = 2024,
        month = sep,
       volume = {972},
       number = {1},
          eid = {26},
        pages = {26},
          doi = {10.3847/1538-4357/ad571b},
archivePrefix = {arXiv},
       eprint = {2404.02874},
 primaryClass = {astro-ph.GA},
       adsurl = {https://ui.adsabs.harvard.edu/abs/2024ApJ...972...26S}
}

@ARTICLE{Delabrouille:2024,
       author = {{Delabrouille}, Jacques},
        title = "{Planck data revisited: Low-noise synchrotron polarization maps from the WMAP and Planck space missions}",
      journal = {\aap},
         year = 2024,
        month = sep,
       volume = {689},
          eid = {A353},
        pages = {A353},
          doi = {10.1051/0004-6361/202450007},
archivePrefix = {arXiv},
       eprint = {2403.18123},
 primaryClass = {astro-ph.CO},
       adsurl = {https://ui.adsabs.harvard.edu/abs/2024A&A...689A.353D}
}

@ARTICLE{Halal:2024a,
       author = {{Halal}, George and {Clark}, S.~E. and {Tahani}, Mehrnoosh},
        title = "{Imprints of the Local Bubble and Dust Complexity on Polarized Dust Emission}",
      journal = {\apj},
         year = 2024,
        month = sep,
       volume = {973},
       number = {1},
          eid = {54},
        pages = {54},
          doi = {10.3847/1538-4357/ad61e0},
archivePrefix = {arXiv},
       eprint = {2404.11009},
 primaryClass = {astro-ph.GA},
       adsurl = {https://ui.adsabs.harvard.edu/abs/2024ApJ...973...54H}
}

@ARTICLE{Lei:2024,
       author = {{Lei}, Minjie and {Clark}, S.~E.},
        title = "{A New Constraint on the Relative Disorder of Magnetic Fields between Neutral Interstellar Medium Phases}",
      journal = {\apj},
         year = 2024,
        month = sep,
       volume = {972},
       number = {1},
          eid = {66},
        pages = {66},
          doi = {10.3847/1538-4357/ad5ade},
archivePrefix = {arXiv},
       eprint = {2312.03846},
 primaryClass = {astro-ph.GA},
       adsurl = {https://ui.adsabs.harvard.edu/abs/2024ApJ...972...66L}
}

@ARTICLE{Watts:2024,
       author = {{Watts}, D.~J. and {Galloway}, M. and {Gjerl{\o}w}, E. and {San}, M. and {Aurlien}, R. and {Basyrov}, A. and {Brilenkov}, M. and {Eriksen}, H.~K. and {Fuskeland}, U. and {Hergt}, L.~T. and {Herman}, D. and {Ihle}, H.~T. and {Lunde}, J.~G.~S. and {N{\ae}ss}, S.~K. and {Stutzer}, N. -O. and {Thommesen}, H. and {Wehus}, I.~K.},
        title = "{Cosmoglobe DR2. I. Global Bayesian analysis of COBE-DIRBE}",
      journal = {arXiv e-prints},
         year = 2024,
        month = aug,
          eid = {arXiv:2408.10952},
        pages = {arXiv:2408.10952},
          doi = {10.48550/arXiv.2408.10952},
archivePrefix = {arXiv},
       eprint = {2408.10952},
 primaryClass = {astro-ph.CO},
       adsurl = {https://ui.adsabs.harvard.edu/abs/2024arXiv240810952W}
}

@ARTICLE{May:2024,
       author = {{May}, Jared L. and {Adler}, Alexandre E. and {Austermann}, Jason E. and {Benton}, Steven J. and {Bihary}, Rick and {Durkin}, Malcolm and {Duff}, Shannon M. and {Filippini}, Jeffrey P. and {Fraisse}, Aurelien A. and {Gascard}, Thomas J.~L.~J. and {Gibbs}, Sho M. and {Gourapura}, Suren and {Gudmundsson}, Jon E. and {Hartley}, John W. and {Hubmayr}, Johannes and {Jones}, William C. and {Li}, Steven and {Nagy}, Johanna M. and {Okun}, Kate and {Padilla}, Ivan L. and {Romualdez}, L. Javier and {Tartakovsky}, Simon and {Vissers}, Michael R.},
        title = "{Instrument Overview of Taurus: A Balloon-borne CMB and Dust Polarization Experiment}",
      journal = {arXiv e-prints},
         year = 2024,
        month = jul,
          eid = {arXiv:2407.01438},
        pages = {arXiv:2407.01438},
          doi = {10.48550/arXiv.2407.01438},
archivePrefix = {arXiv},
       eprint = {2407.01438},
 primaryClass = {astro-ph.IM},
       adsurl = {https://ui.adsabs.harvard.edu/abs/2024arXiv240701438M}
}

@ARTICLE{Wolz:2024,
       author = {{Wolz}, Kevin and {Azzoni}, Susanna and {Herv{\'\i}as-Caimapo}, Carlos and {Errard}, Josquin and {Krachmalnicoff}, Nicoletta and {Alonso}, David and {Baccigalupi}, Carlo and {Baleato Lizancos}, Ant{\'o}n and {Brown}, Michael L. and {Calabrese}, Erminia and {Chluba}, Jens and {Dunkley}, Jo and {Fabbian}, Giulio and {Galitzki}, Nicholas and {Jost}, Baptiste and {Morshed}, Magdy and {Nati}, Federico},
        title = "{The Simons Observatory: Pipeline comparison and validation for large-scale B-modes}",
      journal = {\aap},
         year = 2024,
        month = jun,
       volume = {686},
          eid = {A16},
        pages = {A16},
          doi = {10.1051/0004-6361/202346105},
archivePrefix = {arXiv},
       eprint = {2302.04276},
 primaryClass = {astro-ph.CO},
       adsurl = {https://ui.adsabs.harvard.edu/abs/2024A&A...686A..16W}
}

@ARTICLE{Yao:2024,
       author = {{Yao}, Jian and {Krachmalnicoff}, Nicoletta and {Foschi}, Marianna and {Puglisi}, Giuseppe and {Baccigalupi}, Carlo},
        title = "{FORSE+: Simulating non-Gaussian CMB foregrounds at 3 arcmin in a stochastic way based on a generative adversarial network}",
      journal = {\aap},
         year = 2024,
        month = jun,
       volume = {686},
          eid = {A290},
        pages = {A290},
          doi = {10.1051/0004-6361/202449827},
archivePrefix = {arXiv},
       eprint = {2406.14519},
 primaryClass = {astro-ph.CO},
       adsurl = {https://ui.adsabs.harvard.edu/abs/2024A&A...686A.290Y}
}

@ARTICLE{Edenhofer:2024,
       author = {{Edenhofer}, Gordian and {Zucker}, Catherine and {Frank}, Philipp and {Saydjari}, Andrew K. and {Speagle}, Joshua S. and {Finkbeiner}, Douglas and {En{\ss}lin}, Torsten A.},
        title = "{A parsec-scale Galactic 3D dust map out to 1.25 kpc from the Sun}",
      journal = {\aap},
         year = 2024,
        month = may,
       volume = {685},
          eid = {A82},
        pages = {A82},
          doi = {10.1051/0004-6361/202347628},
archivePrefix = {arXiv},
       eprint = {2308.01295},
 primaryClass = {astro-ph.GA},
       adsurl = {https://ui.adsabs.harvard.edu/abs/2024A&A...685A..82E}
}

@ARTICLE{TheBICEP/KeckCollaboration:2024,
       author = {{The BICEP/Keck Collaboration} and {:} and {Ade}, P.~A.~R. and {Ahmed}, Z. and {Amiri}, M. and {Barkats}, D. and {Basu Thakur}, R. and {Bischoff}, C.~A. and {Beck}, D. and {Bock}, J.~J. and {Boenish}, H. and {Buza}, V. and {Cheshire}, IV, J.~R. and {Connors}, J. and {Cornelison}, J. and {Crumrine}, M. and {Cukierman}, A. and {Denison}, E.~V. and {Dierickx}, M. and {Duband}, L. and {Eiben}, M. and {Elwood}, B. and {Fatigoni}, S. and {Filippini}, J.~P. and {Gao}, M. and {Giannakopoulos}, C. and {Goeckner-Wald}, N. and {Goldfinger}, D.~C. and {Grayson}, J. and {Grimes}, P. and {Hall}, G. and {Halal}, G. and {Halpern}, M. and {Hand}, E. and {Harrison}, S. and {Henderson}, S. and {Hubmayr}, J. and {Hui}, H. and {Irwin}, K.~D. and {Kang}, J. and {Karkare}, K.~S. and {Kefeli}, S. and {Kovac}, J.~M. and {Kuo}, C L. and {Lau}, K. and {Lennox}, A. and {Liu}, T. and {Megerian}, K.~G. and {Minutolo}, L. and {Moncelsi}, L. and {Nakato}, Y. and {Namikawa}, T. and {Nguyen}, H.~T. and {O'Brient}, R. and {Palladino}, S. and {Petroff}, M. and {Polish}, A. and {Precup}, N. and {Prouve}, T. and {Pryke}, C. and {Racine}, B. and {Reintsema}, C.~D. and {Romand}, T. and {Salatino}, M. and {Schillaci}, A. and {Schmitt}, B.~L. and {Singari}, B. and {Soliman}, A. and {St. Germaine}, T. and {Steiger}, A. and {Steinbach}, B. and {Sudiwala}, R.~V. and {Thompson}, K.~L. and {Tucker}, C. and {Turner}, A.~D. and {Verg{\`e}s}, C. and {Vieregg}, A.~G. and {Wandui}, A. and {Weber}, A.~C. and {Willmert}, J. and {Wu}, W.~L.~K. and {Yang}, H. and {Yoon}, K.~W. and {Young}, E. and {Yu}, C. and {Zeng}, L. and {Zhang}, C. and {Zhang}, S.},
        title = "{Constraining Inflation with the BICEP/Keck CMB Polarization Experiments}",
      journal = {arXiv e-prints},
         year = 2024,
        month = may,
          eid = {arXiv:2405.19469},
        pages = {arXiv:2405.19469},
          doi = {10.48550/arXiv.2405.19469},
archivePrefix = {arXiv},
       eprint = {2405.19469},
 primaryClass = {astro-ph.CO},
       adsurl = {https://ui.adsabs.harvard.edu/abs/2024arXiv240519469T}
}

@ARTICLE{Dowell:2024,
       author = {{Dowell}, C. Darren and {Hensley}, Brandon S. and {Sauvage}, Marc},
        title = "{Simulation of the Far-Infrared Polarimetry Approach Envisioned for the PRIMA Mission}",
      journal = {arXiv e-prints},
         year = 2024,
        month = apr,
          eid = {arXiv:2404.17050},
        pages = {arXiv:2404.17050},
          doi = {10.48550/arXiv.2404.17050},
archivePrefix = {arXiv},
       eprint = {2404.17050},
 primaryClass = {astro-ph.IM},
       adsurl = {https://ui.adsabs.harvard.edu/abs/2024arXiv240417050D}
}

@ARTICLE{Pelgrims:2024,
       author = {{Pelgrims}, V. and {Mandarakas}, N. and {Skalidis}, R. and {Tassis}, K. and {Panopoulou}, G.~V. and {Pavlidou}, V. and {Blinov}, D. and {Kiehlmann}, S. and {Clark}, S.~E. and {Hensley}, B.~S. and {Romanopoulos}, S. and {Basyrov}, A. and {Eriksen}, H.~K. and {Falalaki}, M. and {Ghosh}, T. and {Gjerl{\o}w}, E. and {Kypriotakis}, J.~A. and {Maharana}, S. and {Papadaki}, A. and {Pearson}, T.~J. and {Potter}, S.~B. and {Ramaprakash}, A.~N. and {Readhead}, A.~C.~S. and {Wehus}, I.~K.},
        title = "{The first degree-scale starlight-polarization-based tomography map of the magnetized interstellar medium}",
      journal = {\aap},
         year = 2024,
        month = apr,
       volume = {684},
          eid = {A162},
        pages = {A162},
          doi = {10.1051/0004-6361/202349015},
archivePrefix = {arXiv},
       eprint = {2404.10821},
 primaryClass = {astro-ph.GA},
       adsurl = {https://ui.adsabs.harvard.edu/abs/2024A&A...684A.162P}
}

@ARTICLE{Coppi:2024,
       author = {{Coppi}, Gabriele and {Dicker}, Simon and {Aguirre}, James E. and {Austermann}, Jason E. and {Beall}, James A. and {Clark}, Susan E. and {Cox}, Erin G. and {Devlin}, Mark J. and {Fissel}, Laura M. and {Galitzki}, Nicholas and {Hensley}, Brandon S. and {Hubmayr}, Johannes and {Molinari}, Sergio and {Nati}, Federico and {Novak}, Giles and {Schisano}, Eugenio and {Soler}, Juan D. and {Tucker}, Carole E. and {Ullom}, Joel N. and {Vaskuri}, Anna and {Vissers}, Michael R. and {Wheeler}, Jordan D. and {Zannoni}, Mario and {(The Blast Observatory Collaboration)}},
        title = "{The BLAST Observatory: A Sensitivity Study for Far-IR Balloon-borne Polarimeters}",
      journal = {\pasp},
         year = 2024,
        month = mar,
       volume = {136},
       number = {3},
          eid = {035003},
        pages = {035003},
          doi = {10.1088/1538-3873/ad2e11},
archivePrefix = {arXiv},
       eprint = {2401.14370},
 primaryClass = {astro-ph.IM},
       adsurl = {https://ui.adsabs.harvard.edu/abs/2024PASP..136c5003C}
}

@ARTICLE{Halal:2024b,
       author = {{Halal}, George and {Clark}, Susan E. and {Cukierman}, Ari and {Beck}, Dominic and {Kuo}, Chao-Lin},
        title = "{Filamentary Dust Polarization and the Morphology of Neutral Hydrogen Structures}",
      journal = {\apj},
         year = 2024,
        month = jan,
       volume = {961},
       number = {1},
          eid = {29},
        pages = {29},
          doi = {10.3847/1538-4357/ad06aa},
archivePrefix = {arXiv},
       eprint = {2306.10107},
 primaryClass = {astro-ph.GA},
       adsurl = {https://ui.adsabs.harvard.edu/abs/2024ApJ...961...29H}
}

@ARTICLE{CordovaRosado:2024,
       author = {{C{\'o}rdova Rosado}, Rodrigo and {Hensley}, Brandon S. and {Clark}, Susan E. and {Duivenvoorden}, Adriaan J. and {Atkins}, Zachary and {Battistelli}, Elia Stefano and {Choi}, Steve K. and {Dunkley}, Jo and {Herv{\'\i}as-Caimapo}, Carlos and {Li}, Zack and {Louis}, Thibaut and {Naess}, Sigurd and {Page}, Lyman A. and {Partridge}, Bruce and {Sif{\'o}n}, Crist{\'o}bal and {Staggs}, Suzanne T. and {Vargas}, Cristian and {Wollack}, Edward J.},
        title = "{The Atacama Cosmology Telescope: Galactic Dust Structure and the Cosmic PAH Background in Cross-correlation with WISE}",
      journal = {\apj},
         year = 2024,
        month = jan,
       volume = {960},
       number = {2},
          eid = {96},
        pages = {96},
          doi = {10.3847/1538-4357/ad05cd},
archivePrefix = {arXiv},
       eprint = {2307.06352},
 primaryClass = {astro-ph.GA},
       adsurl = {https://ui.adsabs.harvard.edu/abs/2024ApJ...960...96C}
}

@ARTICLE{Carones:2024,
       author = {{Carones}, Alessandro and {Carr{\'o}nDuque}, Javier and {Marinucci}, Domenico and {Migliaccio}, Marina and {Vittorio}, Nicola},
        title = "{Minkowski functionals of CMB polarization intensity with PYNKOWSKI: theory and application to Planck and future data}",
      journal = {\mnras},
         year = 2024,
        month = jan,
       volume = {527},
       number = {1},
        pages = {756-773},
          doi = {10.1093/mnras/stad3002},
archivePrefix = {arXiv},
       eprint = {2211.07562},
 primaryClass = {astro-ph.CO},
       adsurl = {https://ui.adsabs.harvard.edu/abs/2024MNRAS.527..756C}
}

@ARTICLE{Clancy:2023,
       author = {{Clancy}, J. and {Puglisi}, G. and {Clark}, S.~E. and {Coppi}, G. and {Fabbian}, G. and {Herv{\'\i}as-Caimapo}, C. and {Hill}, J.~C. and {Nati}, F. and {Reichardt}, C.~L.},
        title = "{Polarization fraction of Planck Galactic cold clumps and forecasts for the Simons Observatory}",
      journal = {\mnras},
         year = 2023,
        month = sep,
       volume = {524},
       number = {3},
        pages = {3712-3723},
          doi = {10.1093/mnras/stad2099},
archivePrefix = {arXiv},
       eprint = {2303.02788},
 primaryClass = {astro-ph.GA},
       adsurl = {https://ui.adsabs.harvard.edu/abs/2023MNRAS.524.3712C}
}

@ARTICLE{Andersen:2023,
       author = {{Andersen}, K.~J. and {Herman}, D. and {Aurlien}, R. and {Banerji}, R. and {Basyrov}, A. and {Bersanelli}, M. and {Bertocco}, S. and {Brilenkov}, M. and {Carbone}, M. and {Colombo}, L.~P.~L. and {Eriksen}, H.~K. and {Eskilt}, J.~R. and {Foss}, M.~K. and {Franceschet}, C. and {Fuskeland}, U. and {Galeotta}, S. and {Galloway}, M. and {Gerakakis}, S. and {Gjerl{\o}w}, E. and {Hensley}, B. and {Iacobellis}, M. and {Ieronymaki}, M. and {Ihle}, H.~T. and {Jewell}, J.~B. and {Karakci}, A. and {Keih{\"a}nen}, E. and {Keskitalo}, R. and {Lunde}, J.~G.~S. and {Maggio}, G. and {Maino}, D. and {Maris}, M. and {Mennella}, A. and {Paradiso}, S. and {Partridge}, B. and {Reinecke}, M. and {San}, M. and {Stutzer}, N. -O. and {Suur-Uski}, A. -S. and {Svalheim}, T.~L. and {Tavagnacco}, D. and {Thommesen}, H. and {Watts}, D.~J. and {Wehus}, I.~K. and {Zacchei}, A.},
        title = "{BEYONDPLANCK. XIII. Intensity foreground sampling, degeneracies, and priors}",
      journal = {\aap},
         year = 2023,
        month = jul,
       volume = {675},
          eid = {A13},
        pages = {A13},
          doi = {10.1051/0004-6361/202243186},
archivePrefix = {arXiv},
       eprint = {2201.08188},
 primaryClass = {astro-ph.CO},
       adsurl = {https://ui.adsabs.harvard.edu/abs/2023A&A...675A..13A}
}

@ARTICLE{Herman:2023,
       author = {{Herman}, D. and {Hensley}, B. and {Andersen}, K.~J. and {Aurlien}, R. and {Banerji}, R. and {Bersanelli}, M. and {Bertocco}, S. and {Brilenkov}, M. and {Carbone}, M. and {Colombo}, L.~P.~L. and {Eriksen}, H.~K. and {Foss}, M.~K. and {Fuskeland}, U. and {Galeotta}, S. and {Galloway}, M. and {Gerakakis}, S. and {Gjerl{\o}w}, E. and {Iacobellis}, M. and {Ieronymaki}, M. and {Ihle}, H.~T. and {Jewell}, J.~B. and {Karakci}, A. and {Keih{\"a}nen}, E. and {Keskitalo}, R. and {Maggio}, G. and {Maino}, D. and {Maris}, M. and {Paradiso}, S. and {Partridge}, B. and {Reinecke}, M. and {Suur-Uski}, A. -S. and {Svalheim}, T.~L. and {Tavagnacco}, D. and {Thommesen}, H. and {Wehus}, I.~K. and {Zacchei}, A.},
        title = "{BEYONDPLANCK. XV. Limits on large-scale polarized anomalous microwave emission from Planck LFI and WMAP}",
      journal = {\aap},
         year = 2023,
        month = jul,
       volume = {675},
          eid = {A15},
        pages = {A15},
          doi = {10.1051/0004-6361/202243081},
archivePrefix = {arXiv},
       eprint = {2201.03530},
 primaryClass = {astro-ph.CO},
       adsurl = {https://ui.adsabs.harvard.edu/abs/2023A&A...675A..15H}
}

@ARTICLE{Watts:2023b,
       author = {{Watts}, D.~J. and {Galloway}, M. and {Ihle}, H.~T. and {Andersen}, K.~J. and {Aurlien}, R. and {Banerji}, R. and {Basyrov}, A. and {Bersanelli}, M. and {Bertocco}, S. and {Brilenkov}, M. and {Carbone}, M. and {Colombo}, L.~P.~L. and {Eriksen}, H.~K. and {Eskilt}, J.~R. and {Foss}, M.~K. and {Franceschet}, C. and {Fuskeland}, U. and {Galeotta}, S. and {Gerakakis}, S. and {Gjerl{\o}w}, E. and {Hensley}, B. and {Herman}, D. and {Iacobellis}, M. and {Ieronymaki}, M. and {Jewell}, J.~B. and {Karakci}, A. and {Keih{\"a}nen}, E. and {Keskitalo}, R. and {Lunde}, J.~G.~S. and {Maggio}, G. and {Maino}, D. and {Maris}, M. and {Paradiso}, S. and {Partridge}, B. and {Reinecke}, M. and {San}, M. and {Stutzer}, N. -O. and {Suur-Uski}, A. -S. and {Svalheim}, T.~L. and {Tavagnacco}, D. and {Thommesen}, H. and {Wehus}, I.~K. and {Zacchei}, A.},
        title = "{From BEYONDPLANCK to COSMOGLOBE: Preliminary WMAP Q-band analysis}",
      journal = {\aap},
         year = 2023,
        month = jul,
       volume = {675},
          eid = {A16},
        pages = {A16},
          doi = {10.1051/0004-6361/202243410},
archivePrefix = {arXiv},
       eprint = {2202.11979},
 primaryClass = {astro-ph.CO},
       adsurl = {https://ui.adsabs.harvard.edu/abs/2023A&A...675A..16W}
}

@INPROCEEDINGS{Hacar:2023,
       author = {{Hacar}, A. and {Clark}, S.~E. and {Heitsch}, F. and {Kainulainen}, J. and {Panopoulou}, G.~V. and {Seifried}, D. and {Smith}, R.},
        title = "{Initial Conditions for Star Formation: a Physical Description of the Filamentary ISM}",
    booktitle = {Protostars and Planets VII},
         year = 2023,
       editor = {{Inutsuka}, S. and {Aikawa}, Y. and {Muto}, T. and {Tomida}, K. and {Tamura}, M.},
       series = {Astronomical Society of the Pacific Conference Series},
       volume = {534},
        month = jul,
        pages = {153},
          doi = {10.48550/arXiv.2203.09562},
archivePrefix = {arXiv},
       eprint = {2203.09562},
 primaryClass = {astro-ph.GA},
       adsurl = {https://ui.adsabs.harvard.edu/abs/2023ASPC..534..153H}
}

@ARTICLE{Lei:2023,
       author = {{Lei}, Minjie and {Clark}, S.~E.},
        title = "{Probing the Cold Neutral Medium through H I Emission Morphology with the Scattering Transform}",
      journal = {\apj},
         year = 2023,
        month = apr,
       volume = {947},
       number = {2},
          eid = {74},
        pages = {74},
          doi = {10.3847/1538-4357/acc02a},
archivePrefix = {arXiv},
       eprint = {2212.06182},
 primaryClass = {astro-ph.GA},
       adsurl = {https://ui.adsabs.harvard.edu/abs/2023ApJ...947...74L}
}

@ARTICLE{Cukierman:2023,
       author = {{Cukierman}, Ari J. and {Clark}, S.~E. and {Halal}, George},
        title = "{Magnetic Misalignment of Interstellar Dust Filaments}",
      journal = {\apj},
         year = 2023,
        month = apr,
       volume = {946},
       number = {2},
          eid = {106},
        pages = {106},
          doi = {10.3847/1538-4357/acb0c4},
archivePrefix = {arXiv},
       eprint = {2208.07382},
 primaryClass = {astro-ph.GA},
       adsurl = {https://ui.adsabs.harvard.edu/abs/2023ApJ...946..106C}
}

@ARTICLE{LiteBIRDCollaboration:2023,
       author = {{LiteBIRD Collaboration} and {Allys}, E. and {Arnold}, K. and {Aumont}, J. and {Aurlien}, R. and {Azzoni}, S. and {Baccigalupi}, C. and {Banday}, A.~J. and {Banerji}, R. and {Barreiro}, R.~B. and {Bartolo}, N. and {Bautista}, L. and {Beck}, D. and {Beckman}, S. and {Bersanelli}, M. and {Boulanger}, F. and {Brilenkov}, M. and {Bucher}, M. and {Calabrese}, E. and {Campeti}, P. and {Carones}, A. and {Casas}, F.~J. and {Catalano}, A. and {Chan}, V. and {Cheung}, K. and {Chinone}, Y. and {Clark}, S.~E. and {Columbro}, F. and {D'Alessandro}, G. and {de Bernardis}, P. and {de Haan}, T. and {de la Hoz}, E. and {De Petris}, M. and {Torre}, S. Della and {Diego-Palazuelos}, P. and {Dobbs}, M. and {Dotani}, T. and {Duval}, J.~M. and {Elleflot}, T. and {Eriksen}, H.~K. and {Errard}, J. and {Essinger-Hileman}, T. and {Finelli}, F. and {Flauger}, R. and {Franceschet}, C. and {Fuskeland}, U. and {Galloway}, M. and {Ganga}, K. and {Gerbino}, M. and {Gervasi}, M. and {G{\'e}nova-Santos}, R.~T. and {Ghigna}, T. and {Giardiello}, S. and {Gjerl{\o}w}, E. and {Grain}, J. and {Grupp}, F. and {Gruppuso}, A. and {Gudmundsson}, J.~E. and {Halverson}, N.~W. and {Hargrave}, P. and {Hasebe}, T. and {Hasegawa}, M. and {Hazumi}, M. and {Henrot-Versill{\'e}}, S. and {Hensley}, B. and {Hergt}, L.~T. and {Herman}, D. and {Hivon}, E. and {Hlozek}, R.~A. and {Hornsby}, A.~L. and {Hoshino}, Y. and {Hubmayr}, J. and {Ichiki}, K. and {Iida}, T. and {Imada}, H. and {Ishino}, H. and {Jaehnig}, G. and {Katayama}, N. and {Kato}, A. and {Keskitalo}, R. and {Kisner}, T. and {Kobayashi}, Y. and {Kogut}, A. and {Kohri}, K. and {Komatsu}, E. and {Komatsu}, K. and {Konishi}, K. and {Krachmalnicoff}, N. and {Kuo}, C.~L. and {Lamagna}, L. and {Lattanzi}, M. and {Lee}, A.~T. and {Leloup}, C. and {Levrier}, F. and {Linder}, E. and {Luzzi}, G. and {Macias-Perez}, J. and {Maciaszek}, T. and {Maffei}, B. and {Maino}, D. and {Mandelli}, S. and {Mart{\'\i}nez-Gonz{\'a}lez}, E. and {Masi}, S. and {Massa}, M. and {Matarrese}, S. and {Matsuda}, F.~T. and {Matsumura}, T. and {Mele}, L. and {Migliaccio}, M. and {Minami}, Y. and {Moggi}, A. and {Montgomery}, J. and {Montier}, L. and {Morgante}, G. and {Mot}, B. and {Nagano}, Y. and {Nagasaki}, T. and {Nagata}, R. and {Nakano}, R. and {Namikawa}, T. and {Nati}, F. and {Natoli}, P. and {Nerval}, S. and {Noviello}, F. and {Odagiri}, K. and {Oguri}, S. and {Ohsaki}, H. and {Pagano}, L. and {Paiella}, A. and {Paoletti}, D. and {Passerini}, A. and {Patanchon}, G. and {Piacentini}, F. and {Piat}, M. and {Pisano}, G. and {Polenta}, G. and {Poletti}, D. and {Prouv{\'e}}, T. and {Puglisi}, G. and {Rambaud}, D. and {Raum}, C. and {Realini}, S. and {Reinecke}, M. and {Remazeilles}, M. and {Ritacco}, A. and {Roudil}, G. and {Rubino-Martin}, J.~A. and {Russell}, M. and {Sakurai}, H. and {Sakurai}, Y. and {Sasaki}, M. and {Scott}, D. and {Sekimoto}, Y. and {Shinozaki}, K. and {Shiraishi}, M. and {Shirron}, P. and {Signorelli}, G. and {Spinella}, F. and {Stever}, S. and {Stompor}, R. and {Sugiyama}, S. and {Sullivan}, R.~M. and {Suzuki}, A. and {Svalheim}, T.~L. and {Switzer}, E. and {Takaku}, R. and {Takakura}, H. and {Takase}, Y. and {Tartari}, A. and {Terao}, Y. and {Thermeau}, J. and {Thommesen}, H. and {Thompson}, K.~L. and {Tomasi}, M. and {Tominaga}, M. and {Tristram}, M. and {Tsuji}, M. and {Tsujimoto}, M. and {Vacher}, L. and {Vielva}, P. and {Vittorio}, N. and {Wang}, W. and {Watanuki}, K. and {Wehus}, I.~K. and {Weller}, J. and {Westbrook}, B. and {Wilms}, J. and {Winter}, B. and {Wollack}, E.~J. and {Yumoto}, J. and {Zannoni}, M. and {Collaboration LiteB I R D}},
        title = "{Probing cosmic inflation with the LiteBIRD cosmic microwave background polarization survey}",
      journal = {Progress of Theoretical and Experimental Physics},
         year = 2023,
        month = apr,
       volume = {2023},
       number = {4},
          eid = {042F01},
        pages = {042F01},
          doi = {10.1093/ptep/ptac150},
archivePrefix = {arXiv},
       eprint = {2202.02773},
 primaryClass = {astro-ph.IM},
       adsurl = {https://ui.adsabs.harvard.edu/abs/2023PTEP.2023d2F01L}
}

@ARTICLE{Martire:2023,
       author = {{Martire}, F.~A. and {Banday}, A.~J. and {Mart{\'\i}nez-Gonz{\'a}lez}, E. and {Barreiro}, R.~B.},
        title = "{Morphological analysis of the polarized synchrotron emission with WMAP and Planck}",
      journal = {\jcap},
         year = 2023,
        month = apr,
       volume = {2023},
       number = {4},
          eid = {049},
        pages = {049},
          doi = {10.1088/1475-7516/2023/04/049},
archivePrefix = {arXiv},
       eprint = {2301.08041},
 primaryClass = {astro-ph.CO},
       adsurl = {https://ui.adsabs.harvard.edu/abs/2023JCAP...04..049M}
}

@ARTICLE{delaHoz:2023,
       author = {{de la Hoz}, E. and {Barreiro}, R.~B. and {Vielva}, P. and {Mart{\'\i}nez-Gonz{\'a}lez}, E. and {Rubi{\~n}o-Mart{\'\i}n}, J.~A. and {Casaponsa}, B. and {Guidi}, F. and {Ashdown}, M. and {G{\'e}nova-Santos}, R.~T. and {Artal}, E. and {Casas}, F.~J. and {Fern{\'a}ndez-Cobos}, R. and {Fern{\'a}ndez-Torreiro}, M. and {Herranz}, D. and {Hoyland}, R.~J. and {Lasenby}, A.~N. and {L{\'o}pez-Caniego}, M. and {L{\'o}pez-Caraballo}, C.~H. and {Peel}, M.~W. and {Piccirillo}, L. and {Poidevin}, F. and {Rebolo}, R. and {Ruiz-Granados}, B. and {Tramonte}, D. and {Vansyngel}, F. and {Watson}, R.~A.},
        title = "{QUIJOTE scientific results - VIII. Diffuse polarized foregrounds from component separation with QUIJOTE-MFI}",
      journal = {\mnras},
         year = 2023,
        month = mar,
       volume = {519},
       number = {3},
        pages = {3504-3525},
          doi = {10.1093/mnras/stac3020},
archivePrefix = {arXiv},
       eprint = {2301.05117},
 primaryClass = {astro-ph.CO},
       adsurl = {https://ui.adsabs.harvard.edu/abs/2023MNRAS.519.3504D}
}

@ARTICLE{Rubino-Martin:2023,
       author = {{Rubi{\~n}o-Mart{\'\i}n}, J.~A. and {Guidi}, F. and {G{\'e}nova-Santos}, R.~T. and {Harper}, S.~E. and {Herranz}, D. and {Hoyland}, R.~J. and {Lasenby}, A.~N. and {Poidevin}, F. and {Rebolo}, R. and {Ruiz-Granados}, B. and {Vansyngel}, F. and {Vielva}, P. and {Watson}, R.~A. and {Artal}, E. and {Ashdown}, M. and {Barreiro}, R.~B. and {Bilbao-Ahedo}, J.~D. and {Casas}, F.~J. and {Casaponsa}, B. and {Cepeda-Arroita}, R. and {de la Hoz}, E. and {Dickinson}, C. and {Fern{\'a}ndez-Cobos}, R. and {Fern{\'a}ndez-Torreiro}, M. and {Gonz{\'a}lez-Gonz{\'a}lez}, R. and {Hern{\'a}ndez-Monteagudo}, C. and {L{\'o}pez-Caniego}, M. and {L{\'o}pez-Caraballo}, C. and {Mart{\'\i}nez-Gonz{\'a}lez}, E. and {Peel}, M.~W. and {Pel{\'a}ez-Santos}, A.~E. and {Perrott}, Y. and {Piccirillo}, L. and {Razavi-Ghods}, N. and {Scott}, P. and {Titterington}, D. and {Tramonte}, D. and {Vignaga}, R.},
        title = "{QUIJOTE scientific results - IV. A northern sky survey in intensity and polarization at 10-20 GHz with the multifrequency instrument}",
      journal = {\mnras},
         year = 2023,
        month = mar,
       volume = {519},
       number = {3},
        pages = {3383-3431},
          doi = {10.1093/mnras/stac3439},
archivePrefix = {arXiv},
       eprint = {2301.05113},
 primaryClass = {astro-ph.GA},
       adsurl = {https://ui.adsabs.harvard.edu/abs/2023MNRAS.519.3383R}
}

@ARTICLE{Vacher:2023,
       author = {{Vacher}, L. and {Chluba}, J. and {Aumont}, J. and {Rotti}, A. and {Montier}, L.},
        title = "{High precision modeling of polarized signals: Moment expansion method generalized to spin-2 fields}",
      journal = {\aap},
         year = 2023,
        month = jan,
       volume = {669},
          eid = {A5},
        pages = {A5},
          doi = {10.1051/0004-6361/202243913},
archivePrefix = {arXiv},
       eprint = {2205.01049},
 primaryClass = {astro-ph.CO},
       adsurl = {https://ui.adsabs.harvard.edu/abs/2023A&A...669A...5V}
}

@ARTICLE{CCAT-PrimeCollaboration:2023,
       author = {{CCAT-Prime Collaboration} and {Aravena}, Manuel and {Austermann}, Jason E. and {Basu}, Kaustuv and {Battaglia}, Nicholas and {Beringue}, Benjamin and {Bertoldi}, Frank and {Bigiel}, Frank and {Bond}, J. Richard and {Breysse}, Patrick C. and {Broughton}, Colton and {Bustos}, Ricardo and {Chapman}, Scott C. and {Charmetant}, Maude and {Choi}, Steve K. and {Chung}, Dongwoo T. and {Clark}, Susan E. and {Cothard}, Nicholas F. and {Crites}, Abigail T. and {Dev}, Ankur and {Douglas}, Kaela and {Duell}, Cody J. and {D{\"u}nner}, Rolando and {Ebina}, Haruki and {Erler}, Jens and {Fich}, Michel and {Fissel}, Laura M. and {Foreman}, Simon and {Freundt}, R.~G. and {Gallardo}, Patricio A. and {Gao}, Jiansong and {Garc{\'\i}a}, Pablo and {Giovanelli}, Riccardo and {Golec}, Joseph E. and {Groppi}, Christopher E. and {Haynes}, Martha P. and {Henke}, Douglas and {Hensley}, Brandon and {Herter}, Terry and {Higgins}, Ronan and {Hlo{\v{z}}ek}, Ren{\'e}e and {Huber}, Anthony and {Huber}, Zachary and {Hubmayr}, Johannes and {Jackson}, Rebecca and {Johnstone}, Douglas and {Karoumpis}, Christos and {Keating}, Laura C. and {Komatsu}, Eiichiro and {Li}, Yaqiong and {Magnelli}, Benjamin and {Matthews}, Brenda C. and {Mauskopf}, Philip D. and {McMahon}, Jeffrey J. and {Meerburg}, P. Daniel and {Meyers}, Joel and {Muralidhara}, Vyoma and {Murray}, Norman W. and {Niemack}, Michael D. and {Nikola}, Thomas and {Okada}, Yoko and {Puddu}, Roberto and {Riechers}, Dominik A. and {Rosolowsky}, Erik and {Rossi}, Kayla and {Rotermund}, Kaja and {Roy}, Anirban and {Sadavoy}, Sarah I. and {Schaaf}, Reinhold and {Schilke}, Peter and {Scott}, Douglas and {Simon}, Robert and {Sinclair}, Adrian K. and {Sivakoff}, Gregory R. and {Stacey}, Gordon J. and {Stutz}, Amelia M. and {Stutzki}, Juergen and {Tahani}, Mehrnoosh and {Thanjavur}, Karun and {Timmermann}, Ralf A. and {Ullom}, Joel N. and {van Engelen}, Alexander and {Vavagiakis}, Eve M. and {Vissers}, Michael R. and {Wheeler}, Jordan D. and {White}, Simon D.~M. and {Zhu}, Yijie and {Zou}, Bugao},
        title = "{CCAT-prime Collaboration: Science Goals and Forecasts with Prime-Cam on the Fred Young Submillimeter Telescope}",
      journal = {\apjs},
         year = 2023,
        month = jan,
       volume = {264},
       number = {1},
          eid = {7},
        pages = {7},
          doi = {10.3847/1538-4365/ac9838},
archivePrefix = {arXiv},
       eprint = {2107.10364},
 primaryClass = {astro-ph.CO},
       adsurl = {https://ui.adsabs.harvard.edu/abs/2023ApJS..264....7C}
}

@ARTICLE{Delouis:2022,
       author = {{Delouis}, J. -M. and {Allys}, E. and {Gauvrit}, E. and {Boulanger}, F.},
        title = "{Non-Gaussian modelling and statistical denoising of Planck dust polarisation full-sky maps using scattering transforms}",
      journal = {\aap},
         year = 2022,
        month = dec,
       volume = {668},
          eid = {A122},
        pages = {A122},
          doi = {10.1051/0004-6361/202244566},
archivePrefix = {arXiv},
       eprint = {2207.12527},
 primaryClass = {astro-ph.CO},
       adsurl = {https://ui.adsabs.harvard.edu/abs/2022A&A...668A.122D}
}

@ARTICLE{Ghosh:2022,
       author = {{Ghosh}, Shamik and {Liu}, Yang and {Zhang}, Le and {Li}, Siyu and {Zhang}, Junzhou and {Wang}, Jiaxin and {Dou}, Jiazheng and {Chen}, Jiming and {Delabrouille}, Jacques and {Remazeilles}, Mathieu and {Feng}, Chang and {Hu}, Bin and {Huang}, Zhi-Qi and {Liu}, Hao and {Santos}, Larissa and {Zhang}, Pengjie and {Zhang}, Zhaoxuan and {Zhao}, Wen and {Li}, Hong and {Zhang}, Xinmin},
        title = "{Performance forecasts for the primordial gravitational wave detection pipelines for AliCPT-1}",
      journal = {\jcap},
         year = 2022,
        month = oct,
       volume = {2022},
       number = {10},
          eid = {063},
        pages = {063},
          doi = {10.1088/1475-7516/2022/10/063},
archivePrefix = {arXiv},
       eprint = {2205.14804},
 primaryClass = {astro-ph.CO},
       adsurl = {https://ui.adsabs.harvard.edu/abs/2022JCAP...10..063G}
}

@ARTICLE{Eskilt:2022,
       author = {{Eskilt}, Johannes R. and {Komatsu}, Eiichiro},
        title = "{Improved constraints on cosmic birefringence from the WMAP and Planck cosmic microwave background polarization data}",
      journal = {\prd},
         year = 2022,
        month = sep,
       volume = {106},
       number = {6},
          eid = {063503},
        pages = {063503},
          doi = {10.1103/PhysRevD.106.063503},
archivePrefix = {arXiv},
       eprint = {2205.13962},
 primaryClass = {astro-ph.CO},
       adsurl = {https://ui.adsabs.harvard.edu/abs/2022PhRvD.106f3503E}
}

@ARTICLE{Fornazier:2022,
       author = {{Fornazier}, Karin S.~F. and {Abdalla}, Filipe B. and {Remazeilles}, Mathieu and {Vieira}, Jordany and {Marins}, Alessandro and {Abdalla}, Elcio and {Santos}, Larissa and {Delabrouille}, Jacques and {Mericia}, Eduardo and {Landim}, Ricardo G. and {Ferreira}, Elisa G.~M. and {Barosi}, Luciano and {Queiroz}, Amilcar R. and {Villela}, Thyrso and {Wang}, Bin and {Wuensche}, Carlos A. and {Costa}, Andre A. and {Liccardo}, Vincenzo and {Paiva Novaes}, Camila and {Peel}, Michael W. and {dos Santos}, Marcelo V. and {Zhang}, Jiajun},
        title = "{The BINGO project. V. Further steps in component separation and bispectrum analysis}",
      journal = {\aap},
         year = 2022,
        month = aug,
       volume = {664},
          eid = {A18},
        pages = {A18},
          doi = {10.1051/0004-6361/202141707},
archivePrefix = {arXiv},
       eprint = {2107.01637},
 primaryClass = {astro-ph.CO},
       adsurl = {https://ui.adsabs.harvard.edu/abs/2022A&A...664A..18F}
}

@ARTICLE{AstropyCollaboration:2022,
       author = {{Astropy Collaboration} and {Price-Whelan}, Adrian M. and {Lim}, Pey Lian and {Earl}, Nicholas and {Starkman}, Nathaniel and {Bradley}, Larry and {Shupe}, David L. and {Patil}, Aarya A. and {Corrales}, Lia and {Brasseur}, C.~E. and {N{\"o}the}, Maximilian and {Donath}, Axel and {Tollerud}, Erik and {Morris}, Brett M. and {Ginsburg}, Adam and {Vaher}, Eero and {Weaver}, Benjamin A. and {Tocknell}, James and {Jamieson}, William and {van Kerkwijk}, Marten H. and {Robitaille}, Thomas P. and {Merry}, Bruce and {Bachetti}, Matteo and {G{\"u}nther}, H. Moritz and {Aldcroft}, Thomas L. and {Alvarado-Montes}, Jaime A. and {Archibald}, Anne M. and {B{\'o}di}, Attila and {Bapat}, Shreyas and {Barentsen}, Geert and {Baz{\'a}n}, Juanjo and {Biswas}, Manish and {Boquien}, M{\'e}d{\'e}ric and {Burke}, D.~J. and {Cara}, Daria and {Cara}, Mihai and {Conroy}, Kyle E. and {Conseil}, Simon and {Craig}, Matthew W. and {Cross}, Robert M. and {Cruz}, Kelle L. and {D'Eugenio}, Francesco and {Dencheva}, Nadia and {Devillepoix}, Hadrien A.~R. and {Dietrich}, J{\"o}rg P. and {Eigenbrot}, Arthur Davis and {Erben}, Thomas and {Ferreira}, Leonardo and {Foreman-Mackey}, Daniel and {Fox}, Ryan and {Freij}, Nabil and {Garg}, Suyog and {Geda}, Robel and {Glattly}, Lauren and {Gondhalekar}, Yash and {Gordon}, Karl D. and {Grant}, David and {Greenfield}, Perry and {Groener}, Austen M. and {Guest}, Steve and {Gurovich}, Sebastian and {Handberg}, Rasmus and {Hart}, Akeem and {Hatfield-Dodds}, Zac and {Homeier}, Derek and {Hosseinzadeh}, Griffin and {Jenness}, Tim and {Jones}, Craig K. and {Joseph}, Prajwel and {Kalmbach}, J. Bryce and {Karamehmetoglu}, Emir and {Ka{\l}uszy{\'n}ski}, Miko{\l}aj and {Kelley}, Michael S.~P. and {Kern}, Nicholas and {Kerzendorf}, Wolfgang E. and {Koch}, Eric W. and {Kulumani}, Shankar and {Lee}, Antony and {Ly}, Chun and {Ma}, Zhiyuan and {MacBride}, Conor and {Maljaars}, Jakob M. and {Muna}, Demitri and {Murphy}, N.~A. and {Norman}, Henrik and {O'Steen}, Richard and {Oman}, Kyle A. and {Pacifici}, Camilla and {Pascual}, Sergio and {Pascual-Granado}, J. and {Patil}, Rohit R. and {Perren}, Gabriel I. and {Pickering}, Timothy E. and {Rastogi}, Tanuj and {Roulston}, Benjamin R. and {Ryan}, Daniel F. and {Rykoff}, Eli S. and {Sabater}, Jose and {Sakurikar}, Parikshit and {Salgado}, Jes{\'u}s and {Sanghi}, Aniket and {Saunders}, Nicholas and {Savchenko}, Volodymyr and {Schwardt}, Ludwig and {Seifert-Eckert}, Michael and {Shih}, Albert Y. and {Jain}, Anany Shrey and {Shukla}, Gyanendra and {Sick}, Jonathan and {Simpson}, Chris and {Singanamalla}, Sudheesh and {Singer}, Leo P. and {Singhal}, Jaladh and {Sinha}, Manodeep and {Sip{\H{o}}cz}, Brigitta M. and {Spitler}, Lee R. and {Stansby}, David and {Streicher}, Ole and {{\v{S}}umak}, Jani and {Swinbank}, John D. and {Taranu}, Dan S. and {Tewary}, Nikita and {Tremblay}, Grant R. and {de Val-Borro}, Miguel and {Van Kooten}, Samuel J. and {Vasovi{\'c}}, Zlatan and {Verma}, Shresth and {de Miranda Cardoso}, Jos{\'e} Vin{\'\i}cius and {Williams}, Peter K.~G. and {Wilson}, Tom J. and {Winkel}, Benjamin and {Wood-Vasey}, W.~M. and {Xue}, Rui and {Yoachim}, Peter and {Zhang}, Chen and {Zonca}, Andrea and {Astropy Project Contributors}},
        title = "{The Astropy Project: Sustaining and Growing a Community-oriented Open-source Project and the Latest Major Release (v5.0) of the Core Package}",
      journal = {\apj},
         year = 2022,
        month = aug,
       volume = {935},
       number = {2},
          eid = {167},
        pages = {167},
          doi = {10.3847/1538-4357/ac7c74},
archivePrefix = {arXiv},
       eprint = {2206.14220},
 primaryClass = {astro-ph.IM},
       adsurl = {https://ui.adsabs.harvard.edu/abs/2022ApJ...935..167A}
}

@ARTICLE{Grewal:2022,
       author = {{Grewal}, Nisha and {Zuntz}, Joe and {Tr{\"o}ster}, Tilman and {Amon}, Alexandra},
        title = "{Minkowski Functionals in Joint Galaxy Clustering \& Weak Lensing Analyses}",
      journal = {The Open Journal of Astrophysics},
         year = 2022,
        month = aug,
       volume = {5},
       number = {1},
          eid = {13},
        pages = {13},
          doi = {10.21105/astro.2206.03877},
archivePrefix = {arXiv},
       eprint = {2206.03877},
 primaryClass = {astro-ph.CO},
       adsurl = {https://ui.adsabs.harvard.edu/abs/2022OJAp....5E..13G}
}

@ARTICLE{Dalya:2022,
       author = {{D{\'a}lya}, G. and {D{\'\i}az}, R. and {Bouchet}, F.~R. and {Frei}, Z. and {Jasche}, J. and {Lavaux}, G. and {Macas}, R. and {Mukherjee}, S. and {P{\'a}lfi}, M. and {de Souza}, R.~S. and {Wandelt}, B.~D. and {Bilicki}, M. and {Raffai}, P.},
        title = "{GLADE+ : an extended galaxy catalogue for multimessenger searches with advanced gravitational-wave detectors}",
      journal = {\mnras},
         year = 2022,
        month = jul,
       volume = {514},
       number = {1},
        pages = {1403-1411},
          doi = {10.1093/mnras/stac1443},
archivePrefix = {arXiv},
       eprint = {2110.06184},
 primaryClass = {astro-ph.CO},
       adsurl = {https://ui.adsabs.harvard.edu/abs/2022MNRAS.514.1403D}
}

@ARTICLE{Hensley:2022,
       author = {{Hensley}, Brandon S. and {Clark}, Susan E. and {Fanfani}, Valentina and {Krachmalnicoff}, Nicoletta and {Fabbian}, Giulio and {Poletti}, Davide and {Puglisi}, Giuseppe and {Coppi}, Gabriele and {Nibauer}, Jacob and {Gerasimov}, Roman and {Galitzki}, Nicholas and {Choi}, Steve K. and {Ashton}, Peter C. and {Baccigalupi}, Carlo and {Baxter}, Eric and {Burkhart}, Blakesley and {Calabrese}, Erminia and {Chluba}, Jens and {Errard}, Josquin and {Frolov}, Andrei V. and {Herv{\'\i}as-Caimapo}, Carlos and {Huffenberger}, Kevin M. and {Johnson}, Bradley R. and {Jost}, Baptiste and {Keating}, Brian and {McCarrick}, Heather and {Nati}, Federico and {Sathyanarayana Rao}, Mayuri and {van Engelen}, Alexander and {Walker}, Samantha and {Wolz}, Kevin and {Xu}, Zhilei and {Zhu}, Ningfeng and {Zonca}, Andrea},
        title = "{The Simons Observatory: Galactic Science Goals and Forecasts}",
      journal = {\apj},
         year = 2022,
        month = apr,
       volume = {929},
       number = {2},
          eid = {166},
        pages = {166},
          doi = {10.3847/1538-4357/ac5e36},
archivePrefix = {arXiv},
       eprint = {2111.02425},
 primaryClass = {astro-ph.GA},
       adsurl = {https://ui.adsabs.harvard.edu/abs/2022ApJ...929..166H}
}

@ARTICLE{Hervias-Caimapo:2022,
       author = {{Herv{\'\i}as-Caimapo}, Carlos and {Huffenberger}, Kevin M.},
        title = "{Full-sky, Arcminute-scale, 3D Models of Galactic Microwave Foreground Dust Emission Based on Filaments}",
      journal = {\apj},
         year = 2022,
        month = mar,
       volume = {928},
       number = {1},
          eid = {65},
        pages = {65},
          doi = {10.3847/1538-4357/ac54b2},
archivePrefix = {arXiv},
       eprint = {2107.08317},
 primaryClass = {astro-ph.CO},
       adsurl = {https://ui.adsabs.harvard.edu/abs/2022ApJ...928...65H}
}

@ARTICLE{Abazajian:2022,
       author = {{Abazajian}, Kevork and {Addison}, Graeme E. and {Adshead}, Peter and {Ahmed}, Zeeshan and {Akerib}, Daniel and {Ali}, Aamir and {Allen}, Steven W. and {Alonso}, David and {Alvarez}, Marcelo and {Amin}, Mustafa A. and {Anderson}, Adam and {Arnold}, Kam S. and {Ashton}, Peter and {Baccigalupi}, Carlo and {Bard}, Debbie and {Barkats}, Denis and {Barron}, Darcy and {Barry}, Peter S. and {Bartlett}, James G. and {Basu Thakur}, Ritoban and {Battaglia}, Nicholas and {Bean}, Rachel and {Bebek}, Chris and {Bender}, Amy N. and {Benson}, Bradford A. and {Bianchini}, Federico and {Bischoff}, Colin A. and {Bleem}, Lindsey and {Bock}, James J. and {Bocquet}, Sebastian and {Boddy}, Kimberly K. and {Richard Bond}, J. and {Borrill}, Julian and {Bouchet}, Fran{\c{c}}ois R. and {Brinckmann}, Thejs and {Brown}, Michael L. and {Bryan}, Sean and {Buza}, Victor and {Byrum}, Karen and {Hervias Caimapo}, Carlos and {Calabrese}, Erminia and {Calafut}, Victoria and {Caldwell}, Robert and {Carlstrom}, John E. and {Carron}, Julien and {Cecil}, Thomas and {Challinor}, Anthony and {Chang}, Clarence L. and {Chinone}, Yuji and {Sherry Cho}, Hsiao-Mei and {Cooray}, Asantha and {Coulton}, Will and {Crawford}, Thomas M. and {Crites}, Abigail and {Cukierman}, Ari and {Cyr-Racine}, Francis-Yan and {de Haan}, Tijmen and {Delabrouille}, Jacques and {Devlin}, Mark and {Di Valentino}, Eleonora and {Dierickx}, Marion and {Dobbs}, Matt and {Duff}, Shannon and {Dvorkin}, Cora and {Eimer}, Joseph and {Elleflot}, Tucker and {Errard}, Josquin and {Essinger-Hileman}, Thomas and {Fabbian}, Giulio and {Feng}, Chang and {Ferraro}, Simone and {Filippini}, Jeffrey P. and {Flauger}, Raphael and {Flaugher}, Brenna and {Fraisse}, Aurelien A. and {Frolov}, Andrei and {Galitzki}, Nicholas and {Gallardo}, Patricio A. and {Galli}, Silvia and {Ganga}, Ken and {Gerbino}, Martina and {Gluscevic}, Vera and {Goeckner-Wald}, Neil and {Green}, Daniel and {Grin}, Daniel and {Grohs}, Evan and {Gualtieri}, Riccardo and {Gudmundsson}, Jon E. and {Gullett}, Ian and {Gupta}, Nikhel and {Habib}, Salman and {Halpern}, Mark and {Halverson}, Nils W. and {Hanany}, Shaul and {Harrington}, Kathleen and {Hasegawa}, Masaya and {Hasselfield}, Matthew and {Hazumi}, Masashi and {Heitmann}, Katrin and {Henderson}, Shawn and {Hensley}, Brandon and {Hill}, Charles and {Colin Hill}, J. and {Hlo{\v{z}}ek}, Ren{\'e}e and {Patty Ho}, Shuay-Pwu and {Hoang}, Thuong and {Holder}, Gil and {Holzapfel}, William and {Hood}, John and {Hubmayr}, Johannes and {Huffenberger}, Kevin M. and {Hui}, Howard and {Irwin}, Kent and {Jeong}, Oliver and {Johnson}, Bradley R. and {Jones}, William C. and {Hwan Kang}, Jae and {Karkare}, Kirit S. and {Katayama}, Nobuhiko and {Keskitalo}, Reijo and {Kisner}, Theodore and {Knox}, Lloyd and {Koopman}, Brian J. and {Kosowsky}, Arthur and {Kovac}, John and {Kovetz}, Ely D. and {Kuhlmann}, Steve and {Kuo}, Chao-lin and {Kusaka}, Akito and {L{\"a}hteenm{\"a}ki}, Anne and {Lawrence}, Charles R. and {Lee}, Adrian T. and {Lewis}, Antony and {Li}, Dale and {Linder}, Eric and {Loverde}, Marilena and {Lowitz}, Amy and {Lubin}, Phil and {Madhavacheril}, Mathew S. and {Mantz}, Adam and {Marques}, Gabriela and {Matsuda}, Frederick and {Mauskopf}, Philip and {McCarrick}, Heather and {McMahon}, Jeffrey and {Daniel Meerburg}, P. and {Melin}, Jean-Baptiste and {Menanteau}, Felipe and {Meyers}, Joel and {Millea}, Marius and {Mohr}, Joseph and {Moncelsi}, Lorenzo and {Monzani}, Maria and {Mroczkowski}, Tony and {Mukherjee}, Suvodip and {Nagy}, Johanna and {Namikawa}, Toshiya and {Nati}, Federico and {Natoli}, Tyler and {Newburgh}, Laura and {Niemack}, Michael D. and {Nishino}, Haruki and {Nord}, Brian and {Novosad}, Valentine and {O'Brient}, Roger and {Padin}, Stephen and {Palladino}, Steven and {Partridge}, Bruce and {Petravick}, Don and {Pierpaoli}, Elena and {Pogosian}, Levon and {Prabhu}, Karthik and {Pryke}, Clement and {Puglisi}, Giuseppe and {Racine}, Benjamin and {Rahlin}, Alexandra and {Sathyanarayana Rao}, Mayuri and {Raveri}, Marco and {Reichardt}, Christian L. and {Remazeilles}, Mathieu and {Rocha}, Graca and {Roe}, Natalie A. and {Roy}, Anirban and {Ruhl}, John E. and {Salatino}, Maria and {Saliwanchik}, Benjamin and {Schaan}, Emmanuel and {Schillaci}, Alessandro and {Schmitt}, Benjamin and {Schmittfull}, Marcel M. and {Scott}, Douglas and {Sehgal}, Neelima and {Shandera}, Sarah and {Sherwin}, Blake D. and {Shirokoff}, Erik and {Simon}, Sara M. and {Slosar}, An{\v{z}}e and {Spergel}, David and {St. Germaine}, Tyler and {Staggs}, Suzanne T.},
        title = "{CMB-S4: Forecasting Constraints on Primordial Gravitational Waves}",
      journal = {\apj},
         year = 2022,
        month = feb,
       volume = {926},
       number = {1},
          eid = {54},
        pages = {54},
          doi = {10.3847/1538-4357/ac1596},
archivePrefix = {arXiv},
       eprint = {2008.12619},
 primaryClass = {astro-ph.CO},
       adsurl = {https://ui.adsabs.harvard.edu/abs/2022ApJ...926...54A}
}

@ARTICLE{Zonca:2021,
       author = {{Zonca}, Andrea and {Thorne}, Ben and {Krachmalnicoff}, Nicoletta and {Borrill}, Julian},
        title = "{The Python Sky Model 3 software}",
      journal = {The Journal of Open Source Software},
         year = 2021,
        month = nov,
       volume = {6},
       number = {67},
          eid = {3783},
        pages = {3783},
          doi = {10.21105/joss.03783},
archivePrefix = {arXiv},
       eprint = {2108.01444},
 primaryClass = {astro-ph.IM},
       adsurl = {https://ui.adsabs.harvard.edu/abs/2021JOSS....6.3783Z}
}

@ARTICLE{Guan:2021,
       author = {{Guan}, Yilun and {Clark}, Susan E. and {Hensley}, Brandon S. and {Gallardo}, Patricio A. and {Naess}, Sigurd and {Duell}, Cody J. and {Aiola}, Simone and {Atkins}, Zachary and {Calabrese}, Erminia and {Choi}, Steve K. and {Cothard}, Nicholas F. and {Devlin}, Mark and {Duivenvoorden}, Adriaan J. and {Dunkley}, Jo and {D{\"u}nner}, Rolando and {Ferraro}, Simone and {Hasselfield}, Matthew and {Hughes}, John P. and {Koopman}, Brian J. and {Kosowsky}, Arthur B. and {Madhavacheril}, Mathew S. and {McMahon}, Jeff and {Nati}, Federico and {Niemack}, Michael D. and {Page}, Lyman A. and {Salatino}, Maria and {Schaan}, Emmanuel and {Sehgal}, Neelima and {Sif{\'o}n}, Crist{\'o}bal and {Staggs}, Suzanne and {Vavagiakis}, Eve M. and {Wollack}, Edward J. and {Xu}, Zhilei},
        title = "{The Atacama Cosmology Telescope: Microwave Intensity and Polarization Maps of the Galactic Center}",
      journal = {\apj},
         year = 2021,
        month = oct,
       volume = {920},
       number = {1},
          eid = {6},
        pages = {6},
          doi = {10.3847/1538-4357/ac133f},
archivePrefix = {arXiv},
       eprint = {2105.05267},
 primaryClass = {astro-ph.GA},
       adsurl = {https://ui.adsabs.harvard.edu/abs/2021ApJ...920....6G}
}

@ARTICLE{Ade:2021,
       author = {{Ade}, P.~A.~R. and {Ahmed}, Z. and {Amiri}, M. and {Barkats}, D. and {Thakur}, R. Basu and {Bischoff}, C.~A. and {Beck}, D. and {Bock}, J.~J. and {Boenish}, H. and {Bullock}, E. and {Buza}, V. and {Cheshire}, J.~R. and {Connors}, J. and {Cornelison}, J. and {Crumrine}, M. and {Cukierman}, A. and {Denison}, E.~V. and {Dierickx}, M. and {Duband}, L. and {Eiben}, M. and {Fatigoni}, S. and {Filippini}, J.~P. and {Fliescher}, S. and {Goeckner-Wald}, N. and {Goldfinger}, D.~C. and {Grayson}, J. and {Grimes}, P. and {Hall}, G. and {Halal}, G. and {Halpern}, M. and {Hand}, E. and {Harrison}, S. and {Henderson}, S. and {Hildebrandt}, S.~R. and {Hilton}, G.~C. and {Hubmayr}, J. and {Hui}, H. and {Irwin}, K.~D. and {Kang}, J. and {Karkare}, K.~S. and {Karpel}, E. and {Kefeli}, S. and {Kernasovskiy}, S.~A. and {Kovac}, J.~M. and {Kuo}, C.~L. and {Lau}, K. and {Leitch}, E.~M. and {Lennox}, A. and {Megerian}, K.~G. and {Minutolo}, L. and {Moncelsi}, L. and {Nakato}, Y. and {Namikawa}, T. and {Nguyen}, H.~T. and {O'Brient}, R. and {Ogburn}, R.~W. and {Palladino}, S. and {Prouve}, T. and {Pryke}, C. and {Racine}, B. and {Reintsema}, C.~D. and {Richter}, S. and {Schillaci}, A. and {Schwarz}, R. and {Schmitt}, B.~L. and {Sheehy}, C.~D. and {Soliman}, A. and {Germaine}, T. St. and {Steinbach}, B. and {Sudiwala}, R.~V. and {Teply}, G.~P. and {Thompson}, K.~L. and {Tolan}, J.~E. and {Tucker}, C. and {Turner}, A.~D. and {Umilt{\`a}}, C. and {Verg{\`e}s}, C. and {Vieregg}, A.~G. and {Wandui}, A. and {Weber}, A.~C. and {Wiebe}, D.~V. and {Willmert}, J. and {Wong}, C.~L. and {Wu}, W.~L.~K. and {Yang}, H. and {Yoon}, K.~W. and {Young}, E. and {Yu}, C. and {Zeng}, L. and {Zhang}, C. and {Zhang}, S. and {Bicep/Keck Collaboration}},
        title = "{Improved Constraints on Primordial Gravitational Waves using Planck, WMAP, and BICEP/Keck Observations through the 2018 Observing Season}",
      journal = {\prl},
         year = 2021,
        month = oct,
       volume = {127},
       number = {15},
          eid = {151301},
        pages = {151301},
          doi = {10.1103/PhysRevLett.127.151301},
archivePrefix = {arXiv},
       eprint = {2110.00483},
 primaryClass = {astro-ph.CO},
       adsurl = {https://ui.adsabs.harvard.edu/abs/2021PhRvL.127o1301A}
}

@ARTICLE{Clark:2021,
       author = {{Clark}, S.~E. and {Kim}, Chang-Goo and {Hill}, J. Colin and {Hensley}, Brandon S.},
        title = "{The Origin of Parity Violation in Polarized Dust Emission and Implications for Cosmic Birefringence}",
      journal = {\apj},
         year = 2021,
        month = sep,
       volume = {919},
       number = {1},
          eid = {53},
        pages = {53},
          doi = {10.3847/1538-4357/ac0e35},
archivePrefix = {arXiv},
       eprint = {2105.00120},
 primaryClass = {astro-ph.GA},
       adsurl = {https://ui.adsabs.harvard.edu/abs/2021ApJ...919...53C}
}

@ARTICLE{Rahman:2021,
       author = {{Rahman}, Fazlu and {Chingangbam}, Pravabati and {Ghosh}, Tuhin},
        title = "{The nature of non-Gaussianity and statistical isotropy of the 408 MHz Haslam synchrotron map}",
      journal = {\jcap},
         year = 2021,
        month = jul,
       volume = {2021},
       number = {7},
          eid = {026},
        pages = {026},
          doi = {10.1088/1475-7516/2021/07/026},
archivePrefix = {arXiv},
       eprint = {2104.00419},
 primaryClass = {astro-ph.CO},
       adsurl = {https://ui.adsabs.harvard.edu/abs/2021JCAP...07..026R}
}

@ARTICLE{Krachmalnicoff:2021,
       author = {{Krachmalnicoff}, Nicoletta and {Puglisi}, Giuseppe},
        title = "{ForSE: A GAN-based Algorithm for Extending CMB Foreground Models to Subdegree Angular Scales}",
      journal = {\apj},
         year = 2021,
        month = apr,
       volume = {911},
       number = {1},
          eid = {42},
        pages = {42},
          doi = {10.3847/1538-4357/abe71c},
archivePrefix = {arXiv},
       eprint = {2011.02221},
 primaryClass = {astro-ph.CO},
       adsurl = {https://ui.adsabs.harvard.edu/abs/2021ApJ...911...42K}
}

@ARTICLE{Pelgrims:2021,
       author = {{Pelgrims}, V. and {Clark}, S.~E. and {Hensley}, B.~S. and {Panopoulou}, G.~V. and {Pavlidou}, V. and {Tassis}, K. and {Eriksen}, H.~K. and {Wehus}, I.~K.},
        title = "{Evidence for line-of-sight frequency decorrelation of polarized dust emission in Planck data}",
      journal = {\aap},
         year = 2021,
        month = mar,
       volume = {647},
          eid = {A16},
        pages = {A16},
          doi = {10.1051/0004-6361/202040218},
archivePrefix = {arXiv},
       eprint = {2101.09291},
 primaryClass = {astro-ph.CO},
       adsurl = {https://ui.adsabs.harvard.edu/abs/2021A&A...647A..16P}
}

@ARTICLE{Choi:2020,
       author = {{Choi}, Steve K. and {Hasselfield}, Matthew and {Ho}, Shuay-Pwu Patty and {Koopman}, Brian and {Lungu}, Marius and {Abitbol}, Maximilian H. and {Addison}, Graeme E. and {Ade}, Peter A.~R. and {Aiola}, Simone and {Alonso}, David and {Amiri}, Mandana and {Amodeo}, Stefania and {Angile}, Elio and {Austermann}, Jason E. and {Baildon}, Taylor and {Battaglia}, Nick and {Beall}, James A. and {Bean}, Rachel and {Becker}, Daniel T. and {Bond}, J. Richard and {Bruno}, Sarah Marie and {Calabrese}, Erminia and {Calafut}, Victoria and {Campusano}, Luis E. and {Carrero}, Felipe and {Chesmore}, Grace E. and {Cho}, Hsiao-mei and {Clark}, Susan E. and {Cothard}, Nicholas F. and {Crichton}, Devin and {Crowley}, Kevin T. and {Darwish}, Omar and {Datta}, Rahul and {Denison}, Edward V. and {Devlin}, Mark J. and {Duell}, Cody J. and {Duff}, Shannon M. and {Duivenvoorden}, Adriaan J. and {Dunkley}, Jo and {D{\"u}nner}, Rolando and {Essinger-Hileman}, Thomas and {Fankhanel}, Max and {Ferraro}, Simone and {Fox}, Anna E. and {Fuzia}, Brittany and {Gallardo}, Patricio A. and {Gluscevic}, Vera and {Golec}, Joseph E. and {Grace}, Emily and {Gralla}, Megan and {Guan}, Yilun and {Hall}, Kirsten and {Halpern}, Mark and {Han}, Dongwon and {Hargrave}, Peter and {Henderson}, Shawn and {Hensley}, Brandon and {Hill}, J. Colin and {Hilton}, Gene C. and {Hilton}, Matt and {Hincks}, Adam D. and {Hlo{\v{z}}ek}, Ren{\'e}e and {Hubmayr}, Johannes and {Huffenberger}, Kevin M. and {Hughes}, John P. and {Infante}, Leopoldo and {Irwin}, Kent and {Jackson}, Rebecca and {Klein}, Jeff and {Knowles}, Kenda and {Kosowsky}, Arthur and {Lakey}, Vincent and {Li}, Dale and {Li}, Yaqiong and {Li}, Zack and {Lokken}, Martine and {Louis}, Thibaut and {MacInnis}, Amanda and {Madhavacheril}, Mathew and {Maldonado}, Felipe and {Mallaby-Kay}, Maya and {Marsden}, Danica and {Maurin}, Lo{\"\i}c and {McMahon}, Jeff and {Menanteau}, Felipe and {Moodley}, Kavilan and {Morton}, Tim and {Naess}, Sigurd and {Namikawa}, Toshiya and {Nati}, Federico and {Newburgh}, Laura and {Nibarger}, John P. and {Nicola}, Andrina and {Niemack}, Michael D. and {Nolta}, Michael R. and {Orlowski-Sherer}, John and {Page}, Lyman A. and {Pappas}, Christine G. and {Partridge}, Bruce and {Phakathi}, Phumlani and {Prince}, Heather and {Puddu}, Roberto and {Qu}, Frank J. and {Rivera}, Jesus and {Robertson}, Naomi and {Rojas}, Felipe and {Salatino}, Maria and {Schaan}, Emmanuel and {Schillaci}, Alessandro and {Schmitt}, Benjamin L. and {Sehgal}, Neelima and {Sherwin}, Blake D. and {Sierra}, Carlos and {Sievers}, Jon and {Sifon}, Cristobal and {Sikhosana}, Precious and {Simon}, Sara and {Spergel}, David N. and {Staggs}, Suzanne T. and {Stevens}, Jason and {Storer}, Emilie and {Sunder}, Dhaneshwar D. and {Switzer}, Eric R. and {Thorne}, Ben and {Thornton}, Robert and {Trac}, Hy and {Treu}, Jesse and {Tucker}, Carole and {Vale}, Leila R. and {Van Engelen}, Alexander and {Van Lanen}, Jeff and {Vavagiakis}, Eve M. and {Wagoner}, Kasey and {Wang}, Yuhan and {Ward}, Jonathan T. and {Wollack}, Edward J. and {Xu}, Zhilei and {Zago}, Fernando and {Zhu}, Ningfeng},
        title = "{The Atacama Cosmology Telescope: a measurement of the Cosmic Microwave Background power spectra at 98 and 150 GHz}",
      journal = {\jcap},
         year = 2020,
        month = dec,
       volume = {2020},
       number = {12},
          eid = {045},
        pages = {045},
          doi = {10.1088/1475-7516/2020/12/045},
archivePrefix = {arXiv},
       eprint = {2007.07289},
 primaryClass = {astro-ph.CO},
       adsurl = {https://ui.adsabs.harvard.edu/abs/2020JCAP...12..045C}
}

@ARTICLE{Naess:2020,
       author = {{Naess}, Sigurd and {Aiola}, Simone and {Austermann}, Jason E. and {Battaglia}, Nick and {Beall}, James A. and {Becker}, Daniel T. and {Bond}, Richard J. and {Calabrese}, Erminia and {Choi}, Steve K. and {Cothard}, Nicholas F. and {Crowley}, Kevin T. and {Darwish}, Omar and {Datta}, Rahul and {Denison}, Edward V. and {Devlin}, Mark and {Duell}, Cody J. and {Duff}, Shannon M. and {Duivenvoorden}, Adriaan J. and {Dunkley}, Jo and {D{\"u}nner}, Rolando and {Fox}, Anna E. and {Gallardo}, Patricio A. and {Halpern}, Mark and {Han}, Dongwon and {Hasselfield}, Matthew and {Hill}, J. Colin and {Hilton}, Gene C. and {Hilton}, Matt and {Hincks}, Adam D. and {Hlo{\v{z}}ek}, Ren{\'e}e and {Ho}, Shuay-Pwu Patty and {Hubmayr}, Johannes and {Huffenberger}, Kevin and {Hughes}, John P. and {Kosowsky}, Arthur B. and {Louis}, Thibaut and {Madhavacheril}, Mathew S. and {McMahon}, Jeff and {Moodley}, Kavilan and {Nati}, Federico and {Nibarger}, John P. and {Niemack}, Michael D. and {Page}, Lyman and {Partridge}, Bruce and {Salatino}, Maria and {Schaan}, Emmanuel and {Schillaci}, Alessandro and {Schmitt}, Benjamin and {Sherwin}, Blake D. and {Sehgal}, Neelima and {Sif{\'o}n}, Crist{\'o}bal and {Spergel}, David and {Staggs}, Suzanne and {Stevens}, Jason and {Storer}, Emilie and {Ullom}, Joel N. and {Vale}, Leila R. and {Van Engelen}, Alexander and {Van Lanen}, Jeff and {Vavagiakis}, Eve M. and {Wollack}, Edward J. and {Xu}, Zhilei},
        title = "{The Atacama Cosmology Telescope: arcminute-resolution maps of 18 000 square degrees of the microwave sky from ACT 2008-2018 data combined with Planck}",
      journal = {\jcap},
         year = 2020,
        month = dec,
       volume = {2020},
       number = {12},
          eid = {046},
        pages = {046},
          doi = {10.1088/1475-7516/2020/12/046},
archivePrefix = {arXiv},
       eprint = {2007.07290},
 primaryClass = {astro-ph.IM},
       adsurl = {https://ui.adsabs.harvard.edu/abs/2020JCAP...12..046N}
}

@INPROCEEDINGS{Moncelsi:2020,
       author = {{Moncelsi}, L. and {Ade}, P.~A.~R. and {Ahmed}, Z. and {Amiri}, M. and {Barkats}, D. and {Basu Thakur}, R. and {Bischoff}, C.~A. and {Bock}, J.~J. and {Buza}, V. and {Cheshire}, J.~R. and {Connors}, J. and {Cornelison}, J. and {Crumrine}, M. and {Cukierman}, A.~J. and {Denison}, E.~V. and {Dierickx}, M. and {Duband}, L. and {Eiben}, M. and {Fatigoni}, S. and {Filippini}, J.~P. and {Goeckner-Wald}, N. and {Goldfinger}, D. and {Grayson}, J.~A. and {Grimes}, P. and {Hall}, G. and {Halpern}, M. and {Harrison}, S.~A. and {Henderson}, S. and {Hildebrandt}, S.~R. and {Hilton}, G.~C. and {Hubmayr}, J. and {Hui}, H. and {Irwin}, K.~D. and {Kang}, J.~H. and {Karkare}, K.~S. and {Kefeli}, S. and {Kovac}, J.~M. and {Kuo}, C.~L. and {Lau}, K. and {Leitch}, E.~M. and {Megerian}, K.~G. and {Minutolo}, L. and {Nakato}, Y. and {Namikawa}, T. and {Nguyen}, H.~T. and {O'brient}, R. and {Palladino}, S. and {Precup}, N. and {Prouve}, T. and {Pryke}, C. and {Racine}, B. and {Reintsema}, C.~D. and {Schillaci}, A. and {Schmitt}, B.~L. and {Soliman}, A. and {St. Germaine}, T. and {Steinbach}, B. and {Sudiwala}, R.~V. and {Thompson}, K.~L. and {Tucker}, C. and {Turner}, A.~D. and {Umilt{\`a}}, C. and {Vieregg}, A.~G. and {Wandui}, A. and {Weber}, A.~C. and {Wiebe}, D.~V. and {Willmert}, J. and {Wu}, W.~L.~K. and {Yang}, E. and {Yoon}, K.~W. and {Young}, E. and {Yu}, C. and {Zeng}, L. and {Zhang}, C. and {Zhang}, S.},
        title = "{Receiver development for BICEP Array, a next-generation CMB polarimeter at the South Pole}",
    booktitle = {Millimeter, Submillimeter, and Far-Infrared Detectors and Instrumentation for Astronomy X},
         year = 2020,
       editor = {{Zmuidzinas}, Jonas and {Gao}, Jian-Rong},
       series = {Society of Photo-Optical Instrumentation Engineers (SPIE) Conference Series},
       volume = {11453},
        month = dec,
          eid = {1145314},
        pages = {1145314},
          doi = {10.1117/12.2561995},
archivePrefix = {arXiv},
       eprint = {2012.04047},
 primaryClass = {astro-ph.IM},
       adsurl = {https://ui.adsabs.harvard.edu/abs/2020SPIE11453E..14M}
}

@ARTICLE{Minami:2020,
       author = {{Minami}, Yuto and {Komatsu}, Eiichiro},
        title = "{New Extraction of the Cosmic Birefringence from the Planck 2018 Polarization Data}",
      journal = {\prl},
         year = 2020,
        month = nov,
       volume = {125},
       number = {22},
          eid = {221301},
        pages = {221301},
          doi = {10.1103/PhysRevLett.125.221301},
archivePrefix = {arXiv},
       eprint = {2011.11254},
 primaryClass = {astro-ph.CO},
       adsurl = {https://ui.adsabs.harvard.edu/abs/2020PhRvL.125v1301M}
}

@ARTICLE{PlanckCollaboration:2020,
       author = {{Planck Collaboration} and {Akrami}, Y. and {Andersen}, K.~J. and {Ashdown}, M. and {Baccigalupi}, C. and {Ballardini}, M. and {Banday}, A.~J. and {Barreiro}, R.~B. and {Bartolo}, N. and {Basak}, S. and {Benabed}, K. and {Bernard}, J. -P. and {Bersanelli}, M. and {Bielewicz}, P. and {Bond}, J.~R. and {Borrill}, J. and {Burigana}, C. and {Butler}, R.~C. and {Calabrese}, E. and {Casaponsa}, B. and {Chiang}, H.~C. and {Colombo}, L.~P.~L. and {Combet}, C. and {Crill}, B.~P. and {Cuttaia}, F. and {de Bernardis}, P. and {de Rosa}, A. and {de Zotti}, G. and {Delabrouille}, J. and {Di Valentino}, E. and {Diego}, J.~M. and {Dor{\'e}}, O. and {Douspis}, M. and {Dupac}, X. and {Eriksen}, H.~K. and {Fernandez-Cobos}, R. and {Finelli}, F. and {Frailis}, M. and {Fraisse}, A.~A. and {Franceschi}, E. and {Frolov}, A. and {Galeotta}, S. and {Galli}, S. and {Ganga}, K. and {Gerbino}, M. and {Ghosh}, T. and {Gonz{\'a}lez-Nuevo}, J. and {G{\'o}rski}, K.~M. and {Gruppuso}, A. and {Gudmundsson}, J.~E. and {Handley}, W. and {Helou}, G. and {Herranz}, D. and {Hildebrandt}, S.~R. and {Hivon}, E. and {Huang}, Z. and {Jaffe}, A.~H. and {Jones}, W.~C. and {Keih{\"a}nen}, E. and {Keskitalo}, R. and {Kiiveri}, K. and {Kim}, J. and {Kisner}, T.~S. and {Krachmalnicoff}, N. and {Kunz}, M. and {Kurki-Suonio}, H. and {Lasenby}, A. and {Lattanzi}, M. and {Lawrence}, C.~R. and {Le Jeune}, M. and {Levrier}, F. and {Liguori}, M. and {Lilje}, P.~B. and {Lilley}, M. and {Lindholm}, V. and {L{\'o}pez-Caniego}, M. and {Lubin}, P.~M. and {Mac{\'\i}as-P{\'e}rez}, J.~F. and {Maino}, D. and {Mandolesi}, N. and {Marcos-Caballero}, A. and {Maris}, M. and {Martin}, P.~G. and {Mart{\'\i}nez-Gonz{\'a}lez}, E. and {Matarrese}, S. and {Mauri}, N. and {McEwen}, J.~D. and {Meinhold}, P.~R. and {Mennella}, A. and {Migliaccio}, M. and {Mitra}, S. and {Molinari}, D. and {Montier}, L. and {Morgante}, G. and {Moss}, A. and {Natoli}, P. and {Paoletti}, D. and {Partridge}, B. and {Patanchon}, G. and {Pearson}, D. and {Pearson}, T.~J. and {Perrotta}, F. and {Piacentini}, F. and {Polenta}, G. and {Rachen}, J.~P. and {Reinecke}, M. and {Remazeilles}, M. and {Renzi}, A. and {Rocha}, G. and {Rosset}, C. and {Roudier}, G. and {Rubi{\~n}o-Mart{\'\i}n}, J.~A. and {Ruiz-Granados}, B. and {Salvati}, L. and {Savelainen}, M. and {Scott}, D. and {Sirignano}, C. and {Sirri}, G. and {Spencer}, L.~D. and {Suur-Uski}, A. -S. and {Svalheim}, L.~T. and {Tauber}, J.~A. and {Tavagnacco}, D. and {Tenti}, M. and {Terenzi}, L. and {Thommesen}, H. and {Toffolatti}, L. and {Tomasi}, M. and {Tristram}, M. and {Trombetti}, T. and {Valiviita}, J. and {Van Tent}, B. and {Vielva}, P. and {Villa}, F. and {Vittorio}, N. and {Wandelt}, B.~D. and {Wehus}, I.~K. and {Zacchei}, A. and {Zonca}, A.},
        title = "{Planck intermediate results. LVII. Joint Planck LFI and HFI data processing}",
      journal = {\aap},
         year = 2020,
        month = nov,
       volume = {643},
          eid = {A42},
        pages = {A42},
          doi = {10.1051/0004-6361/202038073},
archivePrefix = {arXiv},
       eprint = {2007.04997},
 primaryClass = {astro-ph.CO},
       adsurl = {https://ui.adsabs.harvard.edu/abs/2020A&A...643A..42P}
}

@ARTICLE{Regaldo-SaintBlancard:2020,
       author = {{Regaldo-Saint Blancard}, B. and {Levrier}, F. and {Allys}, E. and {Bellomi}, E. and {Boulanger}, F.},
        title = "{Statistical description of dust polarized emission from the diffuse interstellar medium. A RWST approach}",
      journal = {\aap},
         year = 2020,
        month = oct,
       volume = {642},
          eid = {A217},
        pages = {A217},
          doi = {10.1051/0004-6361/202038044},
archivePrefix = {arXiv},
       eprint = {2007.08242},
 primaryClass = {astro-ph.CO},
       adsurl = {https://ui.adsabs.harvard.edu/abs/2020A&A...642A.217R}
}

@ARTICLE{Harris:2020,
       author = {{Harris}, Charles R. and {Millman}, K. Jarrod and {van der Walt}, St{\'e}fan J. and {Gommers}, Ralf and {Virtanen}, Pauli and {Cournapeau}, David and {Wieser}, Eric and {Taylor}, Julian and {Berg}, Sebastian and {Smith}, Nathaniel J. and {Kern}, Robert and {Picus}, Matti and {Hoyer}, Stephan and {van Kerkwijk}, Marten H. and {Brett}, Matthew and {Haldane}, Allan and {del R{\'\i}o}, Jaime Fern{\'a}ndez and {Wiebe}, Mark and {Peterson}, Pearu and {G{\'e}rard-Marchant}, Pierre and {Sheppard}, Kevin and {Reddy}, Tyler and {Weckesser}, Warren and {Abbasi}, Hameer and {Gohlke}, Christoph and {Oliphant}, Travis E.},
        title = "{Array programming with NumPy}",
      journal = {\nat},
         year = 2020,
        month = sep,
       volume = {585},
       number = {7825},
        pages = {357-362},
          doi = {10.1038/s41586-020-2649-2},
archivePrefix = {arXiv},
       eprint = {2006.10256},
 primaryClass = {cs.MS},
       adsurl = {https://ui.adsabs.harvard.edu/abs/2020Natur.585..357H}
}

@ARTICLE{Huffenberger:2020,
       author = {{Huffenberger}, Kevin M. and {Rotti}, Aditya and {Collins}, David C.},
        title = "{The Power Spectra of Polarized, Dusty Filaments}",
      journal = {\apj},
         year = 2020,
        month = aug,
       volume = {899},
       number = {1},
          eid = {31},
        pages = {31},
          doi = {10.3847/1538-4357/ab9df9},
archivePrefix = {arXiv},
       eprint = {1906.10052},
 primaryClass = {astro-ph.CO},
       adsurl = {https://ui.adsabs.harvard.edu/abs/2020ApJ...899...31H}
}

@ARTICLE{Weiland:2020,
       author = {{Weiland}, J.~L. and {Addison}, G.~E. and {Bennett}, C.~L. and {Halpern}, M. and {Hinshaw}, G.},
        title = "{An Examination of Galactic Polarization with Application to the Planck TB Correlation}",
      journal = {\apj},
         year = 2020,
        month = apr,
       volume = {893},
       number = {2},
          eid = {119},
        pages = {119},
          doi = {10.3847/1538-4357/ab7ea6},
archivePrefix = {arXiv},
       eprint = {1907.02486},
 primaryClass = {astro-ph.CO},
       adsurl = {https://ui.adsabs.harvard.edu/abs/2020ApJ...893..119W}
}

@ARTICLE{Virtanen:2020,
       author = {{Virtanen}, Pauli and {Gommers}, Ralf and {Oliphant}, Travis E. and {Haberland}, Matt and {Reddy}, Tyler and {Cournapeau}, David and {Burovski}, Evgeni and {Peterson}, Pearu and {Weckesser}, Warren and {Bright}, Jonathan and {van der Walt}, St{\'e}fan J. and {Brett}, Matthew and {Wilson}, Joshua and {Millman}, K. Jarrod and {Mayorov}, Nikolay and {Nelson}, Andrew R.~J. and {Jones}, Eric and {Kern}, Robert and {Larson}, Eric and {Carey}, C.~J. and {Polat}, {\.I}lhan and {Feng}, Yu and {Moore}, Eric W. and {VanderPlas}, Jake and {Laxalde}, Denis and {Perktold}, Josef and {Cimrman}, Robert and {Henriksen}, Ian and {Quintero}, E.~A. and {Harris}, Charles R. and {Archibald}, Anne M. and {Ribeiro}, Ant{\^o}nio H. and {Pedregosa}, Fabian and {van Mulbregt}, Paul and {SciPy 1. 0 Contributors}},
        title = "{SciPy 1.0: fundamental algorithms for scientific computing in Python}",
      journal = {Nature Methods},
         year = 2020,
        month = feb,
       volume = {17},
        pages = {261-272},
          doi = {10.1038/s41592-019-0686-2},
archivePrefix = {arXiv},
       eprint = {1907.10121},
 primaryClass = {cs.MS},
       adsurl = {https://ui.adsabs.harvard.edu/abs/2020NatMe..17..261V}
}

@ARTICLE{Hensley:2019,
       author = {{Hensley}, Brandon S. and {Zhang}, Cheng and {Bock}, James J.},
        title = "{An Imprint of the Galactic Magnetic Field in the Diffuse Unpolarized Dust Emission}",
      journal = {\apj},
         year = 2019,
        month = dec,
       volume = {887},
       number = {2},
          eid = {159},
        pages = {159},
          doi = {10.3847/1538-4357/ab5183},
archivePrefix = {arXiv},
       eprint = {1909.07394},
 primaryClass = {astro-ph.GA},
       adsurl = {https://ui.adsabs.harvard.edu/abs/2019ApJ...887..159H}
}

@ARTICLE{Clark:2019,
       author = {{Clark}, S.~E. and {Hensley}, Brandon S.},
        title = "{Mapping the Magnetic Interstellar Medium in Three Dimensions over the Full Sky with Neutral Hydrogen}",
      journal = {\apj},
         year = 2019,
        month = dec,
       volume = {887},
       number = {2},
          eid = {136},
        pages = {136},
          doi = {10.3847/1538-4357/ab5803},
archivePrefix = {arXiv},
       eprint = {1909.11673},
 primaryClass = {astro-ph.GA},
       adsurl = {https://ui.adsabs.harvard.edu/abs/2019ApJ...887..136C}
}

@ARTICLE{Kim:2019,
       author = {{Kim}, Chang-Goo and {Choi}, Steve K. and {Flauger}, Raphael},
        title = "{Dust Polarization Maps from TIGRESS: E/B Power Asymmetry and TE Correlation}",
      journal = {\apj},
         year = 2019,
        month = aug,
       volume = {880},
       number = {2},
          eid = {106},
        pages = {106},
          doi = {10.3847/1538-4357/ab29f2},
archivePrefix = {arXiv},
       eprint = {1901.07079},
 primaryClass = {astro-ph.GA},
       adsurl = {https://ui.adsabs.harvard.edu/abs/2019ApJ...880..106K}
}

@ARTICLE{Alonso:2019,
       author = {{Alonso}, David and {Sanchez}, Javier and {Slosar}, An{\v{z}}e and {LSST Dark Energy Science Collaboration}},
        title = "{A unified pseudo-C$_{{\ensuremath{\ell}}}$ framework}",
      journal = {\mnras},
         year = 2019,
        month = apr,
       volume = {484},
       number = {3},
        pages = {4127-4151},
          doi = {10.1093/mnras/stz093},
archivePrefix = {arXiv},
       eprint = {1809.09603},
 primaryClass = {astro-ph.CO},
       adsurl = {https://ui.adsabs.harvard.edu/abs/2019MNRAS.484.4127A}
}

@ARTICLE{Zonca:2019,
       author = {{Zonca}, Andrea and {Singer}, Leo and {Lenz}, Daniel and {Reinecke}, Martin and {Rosset}, Cyrille and {Hivon}, Eric and {Gorski}, Krzysztof},
        title = "{healpy: equal area pixelization and spherical harmonics transforms for data on the sphere in Python}",
      journal = {The Journal of Open Source Software},
         year = 2019,
        month = mar,
       volume = {4},
       number = {35},
          eid = {1298},
        pages = {1298},
          doi = {10.21105/joss.01298},
       adsurl = {https://ui.adsabs.harvard.edu/abs/2019JOSS....4.1298Z}
}

@ARTICLE{Hanany:2019,
       author = {{Hanany}, Shaul and {Alvarez}, Marcelo and {Artis}, Emmanuel and {Ashton}, Peter and {Aumont}, Jonathan and {Aurlien}, Ragnhild and {Banerji}, Ranajoy and {Barreiro}, R. Belen and {Bartlett}, James G. and {Basak}, Soumen and {Battaglia}, Nick and {Bock}, Jamie and {Boddy}, Kimberly K. and {Bonato}, Matteo and {Borrill}, Julian and {Bouchet}, Fran{\c{c}}ois and {Boulanger}, Fran{\c{c}}ois and {Burkhart}, Blakesley and {Chluba}, Jens and {Chuss}, David and {Clark}, Susan E. and {Cooperrider}, Joelle and {Crill}, Brendan P. and {De Zotti}, Gianfranco and {Delabrouille}, Jacques and {Di Valentino}, Eleonora and {Didier}, Joy and {Dor{\'e}}, Olivier and {Eriksen}, Hans K. and {Errard}, Josquin and {Essinger-Hileman}, Tom and {Feeney}, Stephen and {Filippini}, Jeffrey and {Fissel}, Laura and {Flauger}, Raphael and {Fuskeland}, Unni and {Gluscevic}, Vera and {Gorski}, Krzysztof M. and {Green}, Dan and {Hensley}, Brandon and {Herranz}, Diego and {Hill}, J. Colin and {Hivon}, Eric and {Hlo{\v{z}}ek}, Ren{\'e}e and {Hubmayr}, Johannes and {Johnson}, Bradley R. and {Jones}, William and {Jones}, Terry and {Knox}, Lloyd and {Kogut}, Al and {L{\'o}pez-Caniego}, Marcos and {Lawrence}, Charles and {Lazarian}, Alex and {Li}, Zack and {Madhavacheril}, Mathew and {Melin}, Jean-Baptiste and {Meyers}, Joel and {Murray}, Calum and {Negrello}, Mattia and {Novak}, Giles and {O'Brient}, Roger and {Paine}, Christopher and {Pearson}, Tim and {Pogosian}, Levon and {Pryke}, Clem and {Puglisi}, Giuseppe and {Remazeilles}, Mathieu and {Rocha}, Graca and {Schmittfull}, Marcel and {Scott}, Douglas and {Shirron}, Peter and {Stephens}, Ian and {Sutin}, Brian and {Tomasi}, Maurizio and {Trangsrud}, Amy and {van Engelen}, Alexander and {Vansyngel}, Flavien and {Wehus}, Ingunn K. and {Wen}, Qi and {Xu}, Siyao and {Young}, Karl and {Zonca}, Andrea},
        title = "{PICO: Probe of Inflation and Cosmic Origins}",
      journal = {arXiv e-prints},
         year = 2019,
        month = feb,
          eid = {arXiv:1902.10541},
        pages = {arXiv:1902.10541},
          doi = {10.48550/arXiv.1902.10541},
archivePrefix = {arXiv},
       eprint = {1902.10541},
 primaryClass = {astro-ph.IM},
       adsurl = {https://ui.adsabs.harvard.edu/abs/2019arXiv190210541H}
}

@ARTICLE{Ade:2019,
       author = {{Ade}, Peter and {Aguirre}, James and {Ahmed}, Zeeshan and {Aiola}, Simone and {Ali}, Aamir and {Alonso}, David and {Alvarez}, Marcelo A. and {Arnold}, Kam and {Ashton}, Peter and {Austermann}, Jason and {Awan}, Humna and {Baccigalupi}, Carlo and {Baildon}, Taylor and {Barron}, Darcy and {Battaglia}, Nick and {Battye}, Richard and {Baxter}, Eric and {Bazarko}, Andrew and {Beall}, James A. and {Bean}, Rachel and {Beck}, Dominic and {Beckman}, Shawn and {Beringue}, Benjamin and {Bianchini}, Federico and {Boada}, Steven and {Boettger}, David and {Bond}, J. Richard and {Borrill}, Julian and {Brown}, Michael L. and {Bruno}, Sarah Marie and {Bryan}, Sean and {Calabrese}, Erminia and {Calafut}, Victoria and {Calisse}, Paolo and {Carron}, Julien and {Challinor}, Anthony and {Chesmore}, Grace and {Chinone}, Yuji and {Chluba}, Jens and {Cho}, Hsiao-Mei Sherry and {Choi}, Steve and {Coppi}, Gabriele and {Cothard}, Nicholas F. and {Coughlin}, Kevin and {Crichton}, Devin and {Crowley}, Kevin D. and {Crowley}, Kevin T. and {Cukierman}, Ari and {D'Ewart}, John M. and {D{\"u}nner}, Rolando and {de Haan}, Tijmen and {Devlin}, Mark and {Dicker}, Simon and {Didier}, Joy and {Dobbs}, Matt and {Dober}, Bradley and {Duell}, Cody J. and {Duff}, Shannon and {Duivenvoorden}, Adri and {Dunkley}, Jo and {Dusatko}, John and {Errard}, Josquin and {Fabbian}, Giulio and {Feeney}, Stephen and {Ferraro}, Simone and {Flux{\`a}}, Pedro and {Freese}, Katherine and {Frisch}, Josef C. and {Frolov}, Andrei and {Fuller}, George and {Fuzia}, Brittany and {Galitzki}, Nicholas and {Gallardo}, Patricio A. and {Tomas Galvez Ghersi}, Jose and {Gao}, Jiansong and {Gawiser}, Eric and {Gerbino}, Martina and {Gluscevic}, Vera and {Goeckner-Wald}, Neil and {Golec}, Joseph and {Gordon}, Sam and {Gralla}, Megan and {Green}, Daniel and {Grigorian}, Arpi and {Groh}, John and {Groppi}, Chris and {Guan}, Yilun and {Gudmundsson}, Jon E. and {Han}, Dongwon and {Hargrave}, Peter and {Hasegawa}, Masaya and {Hasselfield}, Matthew and {Hattori}, Makoto and {Haynes}, Victor and {Hazumi}, Masashi and {He}, Yizhou and {Healy}, Erin and {Henderson}, Shawn W. and {Hervias-Caimapo}, Carlos and {Hill}, Charles A. and {Hill}, J. Colin and {Hilton}, Gene and {Hilton}, Matt and {Hincks}, Adam D. and {Hinshaw}, Gary and {Hlo{\v{z}}ek}, Ren{\'e}e and {Ho}, Shirley and {Ho}, Shuay-Pwu Patty and {Howe}, Logan and {Huang}, Zhiqi and {Hubmayr}, Johannes and {Huffenberger}, Kevin and {Hughes}, John P. and {Ijjas}, Anna and {Ikape}, Margaret and {Irwin}, Kent and {Jaffe}, Andrew H. and {Jain}, Bhuvnesh and {Jeong}, Oliver and {Kaneko}, Daisuke and {Karpel}, Ethan D. and {Katayama}, Nobuhiko and {Keating}, Brian and {Kernasovskiy}, Sarah S. and {Keskitalo}, Reijo and {Kisner}, Theodore and {Kiuchi}, Kenji and {Klein}, Jeff and {Knowles}, Kenda and {Koopman}, Brian and {Kosowsky}, Arthur and {Krachmalnicoff}, Nicoletta and {Kuenstner}, Stephen E. and {Kuo}, Chao-Lin and {Kusaka}, Akito and {Lashner}, Jacob and {Lee}, Adrian and {Lee}, Eunseong and {Leon}, David and {Leung}, Jason S. -Y. and {Lewis}, Antony and {Li}, Yaqiong and {Li}, Zack and {Limon}, Michele and {Linder}, Eric and {Lopez-Caraballo}, Carlos and {Louis}, Thibaut and {Lowry}, Lindsay and {Lungu}, Marius and {Madhavacheril}, Mathew and {Mak}, Daisy and {Maldonado}, Felipe and {Mani}, Hamdi and {Mates}, Ben and {Matsuda}, Frederick and {Maurin}, Lo{\"\i}c and {Mauskopf}, Phil and {May}, Andrew and {McCallum}, Nialh and {McKenney}, Chris and {McMahon}, Jeff and {Meerburg}, P. Daniel and {Meyers}, Joel and {Miller}, Amber and {Mirmelstein}, Mark and {Moodley}, Kavilan and {Munchmeyer}, Moritz and {Munson}, Charles and {Naess}, Sigurd and {Nati}, Federico and {Navaroli}, Martin and {Newburgh}, Laura and {Nguyen}, Ho Nam and {Niemack}, Michael and {Nishino}, Haruki and {Orlowski-Scherer}, John and {Page}, Lyman and {Partridge}, Bruce and {Peloton}, Julien and {Perrotta}, Francesca and {Piccirillo}, Lucio and {Pisano}, Giampaolo and {Poletti}, Davide and {Puddu}, Roberto and {Puglisi}, Giuseppe and {Raum}, Chris and {Reichardt}, Christian L. and {Remazeilles}, Mathieu and {Rephaeli}, Yoel and {Riechers}, Dominik and {Rojas}, Felipe and {Roy}, Anirban and {Sadeh}, Sharon and {Sakurai}, Yuki and {Salatino}, Maria and {Sathyanarayana Rao}, Mayuri and {Schaan}, Emmanuel and {Schmittfull}, Marcel and {Sehgal}, Neelima and {Seibert}, Joseph},
        title = "{The Simons Observatory: science goals and forecasts}",
      journal = {\jcap},
         year = 2019,
        month = feb,
       volume = {2019},
       number = {2},
          eid = {056},
        pages = {056},
          doi = {10.1088/1475-7516/2019/02/056},
archivePrefix = {arXiv},
       eprint = {1808.07445},
 primaryClass = {astro-ph.CO},
       adsurl = {https://ui.adsabs.harvard.edu/abs/2019JCAP...02..056A}
}

@ARTICLE{Chiang:2019,
       author = {{Chiang}, Yi-Kuan and {M{\'e}nard}, Brice},
        title = "{Extragalactic Imprints in Galactic Dust Maps}",
      journal = {\apj},
         year = 2019,
        month = jan,
       volume = {870},
       number = {2},
          eid = {120},
        pages = {120},
          doi = {10.3847/1538-4357/aaf4f6},
archivePrefix = {arXiv},
       eprint = {1808.03294},
 primaryClass = {astro-ph.GA},
       adsurl = {https://ui.adsabs.harvard.edu/abs/2019ApJ...870..120C}
}

@ARTICLE{BICEP2Collaboration:2018,
       author = {{BICEP2 Collaboration} and {Keck Array Collaboration} and {Ade}, P.~A.~R. and {Ahmed}, Z. and {Aikin}, R.~W. and {Alexander}, K.~D. and {Barkats}, D. and {Benton}, S.~J. and {Bischoff}, C.~A. and {Bock}, J.~J. and {Bowens-Rubin}, R. and {Brevik}, J.~A. and {Buder}, I. and {Bullock}, E. and {Buza}, V. and {Connors}, J. and {Cornelison}, J. and {Crill}, B.~P. and {Crumrine}, M. and {Dierickx}, M. and {Duband}, L. and {Dvorkin}, C. and {Filippini}, J.~P. and {Fliescher}, S. and {Grayson}, J. and {Hall}, G. and {Halpern}, M. and {Harrison}, S. and {Hildebrandt}, S.~R. and {Hilton}, G.~C. and {Hui}, H. and {Irwin}, K.~D. and {Kang}, J. and {Karkare}, K.~S. and {Karpel}, E. and {Kaufman}, J.~P. and {Keating}, B.~G. and {Kefeli}, S. and {Kernasovskiy}, S.~A. and {Kovac}, J.~M. and {Kuo}, C.~L. and {Larsen}, N.~A. and {Lau}, K. and {Leitch}, E.~M. and {Lueker}, M. and {Megerian}, K.~G. and {Moncelsi}, L. and {Namikawa}, T. and {Netterfield}, C.~B. and {Nguyen}, H.~T. and {O'Brient}, R. and {Ogburn}, R.~W. and {Palladino}, S. and {Pryke}, C. and {Racine}, B. and {Richter}, S. and {Schillaci}, A. and {Schwarz}, R. and {Sheehy}, C.~D. and {Soliman}, A. and {St. Germaine}, T. and {Staniszewski}, Z.~K. and {Steinbach}, B. and {Sudiwala}, R.~V. and {Teply}, G.~P. and {Thompson}, K.~L. and {Tolan}, J.~E. and {Tucker}, C. and {Turner}, A.~D. and {Umilt{\`a}}, C. and {Vieregg}, A.~G. and {Wandui}, A. and {Weber}, A.~C. and {Wiebe}, D.~V. and {Willmert}, J. and {Wong}, C.~L. and {Wu}, W.~L.~K. and {Yang}, H. and {Yoon}, K.~W. and {Zhang}, C.},
        title = "{Constraints on Primordial Gravitational Waves Using Planck, WMAP, and New BICEP2/Keck Observations through the 2015 Season}",
      journal = {\prl},
         year = 2018,
        month = nov,
       volume = {121},
       number = {22},
          eid = {221301},
        pages = {221301},
          doi = {10.1103/PhysRevLett.121.221301},
archivePrefix = {arXiv},
       eprint = {1810.05216},
 primaryClass = {astro-ph.CO},
       adsurl = {https://ui.adsabs.harvard.edu/abs/2018PhRvL.121v1301B}
}

@ARTICLE{Krachmalnicoff:2018,
       author = {{Krachmalnicoff}, N. and {Carretti}, E. and {Baccigalupi}, C. and {Bernardi}, G. and {Brown}, S. and {Gaensler}, B.~M. and {Haverkorn}, M. and {Kesteven}, M. and {Perrotta}, F. and {Poppi}, S. and {Staveley-Smith}, L.},
        title = "{S-PASS view of polarized Galactic synchrotron at 2.3 GHz as a contaminant to CMB observations}",
      journal = {\aap},
         year = 2018,
        month = oct,
       volume = {618},
          eid = {A166},
        pages = {A166},
          doi = {10.1051/0004-6361/201832768},
archivePrefix = {arXiv},
       eprint = {1802.01145},
 primaryClass = {astro-ph.GA},
       adsurl = {https://ui.adsabs.harvard.edu/abs/2018A&A...618A.166K}
}

@ARTICLE{AstropyCollaboration:2018,
       author = {{Astropy Collaboration} and {Price-Whelan}, A.~M. and {Sip{\H{o}}cz}, B.~M. and {G{\"u}nther}, H.~M. and {Lim}, P.~L. and {Crawford}, S.~M. and {Conseil}, S. and {Shupe}, D.~L. and {Craig}, M.~W. and {Dencheva}, N. and {Ginsburg}, A. and {VanderPlas}, J.~T. and {Bradley}, L.~D. and {P{\'e}rez-Su{\'a}rez}, D. and {de Val-Borro}, M. and {Aldcroft}, T.~L. and {Cruz}, K.~L. and {Robitaille}, T.~P. and {Tollerud}, E.~J. and {Ardelean}, C. and {Babej}, T. and {Bach}, Y.~P. and {Bachetti}, M. and {Bakanov}, A.~V. and {Bamford}, S.~P. and {Barentsen}, G. and {Barmby}, P. and {Baumbach}, A. and {Berry}, K.~L. and {Biscani}, F. and {Boquien}, M. and {Bostroem}, K.~A. and {Bouma}, L.~G. and {Brammer}, G.~B. and {Bray}, E.~M. and {Breytenbach}, H. and {Buddelmeijer}, H. and {Burke}, D.~J. and {Calderone}, G. and {Cano Rodr{\'\i}guez}, J.~L. and {Cara}, M. and {Cardoso}, J.~V.~M. and {Cheedella}, S. and {Copin}, Y. and {Corrales}, L. and {Crichton}, D. and {D'Avella}, D. and {Deil}, C. and {Depagne}, {\'E}. and {Dietrich}, J.~P. and {Donath}, A. and {Droettboom}, M. and {Earl}, N. and {Erben}, T. and {Fabbro}, S. and {Ferreira}, L.~A. and {Finethy}, T. and {Fox}, R.~T. and {Garrison}, L.~H. and {Gibbons}, S.~L.~J. and {Goldstein}, D.~A. and {Gommers}, R. and {Greco}, J.~P. and {Greenfield}, P. and {Groener}, A.~M. and {Grollier}, F. and {Hagen}, A. and {Hirst}, P. and {Homeier}, D. and {Horton}, A.~J. and {Hosseinzadeh}, G. and {Hu}, L. and {Hunkeler}, J.~S. and {Ivezi{\'c}}, {\v{Z}}. and {Jain}, A. and {Jenness}, T. and {Kanarek}, G. and {Kendrew}, S. and {Kern}, N.~S. and {Kerzendorf}, W.~E. and {Khvalko}, A. and {King}, J. and {Kirkby}, D. and {Kulkarni}, A.~M. and {Kumar}, A. and {Lee}, A. and {Lenz}, D. and {Littlefair}, S.~P. and {Ma}, Z. and {Macleod}, D.~M. and {Mastropietro}, M. and {McCully}, C. and {Montagnac}, S. and {Morris}, B.~M. and {Mueller}, M. and {Mumford}, S.~J. and {Muna}, D. and {Murphy}, N.~A. and {Nelson}, S. and {Nguyen}, G.~H. and {Ninan}, J.~P. and {N{\"o}the}, M. and {Ogaz}, S. and {Oh}, S. and {Parejko}, J.~K. and {Parley}, N. and {Pascual}, S. and {Patil}, R. and {Patil}, A.~A. and {Plunkett}, A.~L. and {Prochaska}, J.~X. and {Rastogi}, T. and {Reddy Janga}, V. and {Sabater}, J. and {Sakurikar}, P. and {Seifert}, M. and {Sherbert}, L.~E. and {Sherwood-Taylor}, H. and {Shih}, A.~Y. and {Sick}, J. and {Silbiger}, M.~T. and {Singanamalla}, S. and {Singer}, L.~P. and {Sladen}, P.~H. and {Sooley}, K.~A. and {Sornarajah}, S. and {Streicher}, O. and {Teuben}, P. and {Thomas}, S.~W. and {Tremblay}, G.~R. and {Turner}, J.~E.~H. and {Terr{\'o}n}, V. and {van Kerkwijk}, M.~H. and {de la Vega}, A. and {Watkins}, L.~L. and {Weaver}, B.~A. and {Whitmore}, J.~B. and {Woillez}, J. and {Zabalza}, V. and {Astropy Contributors}},
        title = "{The Astropy Project: Building an Open-science Project and Status of the v2.0 Core Package}",
      journal = {\aj},
         year = 2018,
        month = sep,
       volume = {156},
       number = {3},
          eid = {123},
        pages = {123},
          doi = {10.3847/1538-3881/aabc4f},
archivePrefix = {arXiv},
       eprint = {1801.02634},
 primaryClass = {astro-ph.IM},
       adsurl = {https://ui.adsabs.harvard.edu/abs/2018AJ....156..123A}
}

@ARTICLE{Portillo:2018,
       author = {{Portillo}, Stephen K.~N. and {Slepian}, Zachary and {Burkhart}, Blakesley and {Kahraman}, Sule and {Finkbeiner}, Douglas P.},
        title = "{Developing the 3-point Correlation Function for the Turbulent Interstellar Medium}",
      journal = {\apj},
         year = 2018,
        month = aug,
       volume = {862},
       number = {2},
          eid = {119},
        pages = {119},
          doi = {10.3847/1538-4357/aacb80},
archivePrefix = {arXiv},
       eprint = {1711.09907},
 primaryClass = {astro-ph.CO},
       adsurl = {https://ui.adsabs.harvard.edu/abs/2018ApJ...862..119P}
}

@ARTICLE{Martinez-Solaeche:2018,
       author = {{Mart{\'\i}nez-Solaeche}, Gin{\'e}s and {Karakci}, Ata and {Delabrouille}, Jacques},
        title = "{A 3D model of polarized dust emission in the Milky Way}",
      journal = {\mnras},
         year = 2018,
        month = may,
       volume = {476},
       number = {1},
        pages = {1310-1330},
          doi = {10.1093/mnras/sty204},
archivePrefix = {arXiv},
       eprint = {1706.04162},
 primaryClass = {astro-ph.CO},
       adsurl = {https://ui.adsabs.harvard.edu/abs/2018MNRAS.476.1310M}
}

@ARTICLE{Remazeilles:2018,
       author = {{Remazeilles}, M. and {Banday}, A.~J. and {Baccigalupi}, C. and {Basak}, S. and {Bonaldi}, A. and {De Zotti}, G. and {Delabrouille}, J. and {Dickinson}, C. and {Eriksen}, H.~K. and {Errard}, J. and {Fernandez-Cobos}, R. and {Fuskeland}, U. and {Herv{\'\i}as-Caimapo}, C. and {L{\'o}pez-Caniego}, M. and {Martinez-Gonz{\'a}lez}, E. and {Roman}, M. and {Vielva}, P. and {Wehus}, I. and {Achucarro}, A. and {Ade}, P. and {Allison}, R. and {Ashdown}, M. and {Ballardini}, M. and {Banerji}, R. and {Bartlett}, J. and {Bartolo}, N. and {Baumann}, D. and {Bersanelli}, M. and {Bonato}, M. and {Borrill}, J. and {Bouchet}, F. and {Boulanger}, F. and {Brinckmann}, T. and {Bucher}, M. and {Burigana}, C. and {Buzzelli}, A. and {Cai}, Z. -Y. and {Calvo}, M. and {Carvalho}, C. -S. and {Castellano}, G. and {Challinor}, A. and {Chluba}, J. and {Clesse}, S. and {Colantoni}, I. and {Coppolecchia}, A. and {Crook}, M. and {D'Alessandro}, G. and {de Bernardis}, P. and {de Gasperis}, G. and {Diego}, J. -M. and {Di Valentino}, E. and {Feeney}, S. and {Ferraro}, S. and {Finelli}, F. and {Forastieri}, F. and {Galli}, S. and {Genova-Santos}, R. and {Gerbino}, M. and {Gonz{\'a}lez-Nuevo}, J. and {Grandis}, S. and {Greenslade}, J. and {Hagstotz}, S. and {Hanany}, S. and {Handley}, W. and {Hernandez-Monteagudo}, C. and {Hills}, M. and {Hivon}, E. and {Kiiveri}, K. and {Kisner}, T. and {Kitching}, T. and {Kunz}, M. and {Kurki-Suonio}, H. and {Lamagna}, L. and {Lasenby}, A. and {Lattanzi}, M. and {Lesgourgues}, J. and {Lewis}, A. and {Liguori}, M. and {Lindholm}, V. and {Luzzi}, G. and {Maffei}, B. and {Martins}, C.~J.~A.~P. and {Masi}, S. and {Matarrese}, S. and {McCarthy}, D. and {Melin}, J. -B. and {Melchiorri}, A. and {Molinari}, D. and {Monfardini}, A. and {Natoli}, P. and {Negrello}, M. and {Notari}, A. and {Paiella}, A. and {Paoletti}, D. and {Patanchon}, G. and {Piat}, M. and {Pisano}, G. and {Polastri}, L. and {Polenta}, G. and {Pollo}, A. and {Poulin}, V. and {Quartin}, M. and {Rubino-Martin}, J. -A. and {Salvati}, L. and {Tartari}, A. and {Tomasi}, M. and {Tramonte}, D. and {Trappe}, N. and {Trombetti}, T. and {Tucker}, C. and {Valiviita}, J. and {Van de Weijgaert}, R. and {van Tent}, B. and {Vennin}, V. and {Vittorio}, N. and {Young}, K. and {Zannoni}, M.},
        title = "{Exploring cosmic origins with CORE: B-mode component separation}",
      journal = {\jcap},
         year = 2018,
        month = apr,
       volume = {2018},
       number = {4},
          eid = {023},
        pages = {023},
          doi = {10.1088/1475-7516/2018/04/023},
archivePrefix = {arXiv},
       eprint = {1704.04501},
 primaryClass = {astro-ph.CO},
       adsurl = {https://ui.adsabs.harvard.edu/abs/2018JCAP...04..023R}
}

@ARTICLE{Puglisi:2017,
       author = {{Puglisi}, G. and {Fabbian}, G. and {Baccigalupi}, C.},
        title = "{A 3D model for carbon monoxide molecular line emission as a potential cosmic microwave background polarization contaminant}",
      journal = {\mnras},
         year = 2017,
        month = aug,
       volume = {469},
       number = {3},
        pages = {2982-2996},
          doi = {10.1093/mnras/stx1029},
archivePrefix = {arXiv},
       eprint = {1701.07856},
 primaryClass = {astro-ph.CO},
       adsurl = {https://ui.adsabs.harvard.edu/abs/2017MNRAS.469.2982P}
}

@ARTICLE{Thorne:2017,
       author = {{Thorne}, B. and {Dunkley}, J. and {Alonso}, D. and {N{\ae}ss}, S.},
        title = "{The Python Sky Model: software for simulating the Galactic microwave sky}",
      journal = {\mnras},
         year = 2017,
        month = aug,
       volume = {469},
       number = {3},
        pages = {2821-2833},
          doi = {10.1093/mnras/stx949},
archivePrefix = {arXiv},
       eprint = {1608.02841},
 primaryClass = {astro-ph.CO},
       adsurl = {https://ui.adsabs.harvard.edu/abs/2017MNRAS.469.2821T}
}

@ARTICLE{Genova-Santos:2017,
       author = {{G{\'e}nova-Santos}, R. and {Rubi{\~n}o-Mart{\'\i}n}, J.~A. and {Pel{\'a}ez-Santos}, A. and {Poidevin}, F. and {Rebolo}, R. and {Vignaga}, R. and {Artal}, E. and {Harper}, S. and {Hoyland}, R. and {Lasenby}, A. and {Mart{\'\i}nez-Gonz{\'a}lez}, E. and {Piccirillo}, L. and {Tramonte}, D. and {Watson}, R.~A.},
        title = "{QUIJOTE scientific results - II. Polarisation measurements of the microwave emission in the Galactic molecular complexes W43 and W47 and supernova remnant W44}",
      journal = {\mnras},
         year = 2017,
        month = feb,
       volume = {464},
       number = {4},
        pages = {4107-4132},
          doi = {10.1093/mnras/stw2503},
archivePrefix = {arXiv},
       eprint = {1605.04741},
 primaryClass = {astro-ph.GA},
       adsurl = {https://ui.adsabs.harvard.edu/abs/2017MNRAS.464.4107G}
}

@ARTICLE{Miville-Deschenes:2016,
       author = {{Miville-Desch{\^e}nes}, M. -A. and {Duc}, P. -A. and {Marleau}, F. and {Cuillandre}, J. -C. and {Didelon}, P. and {Gwyn}, S. and {Karabal}, E.},
        title = "{Probing interstellar turbulence in cirrus with deep optical imaging: no sign of energy dissipation at 0.01 pc scale}",
      journal = {\aap},
         year = 2016,
        month = aug,
       volume = {593},
          eid = {A4},
        pages = {A4},
          doi = {10.1051/0004-6361/201628503},
archivePrefix = {arXiv},
       eprint = {1605.08360},
 primaryClass = {astro-ph.GA},
       adsurl = {https://ui.adsabs.harvard.edu/abs/2016A&A...593A...4M}
}

@ARTICLE{BICEP2Collaboration:2016,
       author = {{BICEP2 Collaboration} and {Keck Array Collaboration} and {Ade}, P.~A.~R. and {Ahmed}, Z. and {Aikin}, R.~W. and {Alexander}, K.~D. and {Barkats}, D. and {Benton}, S.~J. and {Bischoff}, C.~A. and {Bock}, J.~J. and {Bowens-Rubin}, R. and {Brevik}, J.~A. and {Buder}, I. and {Bullock}, E. and {Buza}, V. and {Connors}, J. and {Crill}, B.~P. and {Duband}, L. and {Dvorkin}, C. and {Filippini}, J.~P. and {Fliescher}, S. and {Grayson}, J. and {Halpern}, M. and {Harrison}, S. and {Hildebrandt}, S.~R. and {Hilton}, G.~C. and {Hui}, H. and {Irwin}, K.~D. and {Kang}, J. and {Karkare}, K.~S. and {Karpel}, E. and {Kaufman}, J.~P. and {Keating}, B.~G. and {Kefeli}, S. and {Kernasovskiy}, S.~A. and {Kovac}, J.~M. and {Kuo}, C.~L. and {Leitch}, E.~M. and {Lueker}, M. and {Megerian}, K.~G. and {Namikawa}, T. and {Netterfield}, C.~B. and {Nguyen}, H.~T. and {O'Brient}, R. and {Ogburn}, IV, R.~W. and {Orlando}, A. and {Pryke}, C. and {Richter}, S. and {Schwarz}, R. and {Sheehy}, C.~D. and {Staniszewski}, Z.~K. and {Steinbach}, B. and {Sudiwala}, R.~V. and {Teply}, G.~P. and {Thompson}, K.~L. and {Tolan}, J.~E. and {Tucker}, C. and {Turner}, A.~D. and {Vieregg}, A.~G. and {Weber}, A.~C. and {Wiebe}, D.~V. and {Willmert}, J. and {Wong}, C.~L. and {Wu}, W.~L.~K. and {Yoon}, K.~W.},
        title = "{BICEP2/Keck Array. VII. Matrix Based E/B Separation Applied to Bicep2 and the Keck Array}",
      journal = {\apj},
         year = 2016,
        month = jul,
       volume = {825},
       number = {1},
          eid = {66},
        pages = {66},
          doi = {10.3847/0004-637X/825/1/66},
archivePrefix = {arXiv},
       eprint = {1603.05976},
 primaryClass = {astro-ph.IM},
       adsurl = {https://ui.adsabs.harvard.edu/abs/2016ApJ...825...66B}
}

@ARTICLE{Clark:2015,
       author = {{Clark}, S.~E. and {Hill}, J. Colin and {Peek}, J.~E.~G. and {Putman}, M.~E. and {Babler}, B.~L.},
        title = "{Neutral Hydrogen Structures Trace Dust Polarization Angle: Implications for Cosmic Microwave Background Foregrounds}",
      journal = {\prl},
         year = 2015,
        month = dec,
       volume = {115},
       number = {24},
          eid = {241302},
        pages = {241302},
          doi = {10.1103/PhysRevLett.115.241302},
archivePrefix = {arXiv},
       eprint = {1508.07005},
 primaryClass = {astro-ph.CO},
       adsurl = {https://ui.adsabs.harvard.edu/abs/2015PhRvL.115x1302C}
}

@INPROCEEDINGS{Lam:2015,
       author = {{Lam}, Siu Kwan and {Pitrou}, Antoine and {Seibert}, Stanley},
        title = "{Numba: A LLVM-based Python JIT Compiler}",
    booktitle = {Proc. Second Workshop on the LLVM Compiler Infrastructure in HPC},
         year = 2015,
        month = nov,
        pages = {1-6},
          doi = {10.1145/2833157.2833162},
       adsurl = {https://ui.adsabs.harvard.edu/abs/2015llvm.confE...1L}
}

@ARTICLE{Remazeilles:2015,
       author = {{Remazeilles}, M. and {Dickinson}, C. and {Banday}, A.~J. and {Bigot-Sazy}, M. -A. and {Ghosh}, T.},
        title = "{An improved source-subtracted and destriped 408-MHz all-sky map}",
      journal = {\mnras},
         year = 2015,
        month = aug,
       volume = {451},
       number = {4},
        pages = {4311-4327},
          doi = {10.1093/mnras/stv1274},
archivePrefix = {arXiv},
       eprint = {1411.3628},
 primaryClass = {astro-ph.IM},
       adsurl = {https://ui.adsabs.harvard.edu/abs/2015MNRAS.451.4311R}
}

@ARTICLE{Tassis:2015,
       author = {{Tassis}, K. and {Pavlidou}, V.},
        title = "{Searching for inflationary B modes: can dust emission properties be extrapolated from 350 GHz to 150 GHz?}",
      journal = {\mnras},
         year = 2015,
        month = jul,
       volume = {451},
        pages = {L90-L94},
          doi = {10.1093/mnrasl/slv077},
archivePrefix = {arXiv},
       eprint = {1410.8136},
 primaryClass = {astro-ph.CO},
       adsurl = {https://ui.adsabs.harvard.edu/abs/2015MNRAS.451L..90T}
}

@ARTICLE{Schmidt:2015,
       author = {{Schmidt}, Samuel J. and {M{\'e}nard}, Brice and {Scranton}, Ryan and {Morrison}, Christopher B. and {Rahman}, Mubdi and {Hopkins}, Andrew M.},
        title = "{Inferring the redshift distribution of the cosmic infrared background}",
      journal = {\mnras},
         year = 2015,
        month = jan,
       volume = {446},
       number = {3},
        pages = {2696-2708},
          doi = {10.1093/mnras/stu2275},
archivePrefix = {arXiv},
       eprint = {1407.0031},
 primaryClass = {astro-ph.CO},
       adsurl = {https://ui.adsabs.harvard.edu/abs/2015MNRAS.446.2696S}
}

@ARTICLE{AstropyCollaboration:2013,
       author = {{Astropy Collaboration} and {Robitaille}, Thomas P. and {Tollerud}, Erik J. and {Greenfield}, Perry and {Droettboom}, Michael and {Bray}, Erik and {Aldcroft}, Tom and {Davis}, Matt and {Ginsburg}, Adam and {Price-Whelan}, Adrian M. and {Kerzendorf}, Wolfgang E. and {Conley}, Alexander and {Crighton}, Neil and {Barbary}, Kyle and {Muna}, Demitri and {Ferguson}, Henry and {Grollier}, Fr{\'e}d{\'e}ric and {Parikh}, Madhura M. and {Nair}, Prasanth H. and {Unther}, Hans M. and {Deil}, Christoph and {Woillez}, Julien and {Conseil}, Simon and {Kramer}, Roban and {Turner}, James E.~H. and {Singer}, Leo and {Fox}, Ryan and {Weaver}, Benjamin A. and {Zabalza}, Victor and {Edwards}, Zachary I. and {Azalee Bostroem}, K. and {Burke}, D.~J. and {Casey}, Andrew R. and {Crawford}, Steven M. and {Dencheva}, Nadia and {Ely}, Justin and {Jenness}, Tim and {Labrie}, Kathleen and {Lim}, Pey Lian and {Pierfederici}, Francesco and {Pontzen}, Andrew and {Ptak}, Andy and {Refsdal}, Brian and {Servillat}, Mathieu and {Streicher}, Ole},
        title = "{Astropy: A community Python package for astronomy}",
      journal = {\aap},
         year = 2013,
        month = oct,
       volume = {558},
          eid = {A33},
        pages = {A33},
          doi = {10.1051/0004-6361/201322068},
archivePrefix = {arXiv},
       eprint = {1307.6212},
 primaryClass = {astro-ph.IM},
       adsurl = {https://ui.adsabs.harvard.edu/abs/2013A&A...558A..33A}
}

@ARTICLE{Houde:2013,
       author = {{Houde}, Martin and {Hezareh}, Talayeh and {Jones}, Scott and {Rajabi}, Fereshte},
        title = "{Non-Zeeman Circular Polarization of Molecular Rotational Spectral Lines}",
      journal = {\apj},
         year = 2013,
        month = feb,
       volume = {764},
       number = {1},
          eid = {24},
        pages = {24},
          doi = {10.1088/0004-637X/764/1/24},
archivePrefix = {arXiv},
       eprint = {1212.2237},
 primaryClass = {astro-ph.GA},
       adsurl = {https://ui.adsabs.harvard.edu/abs/2013ApJ...764...24H}
}

@ARTICLE{Kogut:2012,
       author = {{Kogut}, A.},
        title = "{Synchrotron Spectral Curvature from 22 MHz to 23 GHz}",
      journal = {\apj},
         year = 2012,
        month = jul,
       volume = {753},
       number = {2},
          eid = {110},
        pages = {110},
          doi = {10.1088/0004-637X/753/2/110},
archivePrefix = {arXiv},
       eprint = {1205.4041},
 primaryClass = {astro-ph.GA},
       adsurl = {https://ui.adsabs.harvard.edu/abs/2012ApJ...753..110K}
}

@ARTICLE{Remazeilles:2011,
       author = {{Remazeilles}, Mathieu and {Delabrouille}, Jacques and {Cardoso}, Jean-Fran{\c{c}}ois},
        title = "{Foreground component separation with generalized Internal Linear Combination}",
      journal = {\mnras},
         year = 2011,
        month = nov,
       volume = {418},
       number = {1},
        pages = {467-476},
          doi = {10.1111/j.1365-2966.2011.19497.x},
archivePrefix = {arXiv},
       eprint = {1103.1166},
 primaryClass = {astro-ph.CO},
       adsurl = {https://ui.adsabs.harvard.edu/abs/2011MNRAS.418..467R}
}

@ARTICLE{Jaffe:2011,
       author = {{Jaffe}, T.~R. and {Banday}, A.~J. and {Leahy}, J.~P. and {Leach}, S. and {Strong}, A.~W.},
        title = "{Connecting synchrotron, cosmic rays and magnetic fields in the plane of the Galaxy}",
      journal = {\mnras},
         year = 2011,
        month = sep,
       volume = {416},
       number = {2},
        pages = {1152-1162},
          doi = {10.1111/j.1365-2966.2011.19114.x},
archivePrefix = {arXiv},
       eprint = {1105.5885},
 primaryClass = {astro-ph.GA},
       adsurl = {https://ui.adsabs.harvard.edu/abs/2011MNRAS.416.1152J}
}

@ARTICLE{Silsbee:2011,
       author = {{Silsbee}, Kedron and {Ali-Ha{\"\i}moud}, Yacine and {Hirata}, Christopher M.},
        title = "{Spinning dust emission: the effect of rotation around a non-principal axis}",
      journal = {\mnras},
         year = 2011,
        month = mar,
       volume = {411},
       number = {4},
        pages = {2750-2769},
          doi = {10.1111/j.1365-2966.2010.17882.x},
archivePrefix = {arXiv},
       eprint = {1003.4732},
 primaryClass = {astro-ph.GA},
       adsurl = {https://ui.adsabs.harvard.edu/abs/2011MNRAS.411.2750S}
}

@ARTICLE{vanderWalt:2011,
       author = {{van der Walt}, St{\'e}fan and {Colbert}, S. Chris and {Varoquaux}, Ga{\"e}l},
        title = "{The NumPy Array: A Structure for Efficient Numerical Computation}",
      journal = {Computing in Science and Engineering},
         year = 2011,
        month = mar,
       volume = {13},
       number = {2},
        pages = {22-30},
          doi = {10.1109/MCSE.2011.37},
archivePrefix = {arXiv},
       eprint = {1102.1523},
 primaryClass = {cs.MS},
       adsurl = {https://ui.adsabs.harvard.edu/abs/2011CSE....13b..22V}
}

@BOOK{Draine:2011,
       author = {{Draine}, Bruce T.},
        title = "{Physics of the Interstellar and Intergalactic Medium}",
         year = 2011,
       adsurl = {https://ui.adsabs.harvard.edu/abs/2011piim.book.....D}
}

@ARTICLE{Ali-Haimoud:2009,
       author = {{Ali-Ha{\"\i}moud}, Yacine and {Hirata}, Christopher M. and {Dickinson}, Clive},
        title = "{A refined model for spinning dust radiation}",
      journal = {\mnras},
         year = 2009,
        month = may,
       volume = {395},
       number = {2},
        pages = {1055-1078},
          doi = {10.1111/j.1365-2966.2009.14599.x},
archivePrefix = {arXiv},
       eprint = {0812.2904},
 primaryClass = {astro-ph},
       adsurl = {https://ui.adsabs.harvard.edu/abs/2009MNRAS.395.1055A}
}

@ARTICLE{Shetty:2009,
       author = {{Shetty}, Rahul and {Kauffmann}, Jens and {Schnee}, Scott and {Goodman}, Alyssa A.},
        title = "{The Effect of Noise on the Dust Temperature-Spectral Index Correlation}",
      journal = {\apj},
         year = 2009,
        month = may,
       volume = {696},
       number = {1},
        pages = {676-680},
          doi = {10.1088/0004-637X/696/1/676},
archivePrefix = {arXiv},
       eprint = {0902.0636},
 primaryClass = {astro-ph.GA},
       adsurl = {https://ui.adsabs.harvard.edu/abs/2009ApJ...696..676S}
}

@ARTICLE{Mantz:2008,
       author = {{Mantz}, Hubert and {Jacobs}, Karin and {Mecke}, Klaus},
        title = "{Utilizing Minkowski functionals for image analysis: a marching square algorithm}",
      journal = {Journal of Statistical Mechanics: Theory and Experiment},
         year = 2008,
        month = dec,
       volume = {2008},
       number = {12},
        pages = {12015},
          doi = {10.1088/1742-5468/2008/12/P12015},
       adsurl = {https://ui.adsabs.harvard.edu/abs/2008JSMTE..12..015M}
}

@ARTICLE{Miville-Deschenes:2008,
       author = {{Miville-Desch{\^e}nes}, M. -A. and {Ysard}, N. and {Lavabre}, A. and {Ponthieu}, N. and {Mac{\'\i}as-P{\'e}rez}, J.~F. and {Aumont}, J. and {Bernard}, J.~P.},
        title = "{Separation of anomalous and synchrotron emissions using WMAP polarization data}",
      journal = {\aap},
         year = 2008,
        month = nov,
       volume = {490},
       number = {3},
        pages = {1093-1102},
          doi = {10.1051/0004-6361:200809484},
archivePrefix = {arXiv},
       eprint = {0802.3345},
 primaryClass = {astro-ph},
       adsurl = {https://ui.adsabs.harvard.edu/abs/2008A&A...490.1093M}
}

@ARTICLE{Cortes:2008,
       author = {{Cortes}, P.~C. and {Crutcher}, R.~M. and {Shepherd}, D.~S. and {Bronfman}, L.},
        title = "{Interferometric Mapping of Magnetic Fields: The Massive Star-forming Region G34.4+0.23 MM}",
      journal = {\apj},
         year = 2008,
        month = mar,
       volume = {676},
       number = {1},
        pages = {464-471},
          doi = {10.1086/524355},
archivePrefix = {arXiv},
       eprint = {0801.4316},
 primaryClass = {astro-ph},
       adsurl = {https://ui.adsabs.harvard.edu/abs/2008ApJ...676..464C}
}

@ARTICLE{Hinshaw:2007,
       author = {{Hinshaw}, G. and {Nolta}, M.~R. and {Bennett}, C.~L. and {Bean}, R. and {Dor{\'e}}, O. and {Greason}, M.~R. and {Halpern}, M. and {Hill}, R.~S. and {Jarosik}, N. and {Kogut}, A. and {Komatsu}, E. and {Limon}, M. and {Odegard}, N. and {Meyer}, S.~S. and {Page}, L. and {Peiris}, H.~V. and {Spergel}, D.~N. and {Tucker}, G.~S. and {Verde}, L. and {Weiland}, J.~L. and {Wollack}, E. and {Wright}, E.~L.},
        title = "{Three-Year Wilkinson Microwave Anisotropy Probe (WMAP) Observations: Temperature Analysis}",
      journal = {\apjs},
         year = 2007,
        month = jun,
       volume = {170},
       number = {2},
        pages = {288-334},
          doi = {10.1086/513698},
archivePrefix = {arXiv},
       eprint = {astro-ph/0603451},
 primaryClass = {astro-ph},
       adsurl = {https://ui.adsabs.harvard.edu/abs/2007ApJS..170..288H}
}

@ARTICLE{Hunter:2007,
       author = {{Hunter}, John D.},
        title = "{Matplotlib: A 2D Graphics Environment}",
      journal = {Computing in Science and Engineering},
         year = 2007,
        month = may,
       volume = {9},
       number = {3},
        pages = {90-95},
          doi = {10.1109/MCSE.2007.55},
       adsurl = {https://ui.adsabs.harvard.edu/abs/2007CSE.....9...90H}
}

@ARTICLE{Gorski:2005,
       author = {{G{\'o}rski}, K.~M. and {Hivon}, E. and {Banday}, A.~J. and {Wandelt}, B.~D. and {Hansen}, F.~K. and {Reinecke}, M. and {Bartelmann}, M.},
        title = "{HEALPix: A Framework for High-Resolution Discretization and Fast Analysis of Data Distributed on the Sphere}",
      journal = {\apj},
         year = 2005,
        month = apr,
       volume = {622},
       number = {2},
        pages = {759-771},
          doi = {10.1086/427976},
archivePrefix = {arXiv},
       eprint = {astro-ph/0409513},
 primaryClass = {astro-ph},
       adsurl = {https://ui.adsabs.harvard.edu/abs/2005ApJ...622..759G}
}

@ARTICLE{Tristram:2005,
       author = {{Tristram}, M. and {Mac{\'\i}as-P{\'e}rez}, J.~F. and {Renault}, C. and {Santos}, D.},
        title = "{XSPECT, estimation of the angular power spectrum by computing cross-power spectra with analytical error bars}",
      journal = {\mnras},
         year = 2005,
        month = apr,
       volume = {358},
       number = {3},
        pages = {833-842},
          doi = {10.1111/j.1365-2966.2005.08760.x},
archivePrefix = {arXiv},
       eprint = {astro-ph/0405575},
 primaryClass = {astro-ph},
       adsurl = {https://ui.adsabs.harvard.edu/abs/2005MNRAS.358..833T}
}

@ARTICLE{Greaves:2002,
       author = {{Greaves}, J.~S. and {Holland}, W.~S. and {Dent}, W.~R.~F.},
        title = "{Magnetic Fields in Gas Flows near the Galactic Center}",
      journal = {\apj},
         year = 2002,
        month = oct,
       volume = {578},
       number = {1},
        pages = {224-228},
          doi = {10.1086/342345},
       adsurl = {https://ui.adsabs.harvard.edu/abs/2002ApJ...578..224G}
}

@ARTICLE{Hivon:2002,
       author = {{Hivon}, Eric and {G{\'o}rski}, Krzysztof M. and {Netterfield}, C. Barth and {Crill}, Brendan P. and {Prunet}, Simon and {Hansen}, Frode},
        title = "{MASTER of the Cosmic Microwave Background Anisotropy Power Spectrum: A Fast Method for Statistical Analysis of Large and Complex Cosmic Microwave Background Data Sets}",
      journal = {\apj},
         year = 2002,
        month = mar,
       volume = {567},
       number = {1},
        pages = {2-17},
          doi = {10.1086/338126},
archivePrefix = {arXiv},
       eprint = {astro-ph/0105302},
 primaryClass = {astro-ph},
       adsurl = {https://ui.adsabs.harvard.edu/abs/2002ApJ...567....2H}
}

@ARTICLE{Dame:2001,
       author = {{Dame}, T.~M. and {Hartmann}, Dap and {Thaddeus}, P.},
        title = "{The Milky Way in Molecular Clouds: A New Complete CO Survey}",
      journal = {\apj},
         year = 2001,
        month = feb,
       volume = {547},
       number = {2},
        pages = {792-813},
          doi = {10.1086/318388},
archivePrefix = {arXiv},
       eprint = {astro-ph/0009217},
 primaryClass = {astro-ph},
       adsurl = {https://ui.adsabs.harvard.edu/abs/2001ApJ...547..792D}
}

@ARTICLE{Greaves:1999,
       author = {{Greaves}, J.~S. and {Holland}, W.~S. and {Friberg}, P. and {Dent}, W.~R.~F.},
        title = "{Polarized CO Emission from Molecular Clouds}",
      journal = {\apjl},
         year = 1999,
        month = feb,
       volume = {512},
       number = {2},
        pages = {L139-L142},
          doi = {10.1086/311888},
archivePrefix = {arXiv},
       eprint = {astro-ph/9812428},
 primaryClass = {astro-ph},
       adsurl = {https://ui.adsabs.harvard.edu/abs/1999ApJ...512L.139G}
}

@ARTICLE{Draine:1998a,
       author = {{Draine}, B.~T. and {Lazarian}, A.},
        title = "{Electric Dipole Radiation from Spinning Dust Grains}",
      journal = {\apj},
         year = 1998,
        month = nov,
       volume = {508},
       number = {1},
        pages = {157-179},
          doi = {10.1086/306387},
archivePrefix = {arXiv},
       eprint = {astro-ph/9802239},
 primaryClass = {astro-ph},
       adsurl = {https://ui.adsabs.harvard.edu/abs/1998ApJ...508..157D}
}

@ARTICLE{Draine:1998b,
       author = {{Draine}, B.~T. and {Lazarian}, A.},
        title = "{Diffuse Galactic Emission from Spinning Dust Grains}",
      journal = {\apjl},
         year = 1998,
        month = feb,
       volume = {494},
       number = {1},
        pages = {L19-L22},
          doi = {10.1086/311167},
archivePrefix = {arXiv},
       eprint = {astro-ph/9710152},
 primaryClass = {astro-ph},
       adsurl = {https://ui.adsabs.harvard.edu/abs/1998ApJ...494L..19D}
}

@ARTICLE{Bennett:1994,
       author = {{Bennett}, C.~L. and {Fixsen}, D.~J. and {Hinshaw}, G. and {Mather}, J.~C. and {Moseley}, S.~H. and {Wright}, E.~L. and {Eplee}, Jr., R.~E. and {Gales}, J. and {Hewagama}, T. and {Isaacman}, R.~B. and {Shafer}, R.~A. and {Turpie}, K.},
        title = "{Morphology of the Interstellar Cooling Lines Detected by COBE}",
      journal = {\apj},
         year = 1994,
        month = oct,
       volume = {434},
        pages = {587},
          doi = {10.1086/174761},
archivePrefix = {arXiv},
       eprint = {astro-ph/9311032},
 primaryClass = {astro-ph},
       adsurl = {https://ui.adsabs.harvard.edu/abs/1994ApJ...434..587B}
}

@BOOK{Rybicki:1986,
       author = {{Rybicki}, George B. and {Lightman}, Alan P.},
        title = "{Radiative Processes in Astrophysics}",
         year = 1986,
       adsurl = {https://ui.adsabs.harvard.edu/abs/1986rpa..book.....R}
}

@ARTICLE{Haslam:1982,
       author = {{Haslam}, C.~G.~T. and {Salter}, C.~J. and {Stoffel}, H. and {Wilson}, W.~E.},
        title = "{A 408-MHZ All-Sky Continuum Survey. II. The Atlas of Contour Maps}",
      journal = {\aaps},
         year = 1982,
        month = jan,
       volume = {47},
        pages = {1},
       adsurl = {https://ui.adsabs.harvard.edu/abs/1982A&AS...47....1H}
}

@ARTICLE{Goldreich:1981,
       author = {{Goldreich}, P. and {Kylafis}, N.~D.},
        title = "{On mapping the magnetic field direction in molecular clouds by polarization measurements}",
      journal = {\apjl},
         year = 1981,
        month = jan,
       volume = {243},
        pages = {L75-L78},
          doi = {10.1086/183446},
       adsurl = {https://ui.adsabs.harvard.edu/abs/1981ApJ...243L..75G}
}

@BOOK{Landau:1975,
       author = {{Landau}, Lev Davidovich and {Lifshitz}, E.~M.},
        title = "{The classical theory of fields}",
         year = 1975,
       adsurl = {https://ui.adsabs.harvard.edu/abs/1975ctf..book.....L}
}

@ARTICLE{planck2016-l03, 
author = {{\sorthelp{Planck Collaboration 2018C}}{Planck Collaboration III}},
title = "{\textit{Planck} 2018 results. III. High Frequency Instrument data
 processing}",
journal = {\aap},
archivePrefix = "arXiv",
eprint = {1807.06207},
year = 2020,
volume = 641,
pages = {A3},
doi = {10.1051/0004-6361/201832909}
}

@ARTICLE{planck2016-l04, 
author = {{\sorthelp{Planck Collaboration 2018D}}{Planck Collaboration IV}},
title = "{\textit{Planck} 2018 results. IV. Diffuse component separation}",
journal = {\aap},
archivePrefix = "arXiv",
eprint = {1807.06208},
year = 2020,
volume = 641,
pages = {A4},
doi = {10.1051/0004-6361/201833881}
}

@ARTICLE{planck2016-l11A,
author = {{\sorthelp{Planck Collaboration 2018K}}{Planck Collaboration XI}},
title = "{\textit{Planck} 2018 results. XI. Polarized dust foregrounds}",
journal = {\aap},
archivePrefix = "arXiv",
eprint = {1801.04945},
year = 2020,
volume = 641,
pages = {A11},
doi = {10.1051/0004-6361/201832618}
}

@ARTICLE{planck2016-l11B,
author = {{\sorthelp{Planck Collaboration 2018L}}{Planck Collaboration XII}},
title = "{\textit{Planck} 2018 results. XII. Galactic astrophysics using
 polarized dust emission}",
journal = {\aap},
archivePrefix = "arXiv",
eprint = {1807.06212},
year = 2020,
volume = 641,
pages = {A12},
doi = {10.1051/0004-6361/201833885}
}

@ARTICLE{planck2014-a11,
author = {{\sorthelp{Planck Collaboration 2015I}}{Planck Collaboration IX}},
title = "{\textit{Planck} 2015 results. IX. Diffuse component separation:
 CMB maps}",
journal = {\aap},
archivePrefix = "arXiv",
eprint = {1502.05956},
year = 2016,
volume = 594,
pages = {A9},
doi = {10.1051/0004-6361/201525936}
}

@ARTICLE{planck2014-a12,
author = {{\sorthelp{Planck Collaboration 2015J}}{Planck Collaboration X}},
title = "{\textit{Planck} 2015 results. X. Diffuse component separation:
 Foreground maps}",
journal = {\aap},
archivePrefix = "arXiv",
eprint = {1502.01588},
year = 2016,
volume = 594,
pages = {A10},
doi = {10.1051/0004-6361/201525967}
}

@ARTICLE{planck2013-p03d,
author = {{\sorthelp{Planck Collaboration 2014I}}{Planck Collaboration IX}},
title = "{\textit{Planck} 2013 results. IX. HFI spectral response}",
journal = {\aap},
archivePrefix = "arXiv",
eprint = {1303.5070},
year = 2014,
volume = 571,
pages = {A9},
doi = {10.1051/0004-6361/201321531}
}

@ARTICLE{planck2013-p03a,
author = {{\sorthelp{Planck Collaboration 2014M}}{Planck Collaboration XIII}},
title = "{\textit{Planck} 2013 results. XIII. Galactic CO emission}",
journal = {\aap},
archivePrefix = "arXiv",
eprint = {1303.5073},
year = 2014,
volume = 571,
pages = {A13},
doi = {10.1051/0004-6361/201321553}
}

@ARTICLE{planck2014-XIX,
author = {{\sorthelp{Planck Collaboration IntS}}{Planck Collaboration Int.
 XIX}},
title = "{\textit{Planck} intermediate results. XIX. An overview of the
 polarized thermal emission from Galactic dust}",
journal = {\aap},
archivePrefix = "arXiv",
eprint = {1405.0871},
year = 2015,
volume = 576,
pages = {A104},
doi = {10.1051/0004-6361/201424082}
}

@ARTICLE{planck2014-XXX,
author = {{\sorthelp{Planck Collaboration IntZE}}{Planck Collaboration Int.
 XXX}},
title = "{\textit{Planck} intermediate results. XXX. The angular power spectrum
 of polarized dust emission at intermediate and high Galactic latitudes}",
journal = {\aap},
archivePrefix = "arXiv",
eprint = {1409.5738},
year = 2016,
volume = 586,
pages = {A133},
doi = {10.1051/0004-6361/201425034}
}

@ARTICLE{planck2016-XLVIII,
author = {{\sorthelp{Planck Collaboration IntZW}}{Planck Collaboration Int.
 XLVIII}},
title = "{\textit{Planck} intermediate results. XLVIII. Disentangling Galactic
 dust emission and cosmic infrared background anisotropies}",
journal = {\aap},
archivePrefix = "arXiv",
eprint = {1605.09387},
year = 2016,
volume = 596,
pages = {A109},
doi = {10.1051/0004-6361/201629022}
}

@ARTICLE{planck2016-L,
author = {{\sorthelp{Planck Collaboration IntZY}}{Planck Collaboration Int.
 L}},
title = "{\textit{Planck} intermediate results. L. Evidence of spatial
 variation of the polarized thermal dust spectral energy distribution and
 implications for CMB \textit{B}-mode analysis}",
journal = {\aap},
archivePrefix = "arXiv",
eprint = {1606.07335},
year = 2017,
volume = 599,
pages = {A51},
doi = {10.1051/0004-6361/201629164}
}

@ARTICLE{delabrouille2012,
author = {{Delabrouille}, J. and {Betoule}, M. and {Melin}, J.-B. and
 {Miville-Desch{\^e}nes}, M.-A. and {Gonzalez-Nuevo}, J. and
 {Le Jeune}, M. and {Castex}, G. and {de Zotti}, G. and {Basak}, S. and
 {Ashdown}, M. and {Aumont}, J. and {Baccigalupi}, C. and {Banday}, A. and
 {Bernard}, J.-P. and {Bouchet}, F.~R. and {Clements}, D.~L. and
 {da Silva}, A. and {Dickinson}, C. and {Dodu}, F. and {Dolag}, K. and
 {Elsner}, F. and {Fauvet}, L. and {Fa{\"y}}, G. and {Giardino}, G. and
 {Leach}, S. and {Lesgourgues}, J. and {Liguori}, M. and {Macias-Perez}, J.~F.
 and {Massardi}, M. and {Matarrese}, S. and {Mazzotta}, P. and {Montier}, L.
 and {Mottet}, S. and {Paladini}, R. and {Partridge}, B. and {Piffaretti}, R.
 and {Prezeau}, G. and {Prunet}, S. and {Ricciardi}, S. and {Roman}, M. and
 {Schaefer}, B. and {Toffolatti}, L.},
title = "{The pre-launch Planck Sky Model: a model of sky emission at
 submillimetre to centimetre wavelengths}",
journal = {\aap},
archivePrefix = "arXiv",
eprint = {1207.3675},
primaryClass = "astro-ph.CO",
year = 2013,
month = may,
volume = 553,
eid = {A96},
pages = {A96},
doi = {10.1051/0004-6361/201220019},
adsurl = {http://adsabs.harvard.edu/abs/2013A%26A...553A..96D}
}

\end{document}